\documentclass[11pt,a4paper]{article}
\usepackage[T1]{fontenc}
\usepackage[latin9]{inputenc}
\usepackage{rotating}
\usepackage{booktabs}
\usepackage{amsmath}
\usepackage{amssymb}
\usepackage{graphicx}
\usepackage{esint}

\makeatletter

\providecommand{\tabularnewline}{\\}

\pdfoutput=1 

\usepackage{jheppub}

\usepackage{etoolbox}
    
    \patchcmd{\maketitle}{\@fpheader}{}{}{}

\providecommand{\tabularnewline}{\\}

\usepackage{amsfonts}

\usepackage{bbm}

\title{\boldmath Generalized Black Holes in Three-dimensional Spacetime}

\author[a,b]{Claudio Bunster,}
\author[a,c]{Marc Henneaux,}
\author[a]{Alfredo P\'{e}rez,}
\author[a]{David Tempo,}
\author[a,b]{and Ricardo Troncoso}

\affiliation[a]{Centro de Estudios Cient\'{i}ficos (CECs), Casilla 1469, Valdivia,
Chile.}
\affiliation[b]{Universidad Andr\'{e}s Bello, Av. Republica 440, Santiago, Chile.}
\affiliation[c]{Universit\'{e} Libre de Bruxelles and International Solvay Institutes,
ULB Campus Plaine C.P.231, B-1050 Bruxelles, Belgium.}

\emailAdd{bunster@cecs.cl}
\emailAdd{henneaux@ulb.ac.be}
\emailAdd{aperez@cecs.cl}
\emailAdd{tempo@cecs.cl}
\emailAdd{troncoso@cecs.cl}

\preprint{CECS-PHY-13/07}

\abstract{Three-dimensional spacetime with a negative cosmological constant
has proven to be a remarkably fertile ground for the study of gravity
and higher spin fields. The theory is topological and, since there are no propagating field degrees of freedom, the asymptotic symmetries become all the more crucial. For pure (2+1) gravity they consist of two copies of the Virasoro algebra.
There exists a black hole which may be endowed with all
the corresponding charges. The pure (2+1) gravity theory may
be reformulated in terms of two Chern-Simons connections for $sl\left(2,\mathbb{R}\right)$. This permits an immediate generalization which may be interpreted
as containing gravity and a finite number of higher spin fields. The generalization is achieved by replacing $sl\left(2,\mathbb{R}\right)$
by $sl\left(3,\mathbb{R}\right)$ or, more generally, by $sl\left(N,\mathbb{R}\right)$.
The asymptotic symmetries are then two copies of the so-called $W_{N}$
algebra, which contains the Virasoro algebra as a subalgebra. The
question then arises as to whether there exists a generalization
of the standard pure gravity (2+1) black hole which would be endowed
with all the $W_{N}$ charges. Since the generalized
Chern-Simons theory does not admit a direct metric interpretation, one must define  the black hole in Euclidean
spacetime through its thermal properties, and then continue to Lorentzian
spacetime. The original pioneering proposal of a black hole along this line for $N=3$ turns out, as shown in this paper, to actually belong to the so called ``diagonal embedding''
of $sl\left(2,\mathbb{R}\right)$ in $sl\left(3,\mathbb{R}\right)$,
and it is therefore endowed with charges of lower rather than
higher spins. In contradistinction, we exhibit herein the most general black hole
which belongs to the ``principal embedding''. It is endowed
with higher spin charges, and possesses two copies of $W_{3}$
as its asymptotic symmetries. The most general diagonal embedding black hole is
studied in detail as well, in a way in which its lower spin charges are
clearly displayed. The extension to $N>3$ is also discussed.  A general formula for the entropy of a generalized
black hole is obtained in terms of the on-shell holonomies. The relationship
between the asymptotic symmetries and the chemical potentials is exhibited,
and the equivalence of the different thermodynamical ensembles is
discussed. A self-contained account of the background necessary
to substantiate the claims made in the paper is included.
}

\makeatother

\begin{document}
\maketitle \flushbottom 
\begin{quotation}
\emph{``The field equations and the boundary conditions are inextricably
connected and the latter can in no way be considered less important
than the former'' }\cite{Fock}\emph{.} 
\end{quotation}

\section{Introduction\label{sec:Introduction}}

Three-dimensional spacetime has proven to be a remarkably fertile
ground for the study of gravity and higher spin fields. In spite of
the fact that the gravitational field has no propagating degrees of
freedom its asymptotic structure is extraordinarily rich, much more
so than that of its (3+1) counterpart. In the (2+1) case the symmetry
algebra at space-like infinity of asymptotically anti-de Sitter spaces
consists of two copies of the infinite-dimensional Virasoro algebra
\cite{Brown:1986nw}. On the other hand, in contradistinction, in
(3+1) dimensions the asymptotic algebra is only $so\left(3,2\right)$
\cite{Abbott:1981ff,Henneaux:1985tv}.

For a gauge theory, the asymptotic symmetries are of fundamental importance.
They are the physical symmetries of the theory. These are symmetries
that alter the state of the system when they act on it, and therefore
have a non-trivial physical effect. They are of the same form as ordinary
(``proper'') gauge transformations, but differ from them in that
they do not became the identity at infinity, and have non-vanishing
generators (``global charges'') \cite{Regge:1974zd}, \cite{Benguria:1976in}.
The asymptotic symmetries are invariant under proper gauge transformations,
and they subsist after the physically irrelevant gauge freedom has
been eliminated, for example by means of gauge conditions.

A given solution of the equations of motion is in general not invariant
under all the asymptotic symmetries, rather, it is covariant\emph{
}under them, i.e.\emph{ under the action of an asymptotic symmetry,
a solution is mapped onto another solution which is generically physically
different from the original one}.

When one formulates the theory in terms of an action principle, the
boundary conditions at infinity, which by construction are left invariant
by the asymptotic symmetries, must be given once and for all. They
are not only obeyed by the solutions of the equations of motion, but
also hold ``off-shell'' because they are part of the definition
of the function space on which the action functional is defined. Inequivalent
boundary conditions yield physically distinct theories, even if the
functional form of the action is the same.

The crucial role of the asymptotic symmetries becomes even more dramatic
in the case of a topological theory such as (2+1) gravity, because
then there are no local bulk degrees of freedom, and the entire dynamical
content is captured by holonomies and boundary degrees of freedom.

In view of the above, it was natural to investigate the asymptotic
symmetry algebra in generalizations of (2+1) gravity which included
the (2+1) analog of ``higher spins''%
\footnote{We have written ``higher spins'' with quotation marks because the
notion of ``spin'' needs to be made precise. We are dealing with
massless fields. In the familiar case of ($3+1$)-dimensional spacetime,
``spin'' really means ``helicity'' and labels the representation
of the little group $SO(2)$ of rotations around the spatial momentum
vector. In $2+1$ dimensions there is no little group, and hence the
concept of helicity is empty, so what is meant by spin here is different.
It is the following. The fields fall into finite-dimensional representations
of $sl(2,\mathbb{R})$, which is a subalgebra of the algebra out of
which the Chern-Simons theory is built. These representations are
characterized by a quantum number $s$, the ``$sl(2,\mathbb{R})$-spin'',
which can be an integer or a half-integer. There is another notion
of spin associated with the conformal symmetry at infinity, which
is the conformal weight $J$. The two notions are related through
$J=s+1$, as we review in Appendix \ref{WeightVersusSpin}. Unless
otherwise specified when we use the term ``spin'' below, it will
mean the conformal weight $J$. With this understanding, we shall
dispense with the quotation marks when referring to spin from now
on. As there are two copies of $sl(2,\mathbb{R})$ and two copies
of the Virasoro algebra, one can in fact introduce a spin for each
copy. Note that pure gravity, or gravity with ``lower spin fields''
have only representations with $s\leq1$, i.e. $J\leq2$. Higher spins
means $s>1$ or $J>2$.%
}. These generalizations were constructed starting from the reformulation
of the standard (2+1) Einstein theory in terms of two Chern-Simons
connections for $sl\left(2,\mathbb{R}\right)$ \cite{Achucarro:1987vz,Witten:1988hc},
by replacing $sl\left(2,\mathbb{R}\right)$ by a ``higher spin algebra''
that contained it \cite{Aragone:1983sz,Vasiliev:1986qx,Blencowe,BBS,Vasiliev:1995dn}.
This algebra can be finite-dimensional ($sl\left(N,\mathbb{R}\right)$)
or infinite-dimensional ($hs\left(\lambda\right)$). It was concluded
that the asymptotic symmetry algebra is then enlarged from two copies
of the Virasoro algebra to two copies of the so-called $W$-algebras
each of which has one Virasoro algebra embedded in it \cite{Henneaux-HS,Theisen-HS,GH}.
The enlargement of the gauge algebra preserves the fundamental simplicity
inherent in the absence of propagating degrees of freedom, i.e., the
theories remain topological.

When the Chern-Simons gauge algebra is enlarged to a $W$-algebra,
an interesting feature arises, which is that there are inequivalent
non-trivial embeddings of the gravitational subalgebra $sl(2,\mathbb{R})\oplus sl(2,\mathbb{R})$
in the Chern-Simons gauge algebra%
\footnote{Two embeddings are inequivalent if the matrices representing the $sl(2,\mathbb{R})$
algebra generators in each embedding are not related by a similarity
transformation.%
}. This phenomenon was described in the context of Hamiltonian reduction,
independently of anti-de Sitter gravity \cite{BaisTjinVanDriel,Bilal,WReview,HR4}.
Its relevance in the higher spin context was discussed in \cite{GK,AGKP,Campoleoni-HS,CM,Castro:2012bc,Grumiller}.
The inequivalent embeddings lead to different theories at infinity
with different asymptotic symmetries. What selects the embedding are
the asymptotic conditions.

The existence of inequivalent embeddings appears in the present context
already in the spin-three case, where the Chern-Simon gauge algebra
is $sl(3,\mathbb{R})\oplus sl(3,\mathbb{R})$. For this reason, we
shall consider for definiteness this simplest case. We will indicate
afterwards how our analysis extends to theories containing also spins
$>3$.

For $sl(3,\mathbb{R})$, there are just two inequivalent nontrivial
embeddings of $sl(2,\mathbb{R})$: (i) the ``principal embedding'',
corresponding to the decomposition $sl(3,\mathbb{R})=D_{1}\oplus D_{2}$
under the adjoint action of $sl(2,\mathbb{R})$, where $D_{s}$ is
the $sl(2,\mathbb{R})$ spin-$s$ representation; and (ii) the ``diagonal
embedding'', corresponding to the decomposition $sl(3,\mathbb{R})=D_{1}\oplus2D_{\frac{1}{2}}\oplus D_{0}$.
Only the principal embedding defines a higher-spin theory since the
diagonal embedding contains conformal spin smaller than or equal to
2. The asymptotic symmetry algebra corresponding to the principal
embedding is denoted as $W_{3}$, while its diagonal embedding counterpart
is denoted as $W_{3}^{\left(2\right)}$.

Another important aspect of the richness of (2+1) gravity is the existence
of the (2+1) black hole \cite{BTZ,BHTZ}. In the form in which the
solution is normally exhibited, either in the metric formalism or
in the Chern-Simons one, it contains only two charges, the mass and
the angular momentum, which are related to the Fourier zero modes
$\mathcal{L}_{0}^{\pm}$ of the two Virasoro algebras. One may say
that this is the black hole in the ``rest frame''. One may obtain
from it the most general black hole endowed with all the charges by
acting on the black hole at rest with a generic element of the asymptotic
symmetry algebra. This is just the analog of setting, say, a (3+1)
Kerr black hole in motion by applying to it a boost, and thus endowing
it with linear momentum in addition to mass and angular momentum.
The moving black hole is of course physically different from the one
at rest, which illustrates the fact that the action of the asymptotic
symmetry algebra changes the physical state.

The question naturally arises as to whether there exists a generalized
(2+1) black hole which with is capable of carrying ``hair'' stemming
from the new charges that are present in the $W$-algebras in addition
to the gravitational charges stemming for the Virasoro algebras.

A generalized $sl(3,\mathbb{R})$-black hole was first proposed in
the pioneering work \cite{GK,AGKP} that initiated the study of generalized
(2+1) black holes. It was thought that the black hole in refs. \cite{GK,AGKP}
was associated with the principal embedding, that its asymptotic symmetry
algebra was $W_{3}$, and that hence it was endowed with charges of
spin 2 and 3. However this interpretation leads to conflicting results
when its entropy was evaluated \cite{GK,AGKP,KP,Gaberdiel:2012yb,PTT1,CFPT2,David:2012iu,PTT2,deBoer:2013gz,KU,dBJ2,ACI,CS,CJS,PTTreview}.
The entropy paradox is resolved when one realizes that, as shown in
this paper, the black hole in refs. \cite{GK,AGKP} is actually a
special case of a $W_{3}^{\left(2\right)}$ black hole and it is therefore
endowed, in addition to the gravitational charge of spin 2, with \emph{lower}
spin charges 1 and 3/2 rather than with the higher spin 3 charge.
The black hole in refs. \cite{GK,AGKP} turns out to be a lower spin
black hole in disguise because the chemical potentials were introduced
in a non canonical form in its derivation.

The issue of finding a higher spin black hole therefore remained an
open one. It is settled herein: we present the most general black
hole associated with the principal embedding. It possesses two copies
of $W_{3}$ as its asymptotic symmetry algebra, and is endowed with
charges of spin 2 and 3. In addition we present the most general black
hole associated with the diagonal embedding, of which the black hole
in refs. \cite{GK,AGKP} is a particular case.

The plan of the paper is the following: section \ref{sec:The-Chern-Simons-formulation}
reviews the Chern-Simons formulation of the pure gravity (2+1) black
hole. It is discussed in particular how the black hole is defined
through its thermal properties, and without reference to a metric,
which requires to consider the Euclidean formulation as the more fundamental
one. This point of view is optional for the pure gravity (2+1) black
hole, but it is mandatory for the generalizations considered herein
because then the metric does not appear naturally. A general formula
for the entropy of the black hole is obtained in terms of the on-shell
holonomies. The formula is first derived for $sl(2,\mathbb{R})$ and
then extended to $sl(N,\mathbb{R})$. Next, section \ref{HS} presents
the principal embedding $W_{3}$ black hole, emphasizing the boundary
conditions that define it and showing that through them, the black
hole is endowed with spin-2 and spin-3 charges. Its thermodynamics
is thoroughly studied. Section \ref{W32-section} is devoted to the
corresponding analysis for the diagonal embedding $W_{3}^{\left(2\right)}$
black hole. The quantum mechanical difficulties of the field theory
associated with the diagonal embedding \cite{Castro:2012bc,Grumiller}
are not an obstacle for this semiclassical study, which we deem necessary
for dealing thoroughly with the problem at hand. Section \ref{sec:Extension-to-higher}
outlines how the analysis for $N=3$ is extended to higher $N$. Finally
\ref{sec:Concluding-remarks} is devoted to concluding remarks. Two
Appendices are included. The first one provides the necessary background
to make the analysis of the paper self-contained, whereas the second
one discusses the relationship with previous analysis of generalized
(2+1) black holes. Some of the main properties of the pure gravity
black hole and its generalizations are compared and contrasted in
Table \ref{sec:table}.

Although subjects such as holography and conjectured CFT dualities
are not dealt with in the present paper, it is hoped that the self-contained
discussion presented herein may be useful as a beacon for incursions
into those territories.

\newpage

\section*{\label{sec:table}}



\begin{small}

\centering 

\setlength{\tabcolsep}{12.0 pt}

\begin{sideways}
\begin{tabular}{ccccc}
\multicolumn{5}{c}{Table 1. Pure gravity black hole and its generalizations compared
and contrasted}\tabularnewline
\midrule 
\addlinespace
 &  &  &  & \tabularnewline
 & \multicolumn{2}{c}{Pure gravity (2+1) black hole} & \multicolumn{2}{c}{Generalized (2+1) black holes }\tabularnewline
 &  &  &  & \tabularnewline
 & \multicolumn{2}{c}{} & Principal embedding\;  & \;Diagonal embedding \tabularnewline
\midrule 
\addlinespace
 &  &  &  & \tabularnewline
Formulation  & Metric  & Chern-Simons  & Only Chern-Simons  & Only Chern-Simons \tabularnewline
 &  &  &  & \tabularnewline
Lorentzian field  & \;\; \;\;Spacetime metric\;\; \;\;  & \;\; \;\;Two connections\;\; \;\;  & \;\; \;\;Two connections \;\;\;\;  & \;\;\;\; Two connections\;\;\;\; \tabularnewline
 & $ds_{L}^{2}=g_{\mu\nu}^{L}dx^{\mu}dx^{\nu}$,  & $A^{\pm}=A_{t}^{\pm}dt$  & $A^{\pm}=A_{t}^{\pm}dt$  & $A^{\pm}=A_{t}^{\pm}dt$\tabularnewline
 & signature $(-,+,+)$  & $\;+A_{r}^{\pm}dr+A_{\varphi}^{\pm}d\varphi$  & $\;+A_{r}^{\pm}dr+A_{\varphi}^{\pm}d\varphi$  & $\;+A_{r}^{\pm}dr+A_{\varphi}^{\pm}d\varphi$\tabularnewline
 &  & for $sl\left(2,\mathbb{R}\right)$  & for $sl\left(3,\mathbb{R}\right)$  & for $sl\left(3,\mathbb{R}\right)$\tabularnewline
 &  &  &  & \tabularnewline
Euclidean field  & \;\;Spacetime metric \;\; & \;\;One connection \;\; & \;\;One connection \;\; & \;\;One connection\;\;\tabularnewline
 & $ds_{E}^{2}=g_{\mu\nu}^{E}dx^{\mu}dx^{\nu}$  & $A=A_{\tau}d\tau$  & $A=A_{\tau}d\tau$  & $A=A_{\tau}d\tau$\tabularnewline
 & signature $(+,+,+)$  & $\;+A_{r}dr+A_{\varphi}d\varphi$  & $\;+A_{r}dr+A_{\varphi}d\varphi$  & $\;+A_{r}dr+A_{\varphi}d\varphi$\tabularnewline
 &  & for $sl\left(2,\mathbb{C}\right)$  & for $sl\left(3,\mathbb{C}\right)$  & for $sl\left(3,\mathbb{C}\right)$\tabularnewline
Relation between  &  &  &  & \tabularnewline
Euclidean and  & \;\;$ds_{E}^{2}=ds_{L}^{2}$ \;\; & \;\;$A=A^{+}$ \;\; & \;\;$A=A^{+}$ \;\; & \;\;$A=A^{+}$\;\;\tabularnewline
Lorentzian fields,  &  & $-A^{\dagger}=A^{-}$  & $-A^{\dagger}=A^{-}$  & $-A^{\dagger}=A^{-}$\tabularnewline
with $\tau=it$  &  &  &  & \tabularnewline
 &  &  &  & \tabularnewline
Topology of  & \;\;$\mathbb{R}^{2}\times S^{1}$ \;\; & \;\;$\mathbb{R}^{2}\times S^{1}$ \;\; & \;\;$\mathbb{R}^{2}\times S^{1}$ \;\; & \;\;$\mathbb{R}^{2}\times S^{1}$ \;\;\tabularnewline
Euclidean  & (solid torus)  & (solid torus)  & (solid torus)  & (solid torus)\tabularnewline
spacetime  &  &  &  & \tabularnewline
 &  &  &  & \tabularnewline
Asymptotic  &  &  &  & \tabularnewline
symmetries  & \;\;Two Virasoro \;\; & \;\;Two Virasoro \;\; & \;\;Two $W_{3}$ \;\; & \;\;Two $W_{3}^{\left(2\right)}$\;\;\tabularnewline
of Lorentzian  & algebras  & algebras  & algebras  & algebras\tabularnewline
field  &  &  &  & \tabularnewline
 &  &  &  & \tabularnewline
\bottomrule
\end{tabular}
\end{sideways}

\begin{sideways}
\begin{tabular}{ccccc}
\multicolumn{5}{c}{Table 1 (\emph{continued})}\tabularnewline
\hline 
 &  &  &  & \tabularnewline
 & \multicolumn{2}{c}{Pure gravity (2+1) black hole} & \multicolumn{2}{c}{Generalized (2+1) black holes }\tabularnewline
 &  &  &  & \tabularnewline
 &  &  & Principal embedding\;  & \;Diagonal embedding \tabularnewline
\hline 
 &  &  &  & \tabularnewline
Black hole charges  &  &  &  & \tabularnewline
in the ``rest frame''  & $\mathcal{L}_{0}^{\pm}$  & $\mathcal{L}_{0}^{\pm}$  & $\mathcal{L}_{0}^{\pm},\mathcal{W}_{0}^{\pm}$  & $\hat{\mathcal{L}}_{0}^{\pm},\psi_{0[+]}^{\pm},\psi_{0[-]}^{\pm},\mathcal{U}_{0}^{\pm}$\tabularnewline
and their  & spin 2  & spin 2  & spins $2,3$  & spins $2,\frac{3}{2},\frac{3}{2},1$\tabularnewline
conformal spin  &  &  &  & \tabularnewline
 &  &  &  & \tabularnewline
Black hole  & \;\;$\mathcal{M}=\frac{1}{\ell}\left(\mathcal{L}_{0}^{+}+\mathcal{L}_{0}^{-}\right)$\;\;  & \;\;$\mathcal{M}=\frac{1}{\ell}\left(\mathcal{L}_{0}^{+}+\mathcal{L}_{0}^{-}\right)$\;\;  & \;\;$\mathcal{M}=\frac{1}{\ell}\left(\mathcal{L}_{0}^{+}+\mathcal{L}_{0}^{-}\right)$\;\;  & \;\;$\mathcal{M}=\frac{1}{\ell}\left(\hat{\mathcal{L}}_{0}^{+}+\hat{\mathcal{L}}_{0}^{-}\right)$\;\tabularnewline
mass and inverse  & $\beta=N_{\text{Lor}}\left(\infty\right)$  & $\beta=\frac{\ell}{4\pi}\left(\xi_{0}^{+}+\xi_{0}^{-}\right)$  & $\beta=\frac{\ell}{4\pi}\left(\xi_{0}^{+}+\xi_{0}^{-}\right)$  & $\beta=\frac{\ell}{4\pi}\left(\hat{\xi}_{0}^{+}+\hat{\xi}_{0}^{-}\right)$\tabularnewline
temperature  &  &  &  & \tabularnewline
 &  &  &  & \tabularnewline
Chemical  & $\beta\mu_{\mathcal{J}_{\text{Lor}}}=N_{\text{Lor}}^{\varphi}\left(\infty\right)$  & $\beta\mu_{\mathcal{J}_{\text{Lor}}}=-\frac{1}{4\pi}\left(\xi_{0}^{+}-\xi_{0}^{-}\right)$  & $\beta\mu_{\mathcal{J}_{\text{Lor}}}=-\frac{1}{4\pi}\left(\xi_{0}^{+}-\xi_{0}^{-}\right)$  & $\beta\mu_{\mathcal{J}_{\text{Lor}}}=-\frac{1}{4\pi}\left(\hat{\xi}_{0}^{+}-\hat{\xi}_{0}^{-}\right)$\tabularnewline
potentials in the  &  &  & $\beta\mu_{\mathcal{W}^{\pm}}=\frac{1}{2\pi}\eta_{0}^{\pm}$  & $\beta\mu_{\psi_{\left[a\right]}^{\pm}}=\frac{1}{2\pi}\vartheta_{\left[a\right]0}^{\pm}$\tabularnewline
``rest frame''  &  &  &  & $\beta\mu_{\mathcal{U^{\pm}}}=\frac{1}{2\pi}\nu_{0}^{\pm}$\tabularnewline
 &  &  &  & \tabularnewline
 &  &  &  & \tabularnewline
Black hole  & $\left[\frac{\Theta}{8\pi G}\int_{r_{+}}\sqrt{g_{\varphi\varphi}}d\varphi\right]_{\text{on-shell}}$\;  & \;\;$-\frac{k_{2}}{\pi}\text{Im}\left[\text{tr}\left(\int_{r_{+}}A_{\tau}d\tau\right.\right.$\;  & \;\;$-\frac{k_{3}}{\pi}\text{Im}\left[\text{tr}\left(\int_{r_{+}}A_{\tau}d\tau\right.\right.$\;  & \;\;$-\frac{k_{3}}{\pi}\text{Im}\left[\text{tr}\left(\int_{r_{+}}A_{\tau}d\tau\right.\right.$\tabularnewline
entropy  &  & \;\;$\times\left.\left.\int_{r_{+}}A_{\varphi}d\varphi\right)\right]_{\text{on-shell}}$\;  & \;\;$\times\left.\left.\int_{r_{+}}A_{\varphi}d\varphi\right)\right]_{\text{on-shell}}$\;  & \;\;$\times\left.\left.\int_{r_{+}}A_{\varphi}d\varphi\right)\right]_{\text{on-shell}}$\tabularnewline
 &  &  &  & \tabularnewline
 &  &  &  & \tabularnewline
Regularity  & No conical  & Trivial holonomy  & Trivial holonomy  & Trivial holonomy\tabularnewline
condition on the  & singularity  & for contractible  & for contractible  & for contractible\tabularnewline
Euclidean  &  & $\tau$ cycle  & $\tau$ cycle  & $\tau$ cycle\tabularnewline
horizon  & $\Theta^{\text{on-shell}}=2\pi$\;  & \;\;$\left.e^{\int_{r_{+}}A_{\tau}d\tau}\right|_{\text{on-shell}}=-\mathbbm{1}$\;  & \;\;$\left.e^{\int_{r_{+}}A_{\tau}d\tau}\right|_{\text{on-shell}}=\mathbbm{1}$\;  & \;\;$\left.e^{\int_{r_{+}}A_{\tau}d\tau}\right|_{\text{on-shell}}=\mathbbm{1}$\tabularnewline
 &  &  &  & \tabularnewline
\hline 
\end{tabular}
\end{sideways}



\end{small}

\section{Chern-Simons formulation of (2+1) pure gravity\label{sec:The-Chern-Simons-formulation} }

\setcounter{equation}{0}

\subsection{Action and equations of motion\label{sub:Action-and-equations}}

One may reformulate the standard gravitation theory in 2+1 spacetime
dimensions as a Chern-Simons theory by using, instead of the metric
variables, two independent connections $A^{\pm}$ for $sl\left(2,\mathbb{R}\right)$
\cite{Achucarro:1987vz,Witten:1988hc}. The correspondence between
the connections and metric variables is 
\begin{equation}
A^{\pm}=\left(\omega^{a}\pm\frac{e^{a}}{\ell}\right)X_{a}^{\pm}\;,
\end{equation}
where $\omega^{a}$ and $e^{a}$ are the spin connection and the dreibein
of the metric theory. We will realize the $sl(2,\mathbb{R})$-generators
$X_{a}^{+}$ and $X_{a}^{-}$ by the same $2\times2$ matrices. One
convenient choice for both $X_{a}^{+}$ and $X_{a}^{-}$ is 
\begin{equation}
L_{-1}=\begin{pmatrix}0 & 0\\
1 & 0
\end{pmatrix}\quad;\quad L_{0}=\begin{pmatrix}-\frac{1}{2} & 0\\
0 & \frac{1}{2}
\end{pmatrix}\quad;\quad L_{1}=\begin{pmatrix}0 & -1\\
0 & 0
\end{pmatrix}\;,\label{sl2matrices0}
\end{equation}
which obeys 
\begin{equation}
\left[L_{i},L_{j}\right]=\left(i-j\right)L_{i+j}\quad,\quad i,j=-1,0,1\;.\label{eq:sl2R0}
\end{equation}
More information on our conventions is given in Appendix \ref{Appendix0}.

The action for pure gravity in the Chern-Simons formulation, which
differs from the standard Hilbert action by a boundary term, is given
by 
\begin{equation}
I=I_{\text{CS}}\left[A^{+}\right]-I_{\text{CS}}\left[A^{-}\right]\;,\label{ICStot}
\end{equation}
where 
\begin{equation}
I_{\text{CS}}\left[A^{\pm}\right]=\frac{k_{2}}{4\pi}\int\text{tr}\left[A^{\pm}\wedge dA^{\pm}+\frac{2}{3}A^{\pm}\wedge A^{\pm}\wedge A^{\pm}\right]\;.\label{ICS}
\end{equation}
Here, $k_{2}$ is related to the cosmological constant $\Lambda=-\frac{1}{\ell^{2}}$
and the Newton constant $G$ through the relation $k_{2}=k=\frac{\ell}{4G}$.

The two copies of $sl\left(2,\mathbb{R}\right)$ take the role of
the spacetime diffeomorphisms plus the local rotations of the dreibein.
It is quite interesting that actually the transformation freedom of
the Chern-Simons theory is larger than the one of the metric theory
because, by bona fide $SL\left(2,\mathbb{R}\right)\times SL\left(2,\mathbb{R}\right)$
point dependent transformations one may map a non-degenerate metric
onto a degenerate one without causing any problem. This possibility
is indeed of great practical use, and we will employ it below.

The equations of motion state that the curvature vanishes, 
\begin{equation}
F^{\pm}=dA^{\pm}+A^{\pm}\wedge A^{\pm}=0\;.
\end{equation}
This is the statement in the Chern-Simons language that there are
no propagating degrees of freedom. In spite of this fact, and even
without topological subtleties, the theory is non-trivial because
of the asymptotic structure. There are degrees of freedom at infinity,
the dynamics of which is governed by the asymptotic symmetry algebra.
As it will be discussed immediately below, only ``proper'' gauge
transformations, i.e. those which become the identity at spatial infinity
are bona fide gauge transformations that do not change the physical
state. On the other hand, gauge transformations which approach at
infinity an element of the asymptotic symmetry algebra (improper gauge
transformations) do change the physical state.

One may rewrite the action (\ref{ICStot}), (\ref{ICS}) in Hamiltonian
form as 
\begin{align}
I_{\text{Ham}} & =I_{\text{Ham}}\left[A^{+}\right]-I_{\text{Ham}}\left[A^{-}\right]\;,\label{Ihamtot}
\end{align}
with 
\begin{equation}
I_{\text{Ham}}\left[A^{\pm}\right]=-\frac{k_{2}}{2\pi}\int dtdx^{1}dx^{2}\text{tr}\left(A_{1}^{\pm}\dot{A}_{2}^{\pm}-A_{t}^{\pm}G^{\pm}\right)+B_{\infty}^{\pm}\;,\label{IHam}
\end{equation}

\begin{equation}
G^{\pm}=F_{12}^{\pm}=\partial_{1}A_{2}^{\pm}-\partial_{2}A_{1}^{\pm}+\left[A_{1}^{\pm},A_{2}^{\pm}\right]\;.\label{constraint}
\end{equation}
Equations (\ref{Ihamtot}), (\ref{IHam}), (\ref{constraint}) shows
that $A_{1}^{\pm}$ and $A_{2}^{\pm}$ are canonically conjugate,
and $G^{\pm}$ are the generators of proper gauge transformations
which, ``on-shell'' are constrained to vanish. The boundary term
$B_{\infty}^{\pm}$ must be added because one must allow for ``slowly
decreasing'' field variations at infinity in order for the full asymptotic
symmetry algebra to be able to act. The variation of $B_{\infty}^{\pm}$
cancels the nonvanishing surface terms that one picks up through integration
by parts in the variational principle. The form of $B_{\infty}^{\pm}$
will be given in the next subsection.

\subsection{Asymptotic symmetries\label{sub:Asymptotic-symmetries}}

\subsubsection{Boundary conditions and the most general permissible motion}

The procedure for establishing the boundary conditions for an action
principle is one of trial and error. One starts with an action that,
when extremized, gives the desired equations of motion up to ``surface
terms'' at spatial infinity. To analyze the surface terms at infinity,
one needs to impose boundary conditions. A necessary requirement for
the boundary conditions is that the ``off-shell'' fields admitted
in the action principle should include all ``reasonable'' solutions
of the field equations. Now, in a gauge theory there are constraints
among the canonical variables ($p,q$), which are the generators of
the local gauge symmetries and Lagrange multipliers for them. The
meaning of the Lagrange multipliers is that they are the parameters
per unit of time of a gauge transformation during the time evolution
of the system. Therefore, one first focuses on the solutions of the
constraint equations, and from them one guesses, with the criterion
just given, their boundary conditions. Next one obtains the boundary
conditions for the Lagrange multipliers by demanding that the boundary
conditions for ($p,q$) should be preserved in time.

The Lagrange multipliers are then divided in two classes: if their
value at large spatial distances is such that the surface term at
infinity picked after integration by parts in the variation of the
action vanishes, the corresponding gauge transformation is termed
``proper'', and it corresponds to a bona fide gauge transformation
that does not change the physical state. This normally happens when
the Lagrange multipliers vanish at infinity. On the other hand, when
the Lagrange multipliers do not vanish at infinity, their value corresponds
to the parameter, per unit of time, of a ``global symmetry transformation''
which is included in the evolution of the system, and which does change
the physical state. The coefficient of the asymptotic value of the
Lagrange multiplier in the variation of the Hamiltonian action is,
by definition, the negative of the variation of the charge that generates
the asymptotic symmetry \cite{Regge:1974zd} (see also \cite{Benguria:1976in}).
It should be emphasized that the charges which are the generators
of the asymptotic symmetries are what is called ``a function of state'',
that is, they are defined in terms of the canonical variables on a
$t=const$ surface, and they do not depend on how one continues the
$t=const$ surface into the future. In our particular case, the evolution
from the surface $x^{0}=t$ to the surface $x^{0}=t+\delta t$ is
given by an infinitesimal gauge transformation with gauge parameter
$A_{t}\delta t$. Therefore, at any given time, the definition of
the charges and their value is independent of $A_{t}$, it only depends
on the canonical variables $A_{r}$ and $A_{\varphi}$.

It has been shown \cite{Coussaert:1995zp} that, in the Chern-Simons
formulation one can recast the (off-shell) boundary conditions on
a $t=const$ surface, obtained in the metric formulation in \cite{Brown:1986nw},
in the form 
\begin{equation}
A_{\varphi}^{\pm}\left(r,\varphi\right)\underset{r\rightarrow\infty}{\longrightarrow}L_{\pm1}-\frac{2\pi}{k}\mathcal{L}^{\pm}\left(r,\varphi\right)L_{\mp1}\;,\label{B-H-CONN1}
\end{equation}
with 
\begin{equation}
\mathcal{L}^{\pm}\left(r,\varphi\right)\underset{r\rightarrow\infty}{\longrightarrow}\mathcal{L}^{\pm}\left(\varphi\right)+O\left(\frac{1}{r}\right)\;,\label{BH-CONN2}
\end{equation}
and 
\begin{equation}
A_{r}^{\pm}\underset{r\rightarrow\infty}{\longrightarrow}O\left(\frac{1}{r}\right)\;.\label{b-hcONN3}
\end{equation}

At spatial infinity, the most general time evolution of the spatial
parts of the connections $A_{r}^{\pm},A_{\varphi}^{\pm}$ is an improper
gauge transformation with gauge parameter per unit of time equal to
$A_{t}^{\pm}$. The most general $A_{t}^{\pm}$ which preserves the
boundary conditions (\ref{B-H-CONN1}), (\ref{BH-CONN2}), (\ref{b-hcONN3})
is given by \cite{HPTT} 
\begin{equation}
A_{t}^{\pm}\underset{r\rightarrow\infty}{\longrightarrow}\pm\left(\xi_{\pm}\left(r,\varphi\right)\left(L_{\pm1}-\frac{2\pi}{k}\mathcal{L}^{\pm}\left(r,\varphi\right)L_{\mp1}\right)\mp\xi'_{\pm}\left(r,\varphi\right)L_{0}+\frac{1}{2}\xi''_{\pm}\left(r,\varphi\right)L_{\mp1}\right)\;,\label{BHAt-2}
\end{equation}
where 
\begin{equation}
\xi_{\pm}\left(r,\varphi\right)\underset{r\rightarrow\infty}{\longrightarrow}\xi_{\pm}\left(\varphi\right)+O\left(\frac{1}{r}\right)\;,\label{BHxi-2}
\end{equation}
and the primes are derivatives with respect to $\varphi$. Under this
asymptotic symmetry transformation $\mathcal{L}^{\pm}$ evolve as
\begin{equation}
\dot{\mathcal{L}}^{\text{\ensuremath{\pm}}}=\pm\left(\xi_{\pm}\mathcal{L}^{\pm\prime}+2\mathcal{L}^{\pm}\xi'_{\pm}-\frac{k}{4\pi}\xi'''_{\pm}\right)\;.\label{Lpunto}
\end{equation}

The boundary conditions (\ref{B-H-CONN1}), (\ref{BH-CONN2}), (\ref{b-hcONN3})
hold on all slices $t=const$. As one evolves in time, the functions
$A_{t}^{\pm}$ can be taken at will within the class (\ref{BHAt-2}),
(\ref{BHxi-2}) without spoiling the symmetry, i.e., the parameters
$\xi_{\pm}(t,r,\varphi)$ can be arbitrary functions of time provided
(\ref{BHxi-2}) holds at each $t$.

Usually one takes $\xi_{\pm}$ to tend to unity in (\ref{BHxi-2})
\cite{Coussaert:1995zp,Henneaux-HS,Theisen-HS,Henneaux:1999ib}. This
particular choice of the freedom at infinity has the useful property
that the equations for the Virasoro generators reduce at the boundary
to the chiral equations $\dot{\mathcal{L}}^{\pm}=\pm\mathcal{L}^{\pm\prime}$,
implying $\mathcal{L}^{\pm}=\mathcal{L}^{\pm}(x^{\pm})$. With that
choice, one finds furthermore $A_{\varphi}^{\pm}=\pm A_{t}^{\pm}$,
i.e., $A_{\mp}^{\pm}=0$. Even though the analysis of the symmetries
can be carried through with this convenient and permissible choice
of the Lagrange multipliers, it must be kept in mind, however, that
this is only a particular choice of the global transformation at infinity
included in the evolution of the system. It does not include the most
general permissible motion.

\subsubsection{Virasoro charges}

The boundary terms $B_{\infty}^{\pm}$ appearing in the action (\ref{IHam})
is determined by the requirement that the action should have well
defined functional derivatives with respect to $A_{r}$ and $A_{\varphi}$
when the asymptotic charges $\mathcal{L}^{\pm}\left(\varphi\right)$
are varied and $A_{t}$ includes the most general asymptotic symmetry
\cite{Regge:1974zd}. This gives 
\begin{equation}
B_{\infty}^{\pm}\left[\xi\right]=-\int Q_{\pm}[\xi_{\pm}]dt\,,\label{Binfty-1}
\end{equation}
where 
\begin{equation}
Q_{\pm}[\xi_{\pm}]=\pm\int_{r\rightarrow\infty}\xi_{\pm}\left(\varphi\right)\mathcal{L}^{\pm}\left(\varphi\right)d\varphi.
\end{equation}

If $\xi_{\pm}$ and $\mathcal{L}^{\pm}$ are expanded in Fourier modes,
\begin{eqnarray}
\mathcal{L}^{\pm} & = & \frac{1}{2\pi}\sum_{n}\mathcal{L}_{n}^{\pm}e^{in\varphi},\\
\xi_{\pm} & = & \frac{1}{2\pi}\sum_{n}\xi_{n}^{\pm}e^{in\varphi},
\end{eqnarray}
one finds, by expressing the asymptotic part of the commutator of
two improper gauge transformations in terms of the asymptotic parts
of those two transformations, that the $\mathcal{L}_{n}^{\pm}$ obey,
in terms of the Poisson bracket, the Virasoro algebra with the classical
central charge $c=6k=3\ell/2G$, 
\begin{equation}
i\left\{ \mathcal{L}_{m},\mathcal{L}_{n}\right\} =\left(m-n\right)\mathcal{L}_{m+n}+\frac{k}{2}m^{3}\delta_{m+n,0}.
\end{equation}
These ${\mathcal{L}}_{n}^{\pm}$'s derived along Chern-Simons lines
coincide with the Virasoro generators found in the metric formulation
\cite{Brown:1986nw} (see also formula (\ref{bdrytermvirasoro}) of
Appendix \ref{AppendixA} in this context).

For each of the two Virasoro algebras, $\mathcal{L}_{0}$ and $\mathcal{L}_{\pm1}$
generate the subalgebra $sl\left(2,\mathbb{R}\right)$. They are the
``global'' charges of the gauge group of the Chern-Simons theory,
which then re-emerge as asymptotic symmetries, as it is customary
in gauge theories. However, while in higher dimensions these original
global charges are generically all the asymptotic symmetries, in the
present lower dimensional case the asymptotic symmetry algebra is
much larger since it contains the infinite-dimensional Virasoro algebra
with all the higher $\mathcal{L}_{n}$ modes. The generators $\mathcal{L}_{n}$
with $\left|n\right|\leq1$ transform in the $s=1$ representation
of $sl\left(2,\mathbb{R}\right)$. They constitute the simplest example
of a concept which will reappear later (section \ref{sub:Asymptotic-symmetriesW3});
that of a ``wedge subalgebra'', which is the subalgebra generated
by the modes $\left|n\right|\leq s$ of the asymptotic symmetry algebra.

With the choice $\xi_{\pm}=1$, the boundary term $B_{\infty}^{+}-B_{\infty}^{-}$
in the action reduces to the negative of the integral over time of
the sum $\mathcal{L}_{0}^{+}+\mathcal{L}_{0}^{-}$ of the zero modes
$\mathcal{L}_{0}^{\pm}$ of the Virasoro generators.

Lastly, there is a simple but important conceptual point to be recalled
here. It is the following: although the $\mathcal{L}_{n}^{\pm}$ or
equivalently $\mathcal{L}^{\pm}\left(\varphi\right)$ are the generators
of the symmetry algebra, they do not all remain constant as one moves
from one spacelike slice to the next. What matters, and is true, is
that the action is invariant under the complete asymptotic algebra.
The ``extended Hamiltonian'' is not invariant under the algebra,
it is covariant under it, because is a generic element of the algebra.
The issue at hand has a simple analog for a free relativistic particle.
There, the canonical boost generator $M^{i}=x^{i}\sqrt{\vec{p}^{2}+m^{2}}$
changes as one moves in time. One may define a modified boost generator
$\tilde{M}^{i}=M^{i}-tp^{i}$, which does not change in time but at
the price of being \emph{explicitly time dependent}. This modified
generator is the conserved charge that comes out of Noether theorem.
For the general case, the extended Hamiltonian is a linear combination
$H=Q\left[\lambda\right]=\lambda^{a}Q_{a}$, of the symmetry generators
$Q_{a}$ satisfying $\left[Q_{a},Q_{b}\right]=C_{ab}^{\;\;\: c}Q_{c}$.
One may define $\tilde{Q}\left[\epsilon\right]=\epsilon^{a}Q_{a}$
which will not change in time if $\epsilon^{a}$ obeys the differential
equation $\dot{\epsilon}^{c}+C_{ab}^{\;\;\: c}\epsilon^{a}\lambda^{b}=0$.
The modified charge $\tilde{Q}$ depends explicitly on time and reduces
at $t=0$ to $Q_{b}$ if one imposes the initial condition $\epsilon^{a}\left(t=0\right)=\delta_{b}^{a}$,
for a given $b$.

\subsection{$N=2$, spin 2, black hole in the Chern-Simons formulation\label{sub:(2+1)-pure-gravity}}

\subsubsection{Euclidean approach\label{sub:Euclidean-approach}}

In order to discuss generalizations below, we will recall here the
Chern-Simons formulation of the (2+1) black hole thermodynamics for
pure gravity \cite{Carlip:1994gc}. In doing so, we will avoid reference
to the metric formulation, which is reviewed in Appendix \ref{AppendixA}
and in Table \ref{sec:table}. The reason for this procedure is that
no gauge invariant metric appears to be available in the generalized
case. In the absence of well-defined geometry, the usual way of defining
a black hole in terms of an event horizon out of which not even light
can escape is not available. The only possibility at hand, appears
to be to define the black hole through its thermal properties \cite{GK,AGKP,Ammon:2012wc}.
This requires to formulate the theory ab initio in Euclidean space.
The properties in Lorentzian spacetime are then encrypted in the Euclidean
formulation, and are only revealed after one passes to Lorentzian
spacetime through the inverse of a ``Wick rotation''.

In the Euclidean approach, the two independent $sl\left(2,\mathbb{R}\right)$
connections $A^{\pm}$ are merged into a single complexified $sl(2,\mathbb{C})$-connection
$A$ \cite{Witten} according to the rules 
\begin{eqnarray}
 &  & A^{+}=A,\label{ContinuationA0}\\
 &  & A^{-}=-A^{\dagger}.\label{ContinuationB0}
\end{eqnarray}
These rules are explained in Appendix \ref{Appendix0}. The merging
(\ref{ContinuationA0}) and (\ref{ContinuationB0}) of the two independent
$sl\left(2,\mathbb{R}\right)$ connections is remarkably simple, one
just takes $A^{+}$ and allows it to be complex. The other connection
$A^{-}$ is then related to $A$ according to (\ref{ContinuationB0})
\cite{Witten,Carlip:1994gc,Banados:1998gg,Castro:2011iw}.

It was shown in \cite{Carlip:1994gc}, starting from the metric formulation,
that the topology of the (2+1) black hole is that of the solid torus,
as illustrated in figure 1 of Appendix \ref{AppendixA}. In terms
of ``Schwarzschild-like'' Euclidean coordinates $\left(\tau,r,\varphi\right)$,
where $\tau=it$ is the Euclidean time, both $\tau$ and $\varphi$
are periodic, $0\leq\varphi<2\pi$, $0\leq\tau<1$, and the $\tau$
circles are contractible to a point, whereas the $\varphi$ circles
are not. One defines an (Euclidean) black hole as a solution of the
Chern-Simons field equations with this spacetime topology to which
one may ascribe a nonvanishing entropy.

As discussed in detail in Appendix \ref{AppendixA}, in order to define
black holes through their thermodynamical properties, and in particular
to define their entropy, one needs an action principle and therefore,
one needs families of solutions rather than a single solution. This
situation is quite different from the standard one in Lorentzian spacetime
with the metric, where one can define say, a Schwarzschild black hole
of any given mass by dealing only with it.

The Euclidean action $I_{E}=iI_{\text{Lor}}$ is (see Appendix \ref{Appendix0})
\begin{equation}
I_{E}\left[A\right]=-2\text{Im}\left[I_{\text{CS}}\left[A\right]\right]\;,
\end{equation}
where $I_{\text{CS}}\left[A\right]$ is the standard Chern-Simons
action given by (\ref{ICS}). Its Hamiltonian form reads 
\begin{eqnarray}
I_{E}^{\text{Ham}}\left[A\right] & = & \text{Im}\left[\frac{k_{2}}{\pi}\int d\tau drd\varphi\,\text{tr}\left(A_{r}\dot{A}_{\varphi}-A_{\tau}G\right)\right]+B^{E}\;,\label{IHam-1}\\
G & = & \partial_{r}A_{\varphi}-\partial_{\varphi}A_{r}+\left[A_{r},A_{\varphi}\right]\;,\label{constraint-1}
\end{eqnarray}
where $B^{E}$ stands for the boundary terms both at infinity (already
discussed) and at the horizon (to be discussed in subsection \ref{sub:Thermodynamics}).

\subsubsection{Rest frame}

When investigating the thermodynamics of the four-dimensional Kerr
black hole, one usually assumes that the only nonvanishing charges
are the zeroth component $P^{0}$ of the $4$-momentum (``mass'')
and the angular momentum $J^{\varphi}$. That is, one goes to the
rest frame of the system. There is no more loss of generality in doing
this than the one incurred if one studies the thermodynamics of a
gas in a box assuming that the box is at rest.

The analog situation for the 2+1 black hole is the one in which the
only surviving Virasoro mode are the zero mode of each of the two
Virasoro algebras%
\footnote{The action of the Virasoro group on the Virasoro generators (``coadjoint
orbits'') has been studied in depth in references \cite{Coadj1,Coadj2,Witten:1987ty,Balog:1997zz,Coadj3,Coadj4},
where it has been shown that for a large class of coadjoint orbits
- and in particular for all orbits on which ${\mathcal{L}}_{0}$ is
bounded from below except for one very special orbit -- , one can
indeed always go to a frame (``rest frame'') where $\mathcal{L}(\varphi)$
is constant. We are grateful to Glenn Barnich for useful information
on this point.%
}. When studying black hole thermodynamics we will assume that we are
in that ``rest frame''. Thus, the only extensive parameters present
will be the mass and the angular momentum.

We consider therefore gauge field configurations that behave asymptotically
as in (\ref{B-H-CONN1}), (\ref{BH-CONN2}), (\ref{b-hcONN3}) and
(\ref{BHxi-2}) but with constant ${\mathcal{L}}^{\pm}$ and $\xi_{\pm}$
at infinity, i.e, 
\begin{eqnarray}
 &  & A_{\varphi}^{\pm}\left(r,\varphi\right)\underset{r\rightarrow\infty}{\longrightarrow}L_{\pm1}-\frac{2\pi}{k}\mathcal{L}^{\pm}\left(r,\varphi\right)L_{\mp1}\;,\label{B-H-CONN1b}\\
 &  & \mathcal{L}^{\pm}\left(r,\varphi\right)\underset{r\rightarrow\infty}{\longrightarrow}\frac{1}{2\pi}\mathcal{L}_{0}^{\pm}+O\left(\frac{1}{r}\right)\;,\label{BH-CONN2b}\\
 &  & A_{r}^{\pm}\underset{r\rightarrow\infty}{\longrightarrow}O\left(\frac{1}{r}\right)\;,\label{b-hcONN3b}\\
 &  & A_{t}^{\pm}\underset{r\rightarrow\infty}{\longrightarrow}\pm\xi_{\pm}\left(r,\varphi\right)\left(L_{\pm1}-\frac{2\pi}{k}\mathcal{L}^{\pm}\left(r,\varphi\right)L_{\mp1}\right)\;,\label{BHAt-2b}\\
 &  & \xi_{\pm}\left(r,\varphi\right)\underset{r\rightarrow\infty}{\longrightarrow}\frac{1}{2\pi}\xi_{0}^{\pm}+O\left(\frac{1}{r}\right)\;.\label{BHxi-2b}
\end{eqnarray}
According to the rules of Appendix \ref{Appendix0}, the Euclidean
version of the asymptotic conditions ``in the rest frame'' reads
\begin{eqnarray}
 &  & A_{\varphi}\left(r,\varphi\right)\underset{r\rightarrow\infty}{\longrightarrow}L_{1}-\frac{2\pi}{k}\mathcal{L}\left(r,\varphi\right)L_{-1}\;,\label{B-H-CONN1bE}\\
 &  & \mathcal{L}\left(r,\varphi\right)\underset{r\rightarrow\infty}{\longrightarrow}\frac{1}{2\pi}\mathcal{L}_{0}+O\left(\frac{1}{r}\right)\;,\label{BH-CONN2bE}\\
 &  & A_{r}\underset{r\rightarrow\infty}{\longrightarrow}O\left(\frac{1}{r}\right)\;,\label{b-hcONN3bE}\\
 &  & A_{\tau}\underset{r\rightarrow\infty}{\longrightarrow}-i\xi\left(r,\varphi\right)\left(L_{1}-\frac{2\pi}{k}\mathcal{L}\left(r,\varphi\right)L_{-1}\right)\;,\label{BHAt-2bE}\\
 &  & \xi\left(r,\varphi\right)\underset{r\rightarrow\infty}{\longrightarrow}\frac{1}{2\pi}\xi_{0}+O\left(\frac{1}{r}\right)\;,\label{BHxi-2bE}
\end{eqnarray}
where $t=-i\tau$ and where the complex parameters $\mathcal{L}_{0}$
and $\xi_{0}$ are related to their Lorentzian counterparts through
the continuation rules $\mathcal{L}_{0}^{+}=\mathcal{L}_{0}$, $\mathcal{L}_{0}^{-}=\mathcal{L}_{0}^{*}$,
$\xi_{0}^{+}=\xi_{0}$ and $\xi_{0}^{-}=\xi_{0}^{*}$ (see Appendix
\ref{sub:Eucl-Lor-cont}). Here the $^{*}$ denotes complex conjugation.

The relationship between the only surviving modes ${\mathcal{L}}_{0}^{\pm}$
and the mass and angular momentum is recalled in Appendix \ref{AppendixA}
(formulas \ref{mass-angularmomBTZ}) and reads explicitly 
\begin{align}
\textrm{(Rest mass)} & :=\mathcal{M}_{\text{Lor}}=\frac{1}{\ell}\left(\mathcal{L}_{0}^{+}+\mathcal{L}_{0}^{-}\right)=\frac{M_{\text{Lor}}}{8G},\label{mass-angularmomBTZ-1}\\
\text{(Angular momentum)} & :=\mathcal{J}_{\text{Lor}}=\mathcal{L}_{0}^{+}-\mathcal{L}_{0}^{-}=\frac{J_{\text{Lor}}}{8G}.\label{mass-angularmomBTZ0}
\end{align}
Similarly, one finds the relationship 
\begin{equation}
N\left(\infty\right)=\frac{1}{2\pi}N_{0}=\frac{\ell}{4\pi}\left(\xi_{0}+\xi_{0}^{*}\right),\quad\quad N^{\varphi}\left(\infty\right)=\frac{1}{2\pi}N_{0}^{\varphi}=\frac{i}{4\pi}\left(\xi_{0}-\xi_{0}^{*}\right),\label{eq:Nintermsofxi}
\end{equation}
between the lapse, the shift and the zero modes $\xi_{0}^{\pm}$ at
infinity.

\subsubsection{General flat connection on a solid torus and black hole solution}

The investigation of solutions of the Chern-Simons theory (flat connections
on a solid torus with the required asymptotic behaviour) is most simply
described in the gauge 
\begin{equation}
A_{r}=0\;,\label{Ar}
\end{equation}
for all $r$. It is interesting that this permissible condition which
is so convenient and harmless in the Chern-Simons approach cannot
even be formulated in the metric approach, because it would correspond
to a degenerate metric with $g_{rr}=0$. Once the gauge condition
(\ref{Ar}) is imposed, the fields $A_{\varphi}$ and $A_{\tau}$
do not depend on $r$ and so take the asymptotic form (\ref{B-H-CONN1bE}),
(\ref{BH-CONN2bE}),(\ref{b-hcONN3bE}),(\ref{BHAt-2bE}) and (\ref{BHxi-2bE})
(with $\mathcal{L}$ and $\xi$ constant) everywhere and not just
at infinity. Note that since $A_{\tau}$ is proportional to $A_{\varphi}$
one has 
\begin{equation}
\left[A_{\tau},A_{\varphi}\right]=0\;,
\end{equation}
and the equation $F_{\tau\varphi}=0$ is automatically fulfilled.

Explicitly, in terms of the mass and the angular momentum, the general
flat connection obeying the boundary conditions reads 
\begin{align}
A_{\varphi} & =L_{1}-\frac{1}{2k}\left(\mathcal{M}\ell+i\mathcal{J}\right)L_{-1},\label{aphichicoBTZ-1}\\
A_{\tau} & =-i\xi\left(L_{1}-\frac{1}{2k}\left(\mathcal{M}\ell+i\mathcal{J}\right)L_{-1}\right),\label{atchicoBTZ-1}\\
A_{r} & =0.\label{eq:archico}
\end{align}
The Lorentzian continuation is 
\begin{align}
A_{\varphi}^{\pm} & =L_{\pm1}-\frac{1}{2k}\left(\mathcal{M}_{\text{Lor}}\ell\pm\mathcal{J}_{\text{Lor}}\right)L_{\mp1},\label{aphichicoBTZ}\\
A_{t}^{\pm} & =\pm\xi_{\pm}\left(L_{\pm1}-\frac{1}{2k}\left(\mathcal{M}_{\text{Lor}}\ell\pm\mathcal{J}_{\text{Lor}}\right)L_{\mp1}\right),\label{atchicoBTZ}\\
A_{r}^{\pm} & =0.
\end{align}
The solution is a black hole provided $\mathcal{M}_{\text{Lor}}$
and $\mathcal{J}_{\text{Lor}}$ fulfills the inequality 
\begin{equation}
\vert\mathcal{J}_{\text{Lor}}\vert\leq\mathcal{M}_{\text{Lor}}\ell,\label{JsmallerM}
\end{equation}
that guarantees the existence of an horizon in the metric formulation.
As we shall see, this inequality (\ref{JsmallerM}) also guarantees
on the Euclidean side that the entropy is real and positive - a necessary
condition for having a sensible thermodynamics. On the Lorentzian
side, one can set the coefficients $\xi_{\pm}$ equal to unity by
a gauge transformation, but this cannot be done on the Euclidean side
where one finds that $\xi$ is related to the Virasoro charge $\mathcal{L}$
through the precise relation (\ref{eq:xi}) below and is generically
not equal to unity.

\subsubsection{Boundary conditions at the horizon. Regularity}

It follows from (\ref{atchicoBTZ-1}) that $A_{\tau}\left(r_{+}\right)\neq0$
(the coefficient $\xi$ does not vanish because of (\ref{eq:Nintermsofxi})
and (\ref{eq:betanphi})-(\ref{betamubtz})). Therefore, the integral
of $A_{\tau}$ given by (\ref{atchicoBTZ-1}) over a circle in the
$r-\tau$ plane, centered at $r_{+}$ does not vanish even if the
radius of the circle tends to zero. Since the circle is a contractible
curve within the solid torus, this singularity in the connection reflects
the fact that the connection can be taken as given by (\ref{aphichicoBTZ-1})-(\ref{eq:archico})
\emph{only in a patch excluding the origin $r_{+}$}. To define the
connection at $r_{+}$ one must use another patch, for example a disk
centered at an origin different from $r_{+}$ with its own polar coordinate
system. The regular form of the connection at $r_{+}$, for which
$A_{\tau}\left(r_{+}\right)=0$, would then be obtained from the singular
form (\ref{aphichicoBTZ-1}) - (\ref{eq:archico}) by a ``regularizing
gauge transformation'' which would be singular at $r_{+}$. The form
of that gauge transformation implies the regularity condition that
$A_{\tau}$ must obey when it is written in the form (\ref{atchicoBTZ-1}).

To continue the analysis it is useful to diagonalize (\ref{atchicoBTZ-1}).
One finds 
\begin{equation}
A_{\tau}=-2\pi i\nu\left(\xi,\mathcal{L}\right)L_{0},\label{eq:Ataunu}
\end{equation}
where 
\begin{equation}
\nu\left(\xi,\mathcal{L}\right)=\xi\sqrt{\frac{2\mathcal{L}}{\pi k}}.
\end{equation}
The regularizing gauge transformation that maps $A_{\tau}$ given
by (\ref{eq:Ataunu}) to zero is then implemented by a group element
of the form 
\begin{equation}
g=e^{2\pi i\tau\nu L_{0}},\label{eq:element}
\end{equation}
near $r_{+}$. This transformation is permissible only if $g$ is
periodic in $\tau$ up to a sign, that is, if $g(\tau+1)=\pm g(\tau)$,
or, in terms of $\nu$, 
\begin{equation}
\nu\left(\xi,\mathcal{L}\right)=n,\label{eq:reg}
\end{equation}
where $n$ is an integer. At $r_{+}$ the regularizing gauge transformation
is singular as expected because the angle $\tau$ is not well defined
there. Equation (\ref{eq:reg}) is the regularity condition on the
connection.

When $n$ is an odd integer, the group element $g(\tau)$ is antiperiodic,
$g(\tau=1)=-\mathbbm{1}$. The minus sign is allowed because the (identity
component of the) gauge group of $2+1$ Euclidean gravity is the proper,
orthochronous Lorentz group $SO^{+}(3,1)$, which is isomorphic to
$SL(2,\mathbb{C})/{\mathbb{Z}}_{2}$ where ${\mathbb{Z}}_{2}=\{+\mathbbm{1},-\mathbbm{1}\}$
is the center of $SL(2,\mathbb{C})$ (see Appendix \ref{Appendix0}).
So $-\mathbbm{1}$ is to be identified with the identity. If one had
used instead of the spinor representation (\ref{sl2matrices0}) the
vector representation in terms of the $4\times4$ matrices of $SO^{+}(3,1)$,
one would have obtained $g(\tau=1)=+\mathbbm{1}$ for all even and
odd $n$'s. Note that the computation of the entropy given below does
not depend on the choice of representation. 

One may show directly from (\ref{eq:Nintermsofxi}) and (\ref{eq:betanphi})-(\ref{betamubtz}),
or -better- from the argument given in section \ref{sub:Thermodynamics2}
below, that (\ref{eq:reg}) is equivalent to the regularity condition
\[
\Theta^{\text{reg}}=2\pi,
\]
for the absence of a conical singularity in the metric formulation
(eq. (\ref{thetaigual2pi})) \emph{if and only if} 
\begin{equation}
n=1,\label{eq:nigual1}
\end{equation}
or equivalently 
\begin{equation}
\xi=\sqrt{\frac{\pi k}{2\mathcal{L}}}.\label{eq:xi}
\end{equation}
The regularity condition implies then for the holonomy $H_{\tau}$:
\begin{equation}
H_{\tau}=\left.e^{\int_{r_{+}}A_{\tau}d\tau}\right|_{\text{on-shell}}=\left.e^{A_{\tau}\left(r_{+}\right)}\right|_{\text{on-shell}}=-\mathbbm{1}\;.\label{holonomytau-2}
\end{equation}
We shall adopt the choice $n=1$ and its analog for the generalized
black holes with $N>2$.

It is however interesting to note that from the Chern-Simons point
of view the ``sectors'' with $n\neq1$ would appear to be as regular
as the one with $n=1$, but they would be physically different. For
example, as shown in eq. (\ref{eq:entropyn}) below, for a given value
of the charges the entropy would be $n$ times bigger (which would
seem to exclude $n<0$). We shall not pursue that line of inquiry
any further herein.

Finally, it should be stressed that, just as it happens in the metric
formulation, the regularity analysis is not peculiar to the choice
of the horizon $r_{+}$ as the origin of the polar system of coordinates
$(\tau,r)$. One could have chosen any point in the $r-\tau$ plane
as the origin and the analysis would still go through. In particular,
the regularity condition would still apply. The origin $r_{+}$ is
of great practical advantage in the evaluation of the entropy given
next because then the solution is static, the ``$p\dot{q}$'' bulk
term in the action drops out, and the entropy becomes then expressible
solely as a boundary term at the origin.

\subsubsection{Entropy from contribution to the action at the horizon\label{sub:Thermodynamics}}

The black hole entropy, as well as other thermodynamic functions such
as the Helmholtz and Gibbs free energies, are obtained by evaluating
the appropriate Euclidean action on the black hole solution (on-shell).
Here the word ``appropriate'' means that the chosen action must
be such that if one demands that it be stationary with some boundary
conditions at infinity, then the equations of motion should hold everywhere.
If one fixes at infinity the asymptotic symmetry charges $\mathcal{L}\left(\varphi\right)$,
which corresponds to the microcanonical ensemble, then the value of
the corresponding action is the entropy. If instead one fixes the
asymptotic gauge displacements $\xi\left(\varphi\right)$, which contain
the temperature and the chemical potentials, then the value of the
corresponding action on-shell is $-\beta G$, where $G$ is the Gibbs
free energy. See also Appendix (\ref{sub:Euclidean-action-and}) in
this context.

To construct the desired action, it is simplest to start from the
Hamiltonian form. This is because the black hole solutions will be
time independent since they describe a thermodynamic system in equilibrium.
In that case, the integrand in the first term on the right hand side
of (\ref{IHam-1}) vanishes on-shell because $\dot{A}_{\varphi}=0$,
and the constraint $G=0$ holds. Furthermore, if one works in the
microcanonical ensemble, there is no boundary term at infinity to
be included, i.e., $B_{\infty}=0$ because the charges are fixed at
infinity. This would seem to indicate that the entropy vanishes, but
this is not so, because in the Euclidean case there is an extra condition
that must be fulfilled and is not present in the Lorentzian case,
namely the demand of regularity at the ``Euclidean horizon'' $r_{+}$,
that was just discussed. The action must be such as to imply this
regularity requirement since, if it were violated, the equations of
motion would not hold at the origin. We now address this issue

The Hamiltonian derivation of the (2+1) black hole entropy in the
metric formulation has been given in \cite{Teitelboim_action_and_entropy,BTZ_Gauss_Bonnet,Hawking:1994ii}.
It yields the entropy as a ``boundary term'' \emph{at the horizon}.
We provide here the corresponding derivation for the Chern-Simons
formulation, which yields again the entropy as a contribution to the
action at the horizon%
\footnote{Work relating the black hole entropy to the Chern-Simons action may
be found in \cite{Banados:2012ue,deBoer:2013gz}. It follows different
lines.%
}.

We start by analyzing the variation of the canonical action (\ref{IHam-1}),
(\ref{constraint-1}). When the equations of motion are fulfilled
in the bulk it reads 
\begin{equation}
\delta I_{\text{Ham}}=\delta B_{r_{+}}+\frac{k_{2}}{\pi}\text{Im}\left[\int_{r_{+}}d\tau d\varphi\text{tr}\left[A_{\tau}\delta A_{\varphi}\right]\right].\label{eq:deltaIHamappendix}
\end{equation}
We are not including the boundary term at infinity because it can
be dealt with separately and was already considered above. In the
microcanonical ensemble it is equal to zero anyway.

We will assume that the dynamical fields and the Lagrange multipliers
do not depend on $\tau$ and $\varphi$ near $r_{+}$. This property
holds for the black hole solutions and can in fact be reached by a
suitable gauge transformation. We will not assume however that the
gauge condition $A_{r}=0$ has been imposed, so that the fields can
depend on $r$. Even then the equations of motion imply $\left[A_{\tau},A_{\varphi}\right]=0$,
so that $A_{\tau}$ and $A_{\varphi}$ can be simultaneously diagonalized
on-shell. Eq. (\ref{eq:deltaIHamappendix}) then reduces to

\begin{align}
\delta I_{\text{Ham}} & =\delta B_{r_{+}}+\frac{k_{2}}{\pi}\text{Im}\left[\text{tr}\left[\left(\int_{r_{+}}d\tau A_{\tau}\right)\delta\left(\int_{r_{+}}d\varphi A_{\varphi}\right)\right]\right],\nonumber \\
 & =\delta B_{r_{+}}+2k_{2}\text{Im}\left[\text{tr}\left[A_{\tau}\left(r_{+}\right)\delta A_{\varphi}\left(r_{+}\right)\right]\right].\label{deltaIHamCSBTZ}
\end{align}
This equation is geometrically quite interesting. It may be described
as stating that the holonomy in the $\tau$ cycle is conjugate to
the holonomy in the $\varphi$ cycle, in complete analogy with what
happens in the metric formulation, where the opening angle in the
$r-\tau$ plane is conjugate to the length of the $\varphi$ circle.

We now introduce at the horizon a fixed connection $A_{\tau}^{\text{reg}}$
that fulfills the regularity condition established in the previous
subsection. We want the variational principle $\delta I_{\text{Ham}}=0$
to imply that $A_{\tau}(r_{+})$ coincides with $A_{\tau}^{\text{reg}}(r_{+})$
up to conjugation by some group element $g$, 
\begin{equation}
A_{\tau}^{\text{on-shell}}(r_{+})=gA_{\tau}^{\text{reg}}(r_{+})g^{-1}.\label{Areg}
\end{equation}
Differently put, the eigenvalues $\mu_{k}$ of $A_{\tau}(r_{+})$
should coincide on-shell with those of $A_{\tau}^{\text{reg}}(r_{+})$,
\begin{equation}
\mu_{k}^{\text{on-shell}}=\mu_{k}^{\text{reg}}.
\end{equation}
This requirement will be fulfilled if we choose the boundary term
$B_{r_{+}}$ to be 
\begin{equation}
B_{r_{+}}=-\frac{k_{2}}{\pi}\text{Im}\left[\int_{r_{+}}d\tau d\varphi\left[\sum_{k}\mu_{k}^{\text{reg}}\lambda_{k}\right]\right],
\end{equation}
where the $\lambda_{k}$'s are the eigenvalues of $A_{\varphi}$.
Here, the $\mu_{k}^{\text{reg}}$'s are fixed but the $\lambda_{k}$'s
are varied. Indeed, using the fact that the equations in the bulk
imply that $A_{\varphi}$ and $A_{\tau}$ commute, one then finds
that $\text{tr}\left[A_{\tau}\left(r_{+}\right)\delta A_{\varphi}\left(r_{+}\right)\right]=\sum_{k}\mu_{k}\delta\lambda_{k}$,
so that the condition for the action to be an extremum becomes $\delta I_{\text{Ham}}\equiv2k_{N}\text{Im}\left[\sum_{k}\left(-\mu_{k}^{\text{reg}}+\mu_{k}\right)\delta\lambda_{k}\right]=0$
and leads to the regularity condition (\ref{Areg}) on-shell.

Therefore, in the microcanonical ensemble, the correct action reads
\begin{equation}
I=-\frac{k_{2}}{\pi}\text{Im}\left[\int_{r_{+}}d\tau d\varphi\left[\sum_{k}\mu_{k}^{\text{reg}}\lambda_{k}\right]\right]+\frac{k_{2}}{\pi}\text{Im}\left[\int d\tau drd\varphi\text{tr}\left[A_{r}\dot{A}_{\varphi}-A_{\tau}G\right]\right].\label{Iimproved-1-1}
\end{equation}
For stationary configurations the canonical action vanishes on-shell,
and so equation (\ref{Iimproved-1-1}) with $A_{\tau}$ and $A_{\varphi}$
on-shell gives the entropy, 
\begin{eqnarray}
S & = & -\frac{k_{2}}{\pi}\text{Im}\left[\int_{r_{+}}d\tau d\varphi\text{tr}\left[A_{\tau}A_{\varphi}\right]\right]_{\text{on-shell}}\;,\nonumber \\
 & =- & 2k_{2}\text{Im}\left[\text{tr}\left[A_{\tau}A_{\varphi}\right]\right]_{\text{on-shell}}\;,\label{sataphi}
\end{eqnarray}
which in terms of the Lorentzian gauge fields reads 
\begin{equation}
S=k_{2}\left[\text{tr}\left[A_{t}^{+}A_{\varphi}^{+}\right]-\text{tr}\left[A_{t}^{-}A_{\varphi}^{-}\right]\right]_{\text{on-shell}},
\end{equation}
so that, it manifestly acquires different independent contributions
from each copy of the gauge group.

Eq. (\ref{sataphi}) plays the role in the Chern-Simons formulation
of the celebrated 
\begin{equation}
S=\frac{1}{8\pi G}\Theta^{\text{on-shell}}\int_{r_{+}}\sqrt{g_{\varphi\varphi}^{\text{on-shell}}}d\varphi=\frac{1}{4G}\left(\text{Horizon Area}\right)\;,
\end{equation}
in the metric formulation.

\subsubsection{Entropy as a function of the charges \label{sub:Thermodynamics2}}

In order to know the thermodynamics of the system, it is necessary
to express the entropy in terms of the extensive quantities, which
are the charges defined at infinity. To this end, it is first necessary
to determine $A_{\tau}\left(r_{+}\right)$ from the regularity condition
(\ref{eq:reg}) and then express all the quantities at $r_{+}$ in
terms of the charges at infinity through the solution of the constraint
$G=0$, where $G$ is given by (\ref{constraint-1}) using the boundary
conditions.

One obtains for the entropy 
\begin{equation}
S=2\pi n\sqrt{2\pi k}\left(\sqrt{\mathcal{L}}+\sqrt{\mathcal{L}^{*}}\right).\label{eq:entropyn}
\end{equation}
This expression is identical to the entropy obtained in the metric
formulation if and only if 
\[
n=1,
\]
as it was anticipated in (\ref{eq:nigual1}) above.

Indeed, if one sets $\mathcal{L}=\frac{1}{4\pi}\left(\mathcal{M}\ell+i\mathcal{J}\right)$,
and $\mathcal{L}^{*}=\frac{1}{4\pi}\left(\mathcal{M}\ell-i\mathcal{J}\right)$,
the entropy in terms of the Lorentzian charges $\mathcal{M}=\mathcal{M}_{\text{Lor}}$
and $i\mathcal{J}=\mathcal{J}_{\text{Lor}}$ is given by 
\begin{equation}
S\left(\mathcal{M}_{\text{Lor}},\mathcal{J}_{\text{Lor}}\right)=\pi\ell\sqrt{\frac{\mathcal{M}_{\text{Lor}}}{G}}\left[1+\left(1-\frac{\mathcal{J}_{\text{Lor}}^{2}}{\mathcal{M}_{\text{Lor}}^{2}\ell^{2}}\right)^{\frac{1}{2}}\right]^{\frac{1}{2}},
\end{equation}
an expression which can be rewritten as 
\begin{equation}
S=\frac{1}{4G}\left(2\pi r_{+}\right)\;,
\end{equation}
with 
\begin{equation}
r_{\pm}^{2}=4\ell^{2}G\mathcal{M}_{\text{\text{Lor}}}\left(1\pm\sqrt{1-\frac{\mathcal{J}_{\text{Lor}}^{2}}{\mathcal{M}_{\text{Lor}}^{2}\ell^{2}}}\right)\;.
\end{equation}
The inverse temperature $\beta$, and the chemical potential $\mu_{\mathcal{J}_{\text{Lor}}}$
can be obtained using (\ref{eq:betamicrocan}) and (\ref{eq:betamumicrocan}),
which in terms of $\xi_{\pm}$ read 
\begin{align}
\beta & =\frac{\ell}{2}\left(\xi_{+}+\xi_{-}\right),\\
\beta\mu_{\mathcal{J}_{\text{Lor}}} & =-\frac{1}{2}\left(\xi_{+}-\xi_{-}\right),
\end{align}
where $\xi_{\pm}$ is defined through the Lorentzian continuation
of (\ref{eq:xi}), i.e. $\xi_{\pm}=\sqrt{\frac{\pi k}{2\mathcal{L}^{\pm}}}$.
This justifies the terminology used from the outset of ``temperature''
and ``chemical potential'' for the $\xi$'s. One may rewrite these
expressions as 
\begin{align}
\beta & =\frac{2\pi r_{+}\ell^{2}}{r_{+}^{2}-r_{-}^{2}}\;,\label{betabtz-1}\\
\beta\mu_{\mathcal{J}_{\text{Lor}}} & =\frac{r_{-}}{\ell r_{+}}\;,\label{betamubtz-1}
\end{align}
to make manifest that they coincide with the ones coming from the
metric formalism (eqs. (\ref{betabtz}) and (\ref{betamubtz})).

This confirms that, as it should be the case, both equations $\Theta=2\pi$,
and $N^{\varphi}\left(r_{+}\right)=0$ of the metric formalism, are
captured by demanding that the improved action should have an extremum
under variations of the complex $\mathcal{L}\left(r_{+}\right)$.

We finally note that in order for the entropy to be real and positive,
there is a bound on the charges, namely $\mathcal{L}^{\pm}\geq0$,
which is equivalent to (\ref{JsmallerM}). When the bound is saturated
(``extremal case''), the holonomy along the thermal circle is nontrivial,
signaling a different topology.

\section{$N=3$, $W_{3}$, black hole. Spins 2 and 3}

\label{HS} \setcounter{equation}{0}

The preceding discussion of the pure gravity (2+1) black hole in terms
of a Chern-Simons connection for $sl\left(2,\mathbb{C}\right)$ in
Euclidean spacetime suggests an immediate generalization. One simply
replaces $sl\left(2,\mathbb{C}\right)$ by $sl\left(3,\mathbb{C}\right)$
or, more generally by $sl\left(N,\mathbb{C}\right)$. The asymptotic
symmetry algebra is then complexified $W_{N}$ algebra, which contains
the complexified Virasoro algebra as a subalgebra. In the Lorentzian
continuation one has two copies of the real $W_{N}$ algebra.

There was a pioneering proposal to define a higher spin black hole
in this way \cite{GK,AGKP}, but as it is shown in detail in Appendix
\ref{AppendixB} of this paper, the solution derived in \cite{GK,AGKP}
actually belongs to the so called ``diagonal embedding'' of $sl\left(2,\mathbb{R}\right)$
in $sl\left(3,\mathbb{R}\right)$, with generators having charges
of \emph{lower} spins (1, 3/2 and 2).

In contradistinction, we will exhibit in this section a black hole
which belongs to the other, ``principal embedding''. It has charges
of spin two and three, and possesses two real copies of $W_{3}$ as
its asymptotic symmetry algebra.

\subsection{Principal embedding - Adapted generators\label{sub:Action-and-equationsW3}}

The Lorentzian action for the $sl\left(3,\mathbb{R}\right)\oplus sl\left(3,\mathbb{R}\right)$
theory takes again the form (\ref{ICS}), where now $k_{2}$ is replaced
by $k_{3}=k/4=\ell/16G$. The connections $A^{\pm}$ belong now to
the algebra $sl\left(3,\mathbb{R}\right)$ which we consider in the
basis $\{L_{i},W_{m}\}$, 
\begin{align}
\left[L_{i},L_{j}\right] & =\left(i-j\right)L_{i+j}\;,\\
\left[L_{i},W_{m}\right] & =\left(2i-m\right)W_{i+m}\;,\\
\left[W_{m},W_{n}\right] & =-\frac{1}{3}\left(m-n\right)\left(2m^{2}+2n^{2}-mn-8\right)L_{m+n}.
\end{align}
Here $i,j=-1,0,1$ and $m,n=-2,-1,0,1,2$. The trace in the action
is taken in the defining representation of the algebra, which is formed
by $3\times3$ matrices. An explicit realization of the basis is given
by 
\[
L_{-1}=\begin{pmatrix}0 & -\sqrt{2} & 0\\
0 & 0 & -\sqrt{2}\\
0 & 0 & 0
\end{pmatrix}\quad;\quad L_{0}=\begin{pmatrix}1 & 0 & 0\\
0 & 0 & 0\\
0 & 0 & -1
\end{pmatrix}\quad;\quad L_{1}=\begin{pmatrix}0 & 0 & 0\\
\sqrt{2} & 0 & 0\\
0 & \sqrt{2} & 0
\end{pmatrix}\;,
\]
\begin{equation}
W_{-2}=\begin{pmatrix}0 & 0 & 4\\
0 & 0 & 0\\
0 & 0 & 0
\end{pmatrix}\quad;\quad W_{-1}=\begin{pmatrix}0 & -\sqrt{2} & 0\\
0 & 0 & \sqrt{2}\\
0 & 0 & 0
\end{pmatrix}\quad;\quad W_{0}=\frac{2}{3}\begin{pmatrix}1 & 0 & 0\\
0 & -2 & 0\\
0 & 0 & 1
\end{pmatrix}\;,
\end{equation}
\[
W_{1}=\begin{pmatrix}0 & 0 & 0\\
\sqrt{2} & 0 & 0\\
0 & -\sqrt{2} & 0
\end{pmatrix}\quad;\quad W_{2}=\begin{pmatrix}0 & 0 & 0\\
0 & 0 & 0\\
4 & 0 & 0
\end{pmatrix}\;.
\]
These matrices obey 
\begin{eqnarray}
L_{i}^{\dagger} & = & \left(-1\right)^{i}L_{-i},\\
W_{m}^{\dagger} & = & \left(-1\right)^{m}W_{-m}.
\end{eqnarray}

The basis elements $L_{i}$ generate the $sl\left(2,\mathbb{R}\right)$
subalgebra that is principally embedded. The basis elements $W_{m}$
generate the $sl\left(2,\mathbb{R}\right)$-spin 2 representation,
with $W_{m}$ being a state of spin $-m$ along $L_{0}$.

One may rewrite the action in Hamiltonian form exactly as before (\ref{IHam}),
(\ref{constraint}).

\subsection{Asymptotic symmetries\label{sub:Asymptotic-symmetriesW3}}

The boundary conditions (\ref{B-H-CONN1}), (\ref{BH-CONN2}), (\ref{b-hcONN3})
on a $t=const$ surface are generalized to \cite{Henneaux-HS,Theisen-HS}
\begin{equation}
A_{\varphi}^{\pm}\left(r,\varphi\right)\underset{r\rightarrow\infty}{\longrightarrow}L_{\pm1}-\frac{2\pi}{k}\mathcal{L}^{\pm}\left(r,\varphi\right)L_{\mp1}-\frac{\pi}{2k}\mathcal{W}^{\pm}\left(r,\varphi\right)W_{\mp2}\;,\label{asymptw3phi-1}
\end{equation}
with 
\begin{align}
\mathcal{L}^{\pm}\left(r,\varphi\right) & \underset{r\rightarrow\infty}{\longrightarrow}\mathcal{L}^{\pm}\left(\varphi\right)+O\left(\frac{1}{r}\right)\;,\label{asymptw3l-1}\\
\mathcal{W}^{\pm}\left(r,\varphi\right) & \underset{r\rightarrow\infty}{\longrightarrow}\mathcal{W}^{\pm}\left(\varphi\right)+O\left(\frac{1}{r}\right)\;,\label{asymptw3w-1}
\end{align}
and 
\begin{equation}
A_{r}^{\pm}\underset{r\rightarrow\infty}{\longrightarrow}O\left(\frac{1}{r}\right)\;,\label{asymptw3r-1}
\end{equation}
where the fields $\mathcal{L}^{\pm}\left(r,\varphi\right)$ and $\mathcal{W}^{\pm}\left(r,\varphi\right)$
enter in (\ref{asymptw3phi-1}) along the lowest (highest)-weight
generators of the principal embedding.

A direct computation \cite{HPTT} yields that the most general $A_{t}^{\pm}$
which preserves the boundary conditions (\ref{asymptw3phi-1})-(\ref{asymptw3r-1})
is given by 
\begin{align}
A_{t}^{\pm}\underset{r\rightarrow\infty}{\longrightarrow} & \pm\left[\xi_{\pm}L_{\pm1}+\eta_{\pm}W_{\pm2}\mp\xi_{\pm}^{\prime}L_{0}\mp\eta_{\pm}^{\prime}W_{\pm1}+\frac{1}{2}\left(\xi_{\pm}^{\prime\prime}-\frac{4\pi}{k}\xi_{\pm}\mathcal{L}^{\pm}+\frac{8\pi}{k}\mathcal{W}^{\pm}\eta_{\pm}\right)L_{\mp1}\right.\nonumber \\
 & -\left(\frac{\pi}{2k}\mathcal{W}^{\pm}\xi_{\pm}+\frac{7\pi}{6k}\mathcal{L}^{\pm\prime}\eta_{\pm}^{\prime}+\frac{\pi}{3k}\eta_{\pm}\mathcal{L}^{\pm\prime\prime}+\frac{4\pi}{3k}\mathcal{L}_{\pm}\eta_{\pm}^{\prime\prime}\right.\left.-\frac{4\pi^{2}}{k^{2}}\left(\mathcal{L}^{\pm}\right)^{2}\eta_{\pm}-\frac{1}{24}\eta_{\pm}^{\prime\prime\prime\prime}\right)W_{\mp2}\nonumber \\
 & \left.+\frac{1}{2}\left(\eta_{\pm}^{\prime\prime}-\frac{8\pi}{k}\mathcal{L}^{\pm}\eta_{\pm}\right)W_{0}\mp\frac{1}{6}\left(\eta_{\pm}^{\prime\prime\prime}-\frac{8\pi}{k}\eta_{\pm}\mathcal{L}^{\pm\prime}-\frac{20\pi}{k}\mathcal{L}^{\pm}\eta_{\pm}^{\prime}\right)W_{\mp1}\right]\ ,
\end{align}
where 
\begin{align}
\xi_{\pm}\left(r,\varphi\right)\underset{r\rightarrow\infty}{\longrightarrow} & \xi_{\pm}\left(\varphi\right)+O\left(\frac{1}{r}\right),\nonumber \\
\eta_{\pm}\left(r,\varphi\right)\underset{r\rightarrow\infty}{\longrightarrow} & \eta_{\pm}\left(\varphi\right)+O\left(\frac{1}{r}\right).\label{BHxi-1}
\end{align}
The generalization of equation (\ref{Lpunto}) is 
\begin{align}
\dot{\mathcal{L}}^{\pm} & =\pm\xi_{\pm}\mathcal{L}^{\pm\prime}\mp2\eta_{\pm}\mathcal{W}^{\pm\prime}\mp3\mathcal{W}^{\pm}\eta_{\pm}^{\prime}\pm2\mathcal{L}^{\pm}\xi_{\mathcal{\pm}}^{\prime}\mp\frac{k}{4\pi}\xi_{\mathcal{\pm}}^{\prime\prime\prime}\ ,\\
\dot{\mathcal{W}}^{\pm} & =\pm\xi_{\pm}\mathcal{W}^{\pm\prime}\pm\frac{2}{3}\eta_{\pm}\left(\mathcal{L}^{\pm\prime\prime\prime}-\frac{16\pi}{k}\left(\mathcal{L}^{\pm}\right)^{2\prime}\right)\pm3\mathcal{W}^{\pm}\xi_{\mathcal{\pm}}^{\prime}\pm3\left(\mathcal{L}^{\pm\prime\prime}-\frac{64\pi}{9k}\left(\mathcal{L}^{\pm}\right)^{2}\right)\eta_{\pm}^{\prime}\nonumber \\
 & \pm5\eta_{\pm}^{\prime\prime}\mathcal{L}^{\pm\prime}\pm\frac{10}{3}\mathcal{L}^{\pm}\eta_{\pm}^{\prime\prime\prime}\mp\frac{k}{12\pi}\eta_{\pm}^{\left(5\right)}\ .
\end{align}
In the case where the gauge parameters are chosen as $\xi_{\pm}=1$,
$\eta_{\pm}=0$, these equations reduce to the familiar chiral equations
$\dot{\mathcal{L}}^{\pm}=\pm\mathcal{L}^{\pm\prime}$ and $\dot{\mathcal{W}}^{\pm}=\pm\mathcal{W}^{\pm\prime}$.

The boundary terms (\ref{Binfty-1}) now become 
\begin{equation}
B_{\infty}^{\pm}\left[\xi,\eta\right]=\mp\int\left[\xi_{\pm}\left(\varphi\right)\mathcal{L}^{\pm}\left(\varphi\right)-\eta_{\pm}\left(\varphi\right)\mathcal{W}^{\pm}\left(\varphi\right)\right]d\varphi dt\;.
\end{equation}

Just as before, if $\mathcal{L},\mathcal{W}$ are expanded in Fourier
modes according to 
\begin{align}
\mathcal{L}^{\pm} & =\frac{1}{2\pi}\sum_{n}\mathcal{L}_{n}^{\pm}e^{in\varphi},\\
\mathcal{W}^{\pm} & =\frac{1}{2\pi}\sum_{n}\mathcal{W}_{n}^{\pm}e^{in\varphi},
\end{align}
one finds that the $\mathcal{L}_{n},\mathcal{W}_{n}$ obey, in terms
of the Poisson bracket, the $W_{3}$ algebra with the same classical
central charge $c=6k=3\ell/2G$ as in pure gravity \cite{Henneaux-HS,Theisen-HS},
\begin{align}
i\left\{ \mathcal{L}_{m},\mathcal{L}_{n}\right\}  & =\left(m-n\right)\mathcal{L}_{m+n}+\frac{k}{2}m^{3}\delta_{m+n,0}\ ,\\
i\left\{ \mathcal{L}_{m},\mathcal{W}_{n}\right\}  & =\left(2m-n\right)\mathcal{W}_{m+n}\ ,\label{LWspin3}\\
i\left\{ \mathcal{W}_{m},\mathcal{W}_{n}\right\}  & =\frac{1}{3}\left(m-n\right)\left(2m^{2}-mn+2n^{2}\right)\mathcal{L}_{m+n}+\frac{16}{3k}\left(m-n\right)\Lambda_{m+n}+\frac{k}{6}m^{5}\delta_{m+n,0}\ ,
\end{align}
where 
\begin{equation}
\Lambda_{n}=\sum_{m}\mathcal{L}_{n-m}\mathcal{L}_{m}\ .
\end{equation}
The bracket relation (\ref{LWspin3}) implies that the $\mathcal{W}_{n}$
generators have conformal weight 3.

\subsection{Black hole\label{sub:(2+1)-higher-spinW3}}

To construct the higher spin black hole, which will be endowed with
charges of conformal weight two and three, one works in Euclidean
spacetime keeping the topology as that of a solid torus. One again
defines the thermodynamics in the rest frame where the only nonvanishing
charges are now $\mathcal{L}_{0}$ and $\mathcal{W}_{0}$. The connection
is complexified just as in the pure gravity case, and the rules for
connecting the Euclidean and Lorentzian schemes remain the same. Note,
however, that because $W_{2}^{\dagger}=W_{-2}$, the correspondence
between $\mathcal{W}^{-}$ and $\mathcal{W}$ is $\mathcal{W}^{-}=-\mathcal{W}^{*}$.
Similarly, one has $\eta_{-}=-\eta^{*}$.

The Euclidean connection for the black hole must solve the zero curvature
condition and possess the $W_{3}$-asymptotics just described. It
is explicitly given by 
\begin{eqnarray}
A_{\varphi} & = & L_{1}-\frac{2\pi}{k}\mathcal{L}L_{-1}-\frac{\pi}{2k}\mathcal{W}W_{-2}\;,\label{eq:Aphiapin3}\\
A_{\tau} & = & -i\xi\left(L_{1}-\frac{2\pi}{k}\mathcal{L}L_{-1}-\frac{\pi}{2k}\mathcal{W}W_{-2}\right)\nonumber \\
 &  & -i\eta\left(W_{2}-\frac{4\pi}{k}\mathcal{L}W_{0}+\frac{4\pi^{2}}{k^{2}}\mathcal{L}^{2}W_{-2}+\frac{4\pi}{k}\mathcal{W}L_{-1}\right)\;,\label{eq:Aphitau3}
\end{eqnarray}
where ${\mathcal{L}}$, ${\mathcal{W}}$, $\xi$ and $\eta$ are all
constant. Its Lorentzian continuation is \cite{HPTT} 
\begin{align}
A_{\varphi}^{\pm} & =L_{\pm1}-\frac{2\pi}{k}\mathcal{L}^{\pm}L_{\mp1}-\frac{\pi}{2k}\mathcal{W}^{\pm}W_{\mp2}\;,\label{eq:AphiLorW3}\\
A_{t}^{\pm} & =\pm\left[\xi_{\pm}\left(L_{\pm1}-\frac{2\pi}{k}\mathcal{L}^{\pm}L_{\mp1}-\frac{\pi}{2k}\mathcal{W}^{\pm}W_{\mp2}\right)\right.\nonumber \\
 & \left.+\eta_{\pm}\left(W_{\pm2}+\frac{4\pi}{k}\mathcal{W}^{\pm}L_{\mp1}+\frac{4\pi^{2}}{k^{2}}\left(\mathcal{L}^{\pm}\right)^{2}W_{\mp2}-\frac{4\pi}{k}\mathcal{L}^{\pm}W_{0}\right)\right]\ .\label{eq:ALorspin3}
\end{align}

One sees a new feature, namely that $A_{\tau}$ is not proportional
to $A_{\varphi}$ as in the pure gravity case, but it acquires a new
piece which multiplies the new parameter $\eta$, which is now brought
in together with the new charge $\mathcal{W}$. Even in the presence
of this new piece, one still has 
\begin{equation}
\left[A_{\tau},A_{\varphi}\right]=0\;,\label{eq:comm}
\end{equation}
so that the zero-curvature condition $F_{\tau\varphi}=0$ holds%
\footnote{The fact that $A_{\varphi}$ and $A_{\tau}$ play very different roles
was emphasized in \cite{Banados:2012ue}. It was also stressed there
that the vanishing of the commutator of both connection components
(eq. (\ref{eq:comm})) was the condition for identifying the most
general $A_{\tau}$ compatible with the form (\ref{eq:Aphiapin3})
of $A_{\varphi}$. However, the authors maintained that eqs. (\ref{eq:Aphiapin3})
applied to the black hole in refs. \cite{GK,AGKP}, which is not the
case as explained in detail in Appendix \ref{AppendixB}.%
}.

The statement of regularity at the origin now reads%
\footnote{For a generic $N$ the regularity condition is $H_{\tau}=\left(-1\right)^{N+1}\mathbbm{1}$,
where one employs the representation in terms of smallest matrices
($2\times2$ for $N=2$, $3\times3$ for $N=3$).%
} 
\begin{equation}
H_{\tau}=\left.e^{\int_{r_{+}}A_{\tau}d\tau}\right|_{\text{on-shell}}=\left.e^{A_{\tau}\left(r_{+}\right)}\right|_{\text{on-shell}}=\mathbbm{1}\;.\label{holonomytau-2-1}
\end{equation}
One way to see that one must take the $+$ sign in this expression
is to consider the solution with zero spin-3 parameters ($\mathcal{W}=0$,
$\eta=0$). The connection reduces then exactly to that of the pure
gravity black hole, but with generators $L_{\pm1}$, $L_{0}$ in the
three-dimensional vector representation of $sl(2,\mathbb{R})$ for
which (\ref{holonomytau-2-1}) indeed holds.

We call the above solution a ``higher spin black hole'' not only
because it possesses nonvanishing higher spin charges when ${\mathcal{W}}$
is not equal to zero, but also because it is endowed with well-defined
temperature and entropy, as we shall discuss in the next section.
One could write metrics associated with the above connection that
would have event horizons. But these metrics are gauge-dependent.
The corresponding causal concepts are not invariant under the spin-3
gauge transformations. Studying the geometrical properties of these
metrics might lead therefore to misleading conclusions. For this reason,
we shall not even attempt constructing here a metric associated with
the black hole connection.

The black hole solution described in this paper shares several features
with the proposal in \cite{GK,AGKP}, that gave rise to all the subsequent
study of higher spin black holes. In particular, it has the same temporal
component $A_{\tau}$ of the connection. However, it differs from
it in the angular component $A_{\varphi}$. While (\ref{eq:Aphiapin3})
fulfills the boundary conditions, the connection $A_{\varphi}$ of
\cite{GK,AGKP} has extra terms that violate these boundary conditions.
This is a crucial difference because, as emphasized in the words of
Fock quoted at the beginning of this article, a theory is defined
not only by the equations of motion but also by the boundary conditions.
A configuration that solves the equations of motion without obeying
the boundary conditions is not a solution of the theory.

In the search for black holes endowed with higher spin charges, it
was argued in \cite{GK,AGKP} that in order to introduce chemical
potentials it was necessary to modify the boundary conditions for
both $A_{\varphi}$ and $A_{\tau}$. However, it was indicated in
\cite{HPTT} that this was not the case and that one should rather
keep the boundary conditions for $A_{\varphi}$ unchanged and introduce
the chemical potentials through a modification of $A_{\tau}$ only
as in (\ref{eq:Aphitau3}). Following the latter, canonical, procedure
one indeed obtains a black hole with higher spin charges, the thermodynamics
of which will be discussed next. With the former procedure one obtains
a particular black hole with lower spin charges as shown in Appendix
\ref{AppendixB}. The black hole of refs. \cite{GK,AGKP} is not a
$W_{3}$ black hole, it is a $W_{3}^{\left(2\right)}$ black hole.


\subsection{Thermodynamics}

The previous discussion for the $sl\left(2,\mathbb{C}\right)$ case
extends straightforwardly to $sl\left(3,\mathbb{C}\right)$. This
includes, in particular, the validity of the general formula%
\footnote{For arbitrary $N$, $k_{3}$ is replaced by $k_{N}=6k/N\left(N^{2}-1\right)=3\ell/2GN\left(N^{2}-1\right)$.%
} 
\begin{align}
S & =-2k_{3}\text{Im}\left(\text{tr}\left[A_{\tau}^{\text{on-shell}}\left(r_{+}\right)A_{\varphi}^{\text{on-shell}}\left(r_{+}\right)\right]\right)\;,\label{sataphi-2-1}
\end{align}
for the entropy, where $k_{3}=k/4=\ell/16G$. Given the form of the
connection, one may rewrite equivalently the entropy as 
\begin{align}
S & =4\pi\left[\xi\mathcal{L}-\frac{3}{2}\eta\mathcal{W}+\xi^{*}\mathcal{L}^{*}-\frac{3}{2}\eta^{*}\mathcal{W}^{*}\right]_{\text{on-shell}}\ .\label{sataphi-2-1-22}
\end{align}

The regularity condition (\ref{holonomytau-2-1}) can be easily implemented
by requiring that the eigenvalues of $A_{\tau}^{\text{on-shell}}\left(r_{+}\right)$
be $\lambda_{\tau}=0,\pm2i\pi$, and using them in the characteristic
polynomial of an $sl\left(3,\mathbb{C}\right)$ matrix 
\begin{equation}
\lambda_{\tau}^{3}-\frac{1}{2}\text{tr}\left[A_{\tau}\left(r_{+}\right)^{2}\right]\lambda_{\tau}-\text{det}\left[A_{\tau}\left(r_{+}\right)\right]=0.
\end{equation}
This yields 
\begin{equation}
\text{det}\left[A_{\tau}^{\text{on-shell}}\left(r_{+}\right)\right]=0\quad;\quad\text{tr}\left[A_{\tau}^{\text{on-shell}}\left(r_{+}\right)^{2}\right]+8\pi^{2}=0.\label{eq:RegN3}
\end{equation}
For the black hole connection, these two conditions take the form
\begin{align}
2^{11}\pi^{2}\mathcal{L}^{3}\eta^{3}+3^{3}k^{2}\mathcal{W}\xi^{3}-2^{5}3^{2}\pi k\eta\left(3\mathcal{W}^{2}\eta^{2}-3\eta\xi\mathcal{LW}+2\mathcal{L}^{2}\xi^{2}\right) & =0\ ,\\
\frac{2^{6}}{3k^{2}}\mathcal{L}^{2}\eta^{2}+\frac{2}{\pi k}\xi\left(\mathcal{L}\xi-3\eta\mathcal{W}\right)-1 & =0\ .
\end{align}
The solution to these equations is generically 
\begin{align}
\xi & =\sqrt{\frac{\pi k}{2\mathcal{L}}}\frac{\cos\left(\frac{2\Phi}{3}\right)}{\cos\left(\Phi\right)}\;,\label{eq:xi1}\\
\eta & =\frac{\sqrt{3}k}{8\mathcal{L}}\frac{\sin\left(\frac{\Phi}{3}\right)}{\cos\left(\Phi\right)}\;,\label{eq:eta1}
\end{align}
with 
\begin{equation}
\Phi=\arcsin\left(\frac{3}{8}\sqrt{\frac{3k}{2\pi\mathcal{L}^{3}}}\mathcal{W}\right)\ .\label{PhiLW-1-1}
\end{equation}
When these expressions are inserted in (\ref{sataphi-2-1-22}) one
obtains for the entropy 
\begin{equation}
S=4\pi\sqrt{2\pi k}\text{Re}\left(\sqrt{\mathcal{L}}\cos\left[\frac{1}{3}\arcsin\left(\frac{3}{8}\sqrt{\frac{3k}{2\pi\mathcal{L}^{3}}}\mathcal{W}\right)\right]\right)\;.\label{eq:entropyspin3}
\end{equation}

The Lorentzian continuation of the black hole entropy is then given
by 
\begin{align}
S & =2\pi\sqrt{2\pi k}\left(\sqrt{\mathcal{L}^{+}}\cos\left[\frac{1}{3}\arcsin\left(\frac{3}{8}\sqrt{\frac{3k}{2\pi\left(\mathcal{L}^{+}\right)^{3}}}\mathcal{W}^{+}\right)\right]\right.\nonumber \\
 & \left.+\sqrt{\mathcal{L}^{-}}\cos\left[\frac{1}{3}\arcsin\left(\frac{3}{8}\sqrt{\frac{3k}{2\pi\left(\mathcal{L}^{-}\right)^{3}}}\mathcal{W}^{-}\right)\right]\right).
\end{align}
The arcsine function is multivalued. The branch connected with the
(2+1) pure gravity black hole is the one such that the Lorentzian
continuation of the ``angle $\Phi$'' 
\begin{equation}
\Phi^{\pm}=\arcsin\left(\frac{3}{8}\sqrt{\frac{3k}{2\pi\left(\mathcal{L}^{\pm}\right)^{3}}}\mathcal{W}^{\pm}\right),\label{eq:anglepm}
\end{equation}
lies in the range $-\pi/2<\Phi^{\pm}\leq\pi/2$. The other branches
are disconnected from the (2+1) pure gravity black hole.

In order for the entropy to be real, a bound on the higher spin charges
$\mathcal{W}^{\pm}$ in terms of $\mathcal{L}^{\pm}=\frac{1}{4\pi}\left(\mathcal{M}_{\text{Lor}}\ell\pm\mathcal{J}_{\text{Lor}}\right)$
should be obeyed, 
\begin{equation}
\left|\mathcal{W}^{\pm}\right|\leq\frac{8}{3}\sqrt{\frac{2\pi}{3k}}\left(\mathcal{L}^{\pm}\right)^{3/2},
\end{equation}
(in addition to $\mathcal{L}^{\pm}\geqslant0$). When at least one
of the bounds is saturated, the configuration is ``extremal'', in
the sense that the corresponding holonomy along the thermal circle
becomes nontrivial and the topology is different.


To determine the temperature and the chemical potentials in the microcanonical
ensemble, we use the relations 
\begin{align}
\beta & =\left(\frac{\partial S}{\partial\mathcal{M}_{\text{Lor}}}\right)_{\mathcal{J}_{\text{Lor}},\mathcal{W}_{0\pm}},\label{eq:betamicrocan-1}\\
\beta\mu_{\mathcal{J}_{\text{Lor}}} & =-\left(\frac{\partial S}{\partial\mathcal{J}_{\text{Lor}}}\right)_{\mathcal{M}_{\text{Lor}},\mathcal{W}_{0\pm}},\label{eq:betamumicrocan-1}\\
\beta\mu_{\mathcal{W}_{\pm}} & =-\left(\frac{\partial S}{\partial\mathcal{W}_{0\pm}}\right)_{\mathcal{M}_{\text{Lor}},\mathcal{J}_{\text{Lor}}},\label{eq:betamuWmicrocan-1}
\end{align}
with $\mathcal{M}_{\text{Lor}}=\frac{2\pi}{\ell}\left(\mathcal{L}^{+}+\mathcal{L}^{-}\right)$
and $\mathcal{J}_{\text{Lor}}=2\pi\left(\mathcal{L}^{+}-\mathcal{L}^{-}\right)$
as above, and where $\mathcal{W}_{0}^{\pm}=2\pi\mathcal{W}^{\pm}$
are the spin-3 charges.

Note that the charges come in pairs, with one charge for each chirality
in each pair. One can alternatively define charges that are even (sum)
or odd (difference) under chirality. The even charges might be thought
of as electric while the odd ones as magnetic. On the Euclidean side
they correspond to the real and imaginary parts of the Euclidean charges,
the former being invariant under complex conjugation while the latter
reversing sign.

One finds 
\begin{align}
\beta & =\frac{\ell}{2}\left(\xi_{+}+\xi_{-}\right),\\
\beta\mu_{\mathcal{J}_{\text{Lor}}} & =-\frac{1}{2}\left(\xi_{+}-\xi_{-}\right),\\
\beta\mu_{\mathcal{W}^{\pm}} & =\eta_{\pm},
\end{align}
where $\xi_{\pm}$, $\eta_{\pm}$ are given in terms of the charges
$\mathcal{L}^{\pm}$ and $\mathcal{W}^{\pm}$ by the same expressions
(\ref{eq:xi1}), (\ref{eq:eta1}) and (\ref{PhiLW-1-1}) giving $\xi$
and $\eta$ in terms of $\mathcal{L}$ and $\mathcal{W}$. This shows
that indeed, the parameters introduced in the temporal components
of the connection have the anticipated physical interpretation of
being the temperature and chemical potentials.

\section{$N=3,$ $W_{3}^{\left(2\right)}$, black hole. Spins 1, 3/2, 2}

\setcounter{equation}{0} \label{W32-section}

While we dealt with up to now exclusively with the principal embedding
of $sl(2,\mathbb{R})$ in $sl(N,\mathbb{R})$, which has the property
of yielding solutions carrying higher spin charges up to spin $N$,
it is also of interest to consider other embeddings. This is done
in this section.

We consider explicitly again the case $N=3$. In that case, the only
other non-trivial embedding is the so-called diagonal embedding leading
to two real copies of the Bershadsky-Polyakov algebra $W_{3}^{\left(2\right)}$
at infinity. In this section we exhibit the corresponding black hole
which besides the spin $2$ charges is endowed only with lower spin
charges, namely $U\left(1\right)$ and spin $\frac{3}{2}$ charges.
We will also discuss its thermodynamics. The quantum mechanical difficulties
of the field theory associated with the diagonal embedding, such as
the presence of negative norm states \cite{Castro:2012bc}, are not
an obstacle for this semiclassical study, which we deem necessary
for dealing thoroughly with the problem at hand.

A significant consequence, presented in Appendix \ref{AppendixB},
of the analysis of this section is the following: the black hole in
refs. \cite{GK,AGKP}, which was claimed to be higher spin black hole
associated with the principal embedding is, rather a lower spin black
hole associated with the diagonal embedding. What was aimed to be
a $W_{3}$ black hole became instead a $W_{3}^{\left(2\right)}$ black
hole because of the non-canonical way in which the chemical potentials
were introduced. Once this is realized, the ``entropy paradox''
that created controversy in the literature around the black hole in
refs. \cite{GK,AGKP} is resolved.

\subsection{Diagonal embedding - Adapted generators}

It will be convenient in this section to adopt a basis of $sl(3,\mathbb{R})$
generators adapted to the diagonal embedding. These are 
\begin{align}
\hat{L}_{\pm1} & =\pm\frac{1}{4}W_{\pm2}\ \ ,\ \ \hat{L}_{0}=\frac{1}{2}L_{0}\ \ ,\ \ J_{0}=\frac{1}{2}W_{0}\ ,\\
G_{\pm1/2}^{\left[+\right]} & =\frac{1}{2\sqrt{2}}\left(\pm L_{\pm1}-W_{\pm1}\right)\ \ ,\ \ G_{\pm1/2}^{\left[-\right]}=\frac{1}{2\sqrt{2}}\left(L_{\pm1}\pm W_{\pm1}\right)\ ,
\end{align}
and the $sl\left(3,\mathbb{R}\right)$ commutation relations read
in this basis 
\begin{align}
\left[\hat{L}_{i},\hat{L}_{j}\right] & =\left(i-j\right)\hat{L}_{i+j}\ \ ,\ \ \left[\hat{L}_{i},J_{0}\right]=0\ ,\nonumber \\
\left[\hat{L}_{i},G_{m}^{\left[a\right]}\right] & =\left(\frac{i}{2}-m\right)G_{i+m}^{\left[a\right]}\ \ ,\ \ \left[J_{0},G_{m}^{\left[a\right]}\right]=aG_{m}^{\left[a\right]}\ ,\label{sl(3,R)-Diagonal}\\
\left[G_{m}^{\left[+\right]},G_{n}^{\left[-\right]}\right] & =\hat{L}_{m+n}-\frac{3}{2}\left(m-n\right)J_{0}\ ,\nonumber 
\end{align}
with $i=-1$, $0$, $1$, $m=-1/2$, $1/2$, and $a=-1$, $1$. The
basis elements $\hat{L}_{i}$ generate the $sl(2,\mathbb{R})$ subalgebra
that is diagonally embedded. Note that the $G_{m}^{\left[a\right]}$'s
transform in two independent $sl(2,\mathbb{R})$-spin $\frac{1}{2}$
representations, while $J_{0}$ has $sl(2,\mathbb{R})$-spin $s=0$.
The corresponding generators in the asymptotic conformal field theory
have respective conformal weights $\frac{3}{2}$ and $1$, and are
all bosonic since their algebra involves only commutators.

With the above choice, the explicit realization of the generators
is given by 
\[
\hat{L}_{-1}=\begin{pmatrix}0 & 0 & -1\\
0 & 0 & 0\\
0 & 0 & 0
\end{pmatrix}\quad;\quad\hat{L}_{0}=\begin{pmatrix}\frac{1}{2} & 0 & 0\\
0 & 0 & 0\\
0 & 0 & -\frac{1}{2}
\end{pmatrix}\quad;\quad\hat{L}_{1}=\begin{pmatrix}0 & 0 & 0\\
0 & 0 & 0\\
1 & 0 & 0
\end{pmatrix},
\]
\begin{equation}
J_{0}=\frac{1}{3}\begin{pmatrix}1 & 0 & 0\\
0 & -2 & 0\\
0 & 0 & 1
\end{pmatrix}\quad;\quad G_{+1/2}^{\left[+\right]}=\begin{pmatrix}0 & 0 & 0\\
0 & 0 & 0\\
0 & 1 & 0
\end{pmatrix}\quad;\quad G_{-1/2}^{\left[+\right]}=\begin{pmatrix}0 & 1 & 0\\
0 & 0 & 0\\
0 & 0 & 0
\end{pmatrix},
\end{equation}
\[
G_{+1/2}^{\left[-\right]}=\begin{pmatrix}0 & 0 & 0\\
1 & 0 & 0\\
0 & 0 & 0
\end{pmatrix}\quad;\quad G_{-1/2}^{\left[-\right]}=\begin{pmatrix}0 & 0 & 0\\
0 & 0 & -1\\
0 & 0 & 0
\end{pmatrix}.
\]
These matrices obey 
\begin{equation}
\hat{L}_{i}^{\dagger}=\left(-1\right)^{i}\hat{L}_{-i}\ \ ,\ \ J_{0}^{\dagger}=J_{0}\ \ ,\ \ \left(G_{m}^{\left[a\right]}\right)^{\dagger}=\left(-1\right)^{m+\frac{a}{2}}G_{-m}^{\left[-a\right]}\ ,\label{eq:hermpropw32}
\end{equation}

\subsection{Asymptotic symmetries}

Asymptotic conditions with two copies of $W_{3}^{\left(2\right)}$
symmetry have been previously discussed in \cite{AGKP}, \cite{Campoleoni-HS}.
They follow the lines of Hamiltonian reduction \cite{HR1}. Here we
improve $A_{t}$ so as to include the most general motion compatible
with the given $A_{\varphi}$.

The asymptotic form of the spatial connection can be chosen to have
dynamical components only along the lowest (highest)-weight generators,
i.e. on a $t=const$ surface, 
\begin{equation}
A_{\varphi}^{\pm}\underset{r\rightarrow\infty}{\longrightarrow}\hat{L}_{\pm1}-\frac{8\pi}{k}\left[\left(\mathcal{\hat{L}}^{\pm}\left(r,\varphi\right)-\frac{6\pi}{k}\left(\mathcal{U}^{\pm}\left(r,\varphi\right)\right)^{2}\right)\hat{L}_{\mp1}+\frac{3}{2}\mathcal{U}^{\pm}\left(r,\varphi\right)J_{0}+\psi_{\left[a\right]}^{\pm}\left(r,\varphi\right)G_{\mp1/2}^{\left[a\right]}\right]\ ,\label{aThetaW3(2)}
\end{equation}
with 
\begin{align}
\mathcal{\hat{L}}^{\pm}\left(r,\varphi\right)\underset{r\rightarrow\infty}{\longrightarrow} & \mathcal{\hat{L}}^{\pm}\left(\varphi\right)+O\left(\frac{1}{r}\right),\\
\mathcal{U}^{\pm}\left(r,\varphi\right)\underset{r\rightarrow\infty}{\longrightarrow} & \mathcal{U}^{\pm}\left(\varphi\right)+O\left(\frac{1}{r}\right),\\
\psi_{\left[a\right]}^{\pm}\left(r,\varphi\right)\underset{r\rightarrow\infty}{\longrightarrow} & \psi_{\left[a\right]}^{\pm}\left(\varphi\right)+O\left(\frac{1}{r}\right),
\end{align}
and 
\begin{equation}
A_{r}^{\pm}\underset{r\rightarrow\infty}{\longrightarrow}O\left(\frac{1}{r}\right)\;.\label{eq:ArW32}
\end{equation}
As it is standard in this asymptotic analysis context \cite{Henneaux:1999ib},
we have redefined the Virasoro generators by including the square
of the $U(1)$-currents $\mathcal{U}^{\pm}$, as it is necessary for
these $U(1)$-currents to have conformal weight one with respect to
these redefined generators.

The most general $A_{t}^{\pm}$ which preserves the asymptotic form
of (\ref{aThetaW3(2)}) is given by

\begin{align}
A_{t}^{\pm} & \underset{r\rightarrow\infty}{\longrightarrow}\pm\hat{\xi}_{\pm}\hat{L}_{\pm1}\pm\left[-\frac{8\pi}{k}\hat{\xi}_{\pm}\left(\mathcal{\hat{L}}^{\pm}-\frac{6\pi}{k}\left(\mathcal{U}^{\pm}\right)^{2}\right)+\frac{4\pi}{k}\vartheta_{\left[a\right]}^{\pm}\psi_{\left[a\right]}^{\pm}+\frac{1}{2}\hat{\xi}_{\pm}^{\prime\prime}\right]\hat{L}_{\mp1}\nonumber \\
 & \pm\left(\nu_{\pm}-\frac{12\pi}{k}\hat{\xi}_{\pm}\mathcal{U}^{\pm}\right)J_{0}+a\vartheta_{\left[a\right]}^{\pm}G_{\pm1/2}^{\left[-a\right]}-\hat{\xi}_{\pm}^{\prime}\hat{L}_{0}\nonumber \\
 & \mp\left(\frac{12\pi}{k}\vartheta_{\left[-a\right]}^{\pm}\mathcal{U}^{\pm}+\frac{8\pi}{k}\hat{\xi}_{\pm}\psi_{\left[a\right]}^{\pm}-a\vartheta_{\left[-a\right]}^{\pm\prime}\right)G_{\mp1/2}^{\left[a\right]}\ ,\label{LambdaW3(2)}
\end{align}
where 
\begin{align}
\hat{\xi}_{\pm}\left(r,\varphi\right)\underset{r\rightarrow\infty}{\longrightarrow} & \hat{\xi}_{\pm}\left(\varphi\right)+O\left(\frac{1}{r}\right),\\
\nu_{\pm}^{\pm}\left(r,\varphi\right)\underset{r\rightarrow\infty}{\longrightarrow} & \nu_{\pm}^{\pm}\left(\varphi\right)+O\left(\frac{1}{r}\right),\\
\vartheta_{\left[a\right]}^{\pm}\left(r,\varphi\right)\underset{r\rightarrow\infty}{\longrightarrow} & \vartheta_{\left[a\right]}^{\pm}\left(\varphi\right)+O\left(\frac{1}{r}\right).\label{eq:thetar}
\end{align}
The field equations are then given by 
\begin{align}
\dot{\hat{\mathcal{L}}}^{\pm} & =\pm2\hat{\xi}_{\pm}^{\prime}\mathcal{\hat{L}}^{\pm}\pm\hat{\xi}_{\pm}\mathcal{\hat{L}}^{\pm\prime}\mp\frac{k}{16\pi}\hat{\xi}_{\pm}^{\prime\prime\prime}\mp\mathcal{U}^{\pm}\nu_{\pm}^{\prime}\mp\frac{3}{2}\vartheta_{\left[a\right]}^{\pm\prime}\psi_{\left[a\right]}^{\pm}\mp\frac{1}{2}\vartheta_{\left[a\right]}^{\pm}\psi_{\left[a\right]}^{\pm\prime}\ ,\nonumber \\
\dot{\mathcal{U}}^{\pm} & =\pm\hat{\xi}_{\pm}^{\prime}\mathcal{U}^{\pm}\pm\hat{\xi}_{\pm}\mathcal{U}^{\pm\prime}\pm a\vartheta_{\left[a\right]}^{\pm}\psi_{\left[a\right]}^{\pm}\mp\frac{k}{12\pi}\nu_{\pm}^{\prime}\ ,\label{FEaW3(2)CPs}\\
\dot{\psi}_{\left[a\right]}^{\pm} & =\pm\frac{3}{2}\hat{\xi}_{\pm}^{\prime}\psi_{\left[a\right]}^{\pm}\pm\hat{\xi}_{\pm}\psi_{\left[a\right]}^{\pm\prime}\mp a\nu_{\pm}\psi_{\left[a\right]}^{\pm}\mp a\vartheta_{\left[-a\right]}^{\pm}\left(\frac{24\pi}{k}\left(\mathcal{U}^{\pm}\right)^{2}-\mathcal{\hat{L}}^{\pm}-\frac{3}{2}a\mathcal{U}^{\pm\prime}\right)\nonumber \\
 & \pm3\mathcal{U}^{\pm}\vartheta_{\left[-a\right]}^{\pm\prime}\mp\frac{k}{8\pi}a\vartheta_{\left[-a\right]}^{\pm\prime\prime}\ .\nonumber 
\end{align}

Note that if one takes the gauge parameters as $\hat{\xi}_{\pm}=1$,
$\nu_{\pm}=0$ and $\vartheta_{\left[a\right]}^{\pm}=0$, the equations
reduce again to the chiral equations $\dot{\hat{\mathcal{L}}}^{\pm}=\pm\mathcal{\hat{L}}^{\pm\prime}$,
$\dot{\mathcal{U}}^{\pm}=\pm\mathcal{U}^{\pm\prime}$ and $\dot{\psi}_{\left[a\right]}^{\pm}=\pm\psi_{\left[a\right]}^{\pm\prime}$.

The boundary terms (\ref{Binfty-1}) take the form 
\begin{equation}
B_{\infty}^{\pm}\left[\hat{\xi},\nu,\vartheta_{\left[a\right]}\right]=\mp\int\left[\hat{\xi}_{\pm}\left(\varphi\right)\mathcal{\hat{L}}^{\pm}\left(\varphi\right)-\nu_{\pm}\left(\varphi\right)\mathcal{U}^{\pm}\left(\varphi\right)-\vartheta_{\left[a\right]}^{\pm}\left(\varphi\right)\psi_{\left[a\right]}^{\pm}\left(\varphi\right)\right]d\varphi dt\;.\label{eq:Binftyw32}
\end{equation}

It is straightforward to verify that the global charges span two copies
of the $W_{3}^{\left(2\right)}$ algebra. In terms of Fourier modes,
$X=\frac{1}{2\pi}\sum_{m}X_{m}e^{im\varphi}$, this algebra explicitly
reads 
\begin{align}
i\left\{ \mathcal{\hat{L}}_{m},\mathcal{\hat{L}}_{n}\right\}  & =\left(m-n\right)\mathcal{\hat{L}}_{m+n}+\frac{k}{8}m^{3}\delta_{m+n,0}\ ,\nonumber \\
i\left\{ \mathcal{\hat{L}}_{m},\mathcal{U}_{n}\right\}  & =-n\mathcal{U}_{m+n}\ ,\nonumber \\
i\left\{ \mathcal{U}_{m},\mathcal{U}_{n}\right\}  & =\frac{k}{6}m\delta_{m+n,0}\ ,\nonumber \\
i\left\{ \mathcal{\hat{L}}_{m},\psi_{n}^{\left[a\right]}\right\}  & =\left(\frac{1}{2}m-n\right)\psi_{m+n}^{\left[a\right]}\ ,\label{W3(2)-Algebra}\\
\left\{ \mathcal{U}_{n},\psi_{m}^{\left[a\right]}\right\}  & =a\psi_{m+n}^{\left[a\right]}\ ,\nonumber \\
\left\{ \psi_{m}^{\left[+\right]},\psi_{n}^{\left[-\right]}\right\}  & =\mathcal{\hat{L}}_{m+n}-\frac{12}{k}\Lambda_{m+n}+\frac{3i}{2}\left(m-n\right)\mathcal{U}_{m+n}+\frac{k}{4}m^{2}\delta_{m+n,0}\ ,\nonumber 
\end{align}
where 
\begin{equation}
\Lambda_{n}:={\sum\limits _{m}}\mathcal{U}_{n-m}\mathcal{U}_{m}\ .
\end{equation}

If the spinors $\psi_{p}^{\left[\pm\right]}$ are assumed to fulfill
antiperiodic (Neveu-Schwarz) boundary conditions, then $p$ has to
be a half-integer. In this case, it is apparent that the wedge subalgebra
corresponds to $sl\left(3,\mathbb{R}\right)$ in the basis of eq.
(\ref{sl(3,R)-Diagonal}). It is worth pointing out that, in full
analogy with what occurs for the super Virasoro algebra with $\mathcal{N}=2$
\cite{SS}, representations of the $W_{3}^{\left(2\right)}$ algebra
with spinors obeying periodic (Ramond), or antiperiodic boundary conditions,
are equivalent \cite{Bershadsky}. This is because the $U\left(1\right)$
gauge transformations provide an automorphism that can be used to
``gauge away\textquotedblright{} the corresponding phase in the boundary
conditions for the spinors. Therefore, the generators of the algebra
with periodic boundary conditions can be expressed in terms of those
with antiperiodic boundary conditions.

Note that, as observed earlier in \cite{AGKP}, the central charge
is given by $\frac{c}{4}$, where $c=6k=\frac{3\ell}{2G}$ is the
standard one \cite{Brown:1986nw}.

\subsection{Black hole}

The asymptotic conditions (\ref{aThetaW3(2)})-(\ref{eq:thetar})
include black hole solutions carrying, apart from the mass and the
angular momentum, independent $U\left(1\right)$ and spinorial charges.
These solutions are characterized, for the black hole ``at rest'',
by constant coefficients $\mathcal{\hat{L}}^{\pm}$, $\mathcal{U}^{\pm}$,
$\psi_{\left[a\right]}^{\pm}$, $\hat{\xi}_{\pm}$, $\nu_{\pm}$,
$\vartheta_{\left[a\right]}^{\pm}$, a situation that will be assumed
from now on. The constants $\mathcal{\hat{L}}^{\pm}$, $\mathcal{U}^{\pm}$
and $\psi_{\left[a\right]}^{\pm}$ define the charges, while the constants
$\hat{\xi}_{\pm}$, $\nu_{\pm}$, $\vartheta_{\left[a\right]}^{\pm}$
are the corresponding chemical potentials. The black hole corresponds
to the range of the parameters that yields a real positive entropy.

The Euclidean continuation proceeds as before (see Appendix \ref{Appendix0}).
Hence, since the $sl\left(3,\mathbb{R}\right)$ generators fulfill
the relations (\ref{eq:hermpropw32}), the continuation rules imply
now the correspondence 
\begin{align}
\mathcal{\hat{L}} & =\mathcal{\hat{L}}^{+}\ \ ,\ \ \mathcal{U}=\mathcal{U}^{+}\ \ ,\ \ \psi_{\left[a\right]}=\psi_{\left[a\right]}^{+}\ ,\\
\xi & =\xi_{+}\ \ ,\ \ \nu=\nu_{+}\ \ ,\ \ \vartheta_{\left[a\right]}=\vartheta_{\left[a\right]}^{+}\ ,
\end{align}
and 
\begin{align}
\mathcal{\hat{L}}^{*} & =\mathcal{\hat{L}}^{-}\ \ ,\ \ \mathcal{U}^{*}=-\mathcal{U}^{-}\ \ ,\ \ \psi_{\left[a\right]}^{*}=-a\psi_{\left[-a\right]}^{-}\ ,\\
\xi^{*} & =\xi_{-}\ \ ,\ \ \nu^{*}=-\nu_{-}\ \ ,\ \ \vartheta_{\left[a\right]}^{*}=-a\vartheta_{\left[-a\right]}^{-}\ .
\end{align}
The Euclidean black hole then reads 
\begin{align}
A_{\varphi} & =\hat{L}_{1}-\frac{8\pi}{k}\left[\left(\mathcal{\hat{L}}-\frac{6\pi}{k}\mathcal{U}^{2}\right)\hat{L}_{-1}+\frac{3}{2}\mathcal{U}J_{0}+\psi_{\left[a\right]}G_{-1/2}^{\left[a\right]}\right],\label{eq:aphiw32}\\
A_{\tau} & =-i\left[\hat{\xi}\left(\hat{L}_{1}-\frac{8\pi}{k}\left[\left(\mathcal{\hat{L}}-\frac{6\pi}{k}\mathcal{U}^{2}\right)\hat{L}_{-1}+\frac{3}{2}\mathcal{U}J_{0}+\psi_{\left[a\right]}G_{-1/2}^{\left[a\right]}\right]\right)\right.\nonumber \\
 & \left.+\nu J_{0}+\vartheta_{\left[a\right]}\left(aG_{1/2}^{\left[-a\right]}-\frac{12\pi}{k}\mathcal{U}G_{-1/2}^{\left[-a\right]}+\frac{4\pi}{k}\psi_{\left[a\right]}\hat{L}_{-1}\right)\right]\ ,\label{eq:atw32}
\end{align}
with 
\begin{equation}
a\vartheta_{\left[a\right]}\psi_{\left[a\right]}=0\ ;\ \ \vartheta_{\left[-a\right]}\left(\frac{24\pi}{k}\mathcal{U}^{2}-\mathcal{\hat{L}}\right)+\nu\psi_{\left[a\right]}=0.\ \label{OnShellConds-W32}
\end{equation}
The fields $\mathcal{\hat{L}}$, $\mathcal{U}$, $\psi_{\left[a\right]}$,$\ $and
the chemical potentials $\hat{\xi}$, $\nu$, $\vartheta_{\left[a\right]}$
are complex constants. The algebraic constraints (\ref{OnShellConds-W32})
are a new feature of the diagonal embedding, which does not appear
in the principal embedding. They are necessary to guarantee $F_{t\varphi}^{\pm}=0$.
The constraints (\ref{OnShellConds-W32}) will turn out to be important
when discussing black holes below.

\subsection{Thermodynamics}

The black hole entropy can be readily obtained from the general expression
in eq. (\ref{sataphi}), which for this case reduces to 
\begin{align}
S & =8\pi\operatorname{Re}\left[\hat{\xi}\mathcal{\hat{L}}-\frac{1}{2}\nu\mathcal{U}-\frac{3}{4}\mathcal{\vartheta}_{\left[a\right]}\psi_{\left[a\right]}\right]_{\text{on-shell}}\ ,\label{SmarrLaw-Entropy}
\end{align}

The chemical potentials are related to the charges through: (i) the
regularity conditions that the holonomy along the thermal circle is
trivial (\ref{eq:RegN3}), i.e., 
\begin{equation}
\det\left[A_{\tau}\right]=0\ \ ;\ \ \text{tr}\left[\left(A_{\tau}\right)^{2}\right]+8\pi^{2}=0,\label{Holo-W3(2)}
\end{equation}
\textit{and,} (ii) the constraints (\ref{OnShellConds-W32}) are welcome
features since the two regularity conditions by themselves form an
undetermined system of equations for the four chemical potentials.

When fully developed, the conditions (\ref{Holo-W3(2)}) read 
\begin{align}
0 & =-\left(\mathcal{U}^{3}-\frac{k}{8\pi}\left(\mathcal{U\hat{L}}+\psi_{\left[-\right]}\psi_{\left[+\right]}\right)\right)\hat{\xi}^{3}+\frac{k}{8\pi}\left(\left(\mathcal{U}^{2}-\frac{k}{12\pi}\mathcal{\hat{L}}\right)\nu+\frac{3}{2}\mathcal{U}\vartheta_{\left[a\right]}\psi_{\left[a\right]}\right)\hat{\xi}^{2}\nonumber \\
 & -\frac{k^{2}}{192\pi^{2}}\left(\mathcal{U}\nu^{2}-\frac{72\pi}{k}\left(\mathcal{U}^{2}+\frac{k}{24\pi}\mathcal{\hat{L}}\right)\vartheta_{\left[-\right]}\vartheta_{\left[+\right]}\right)\hat{\xi}+\frac{1}{4}\left(\frac{k}{12\pi}\nu\right)^{3}\label{Holo-det-W32}\\
 & -\frac{k^{2}}{64\pi^{2}}\left(\mathcal{U}\nu+\frac{1}{2}\vartheta_{\left[a\right]}\psi_{\left[a\right]}\right)\vartheta_{\left[-\right]}\vartheta_{\left[+\right]}\ ,\nonumber 
\end{align}
and 
\begin{equation}
\mathcal{\hat{L}}\hat{\xi}^{2}-\left(\mathcal{U}\nu+\frac{3}{2}\vartheta_{\left[a\right]}\psi_{\left[a\right]}\right)\hat{\xi}-3\mathcal{U}\vartheta_{\left[-\right]}\vartheta_{\left[+\right]}+\frac{k}{24\pi}\nu^{2}-\frac{1}{2}\pi k=0\ ,\label{Holo-quadratic-W32}
\end{equation}
respectively. Together with the equations (\ref{OnShellConds-W32}),
they form a nonlinear system which admit various branches of solutions.

We will focus hereafter on the generic case, for which the charges
as well as the chemical potentials are not fine tuned. In this case,
it is useful to parametrize the chemical potentials according to 
\begin{align}
\hat{\xi} & =\sqrt{\frac{\pi k}{2\mathcal{\hat{L}}}}\left(\frac{\cos\left(\frac{2\Phi}{3}\right)}{\cos\left(\Phi\right)}+\mathcal{U}\sqrt{\frac{24\pi}{k\mathcal{\hat{L}}}}\frac{\sin\left(\frac{\Phi}{3}\right)}{\cos\left(\Phi\right)}\right)\ ,\label{eq:xihatintermscharges}\\
\nu & =-2\sqrt{3}\pi\left(1-\frac{24\pi}{k}\frac{\mathcal{U}^{2}}{\mathcal{\hat{L}}}\right)\frac{\sin\left(\frac{\Phi}{3}\right)}{\cos\left(\Phi\right)}\ ,\label{CPs-W3(2)}\\
\vartheta_{\left[a\right]} & =-2\sqrt{3}\pi\left(\frac{\psi_{\left[-a\right]}}{\mathcal{\hat{L}}}\right)\frac{\sin\left(\frac{\Phi}{3}\right)}{\cos\left(\Phi\right)}\ ,
\end{align}
so that the field equations (\ref{OnShellConds-W32}) and the conditions
(\ref{Holo-det-W32}), (\ref{Holo-quadratic-W32}) are solved provided
\begin{align}
\Phi & =\arcsin\left[24\sqrt{\frac{6\pi^{3}}{k^{3}\mathcal{\hat{L}}^{3}}}\left(\mathcal{U}^{3}-\frac{k}{8\pi}\left(\mathcal{U\hat{L}}+\psi_{\left[-\right]}\psi_{\left[+\right]}\right)\right)\right]\mathcal{\ }.\label{Phi-W32}
\end{align}

By virtue of (\ref{eq:xihatintermscharges})-(\ref{Phi-W32}), the
entropy (\ref{SmarrLaw-Entropy}) can be manifestly expressed in terms
of the global charges as 
\begin{equation}
S=4\pi\sqrt{2\pi k}\text{Re}\left[\sqrt{\mathcal{\hat{L}}}\cos\left(\frac{1}{3}\arcsin\left[24\sqrt{\frac{6\pi^{3}}{k^{3}\mathcal{\hat{L}}^{3}}}\left(\mathcal{U}^{3}-\frac{k}{8\pi}\left(\mathcal{U\hat{L}}+\psi_{\left[-\right]}\psi_{\left[+\right]}\right)\right)\right]\right)\right].\label{eq:SW32}
\end{equation}
In terms of the Lorentzian charges the entropy then reads 
\begin{equation}
S=2\pi\sqrt{2\pi k\mathcal{\hat{L}}^{+}}\cos\left(\frac{\Phi_{+}}{3}\right)+2\pi\sqrt{2\pi k\mathcal{\hat{L}}^{-}}\cos\left(\frac{\Phi_{-}}{3}\right)\ ,\label{S-W3(2)-Lorentzian}
\end{equation}
with 
\begin{equation}
\Phi_{\pm}:=\arcsin\left[24\sqrt{\frac{6\pi^{3}}{k^{3}\left(\mathcal{\hat{L}}^{\pm}\right)^{3}}}\left(\left(\mathcal{U}^{\pm}\right)^{3}-\frac{k}{8\pi}\left(\mathcal{U}^{\pm}\mathcal{\hat{L}}^{\pm}+\psi_{\left[+\right]}^{\pm}\psi_{\left[-\right]}^{\pm}\right)\right)\right]\ ,
\end{equation}
where the ``angular variables'' $\Phi_{\pm}$\ range as $-\frac{3\pi}{2}<\Phi_{\pm}<\frac{3\pi}{2}$.
Note that the branch that is connected with the pure gravity black
hole corresponds to $-\frac{\pi}{2}<\Phi_{\pm}<\frac{\pi}{2}$.

The black hole entropy (\ref{S-W3(2)-Lorentzian}) is well-defined
provided the global charges of the black hole fulfill $\mathcal{\hat{L}}^{\pm}\geq0$,
as well as $\sin^{2}(\Phi_{\pm})\leq1$, i.e., 
\begin{equation}
\left\vert \left(\mathcal{U}^{\pm}\right)^{3}-\frac{k}{8\pi}\left(\mathcal{U}^{\pm}\mathcal{\hat{L}}^{\pm}+\psi_{\left[+\right]}^{\pm}\psi_{\left[-\right]}^{\pm}\right)\right\vert \leq\frac{1}{24\sqrt{6}}\left(\frac{k\mathcal{\hat{L}}^{\pm}}{\pi}\right)^{\frac{3}{2}}\ .\label{W32-bound}
\end{equation}
When some of these bounds are saturated, the solution becomes extremal
and the corresponding holonomy along the thermal circle becomes nontrivial
corresponding to a change in the topology. Beyond the bounds, the
solution is not a black hole since one cannot associate with it a
real positive entropy.

Note that the charges also come in electric-magnetic pairs, just as
in the principal embedding.

To determine the temperature and the chemical potentials in the microcanonical
ensemble, we use the relations 
\begin{align}
\beta & =\left(\frac{\partial S}{\partial\mathcal{M}_{\text{Lor}}}\right)_{\mathcal{J}_{\text{Lor}},\mathcal{U}_{0}^{\pm},\psi_{[a]0}^{\pm}},\label{eq:betamicrocan-1-1}\\
\beta\mu_{\mathcal{J}_{\text{Lor}}} & =-\left(\frac{\partial S}{\partial\mathcal{J}_{\text{Lor}}}\right)_{\mathcal{M}_{\text{Lor}},\mathcal{U}_{0}^{\pm},\psi_{[a]0}^{\pm}},\label{eq:betamumicrocan-1-1}\\
\beta\mu_{\mathcal{U}^{\pm}} & =-\left(\frac{\partial S}{\partial\mathcal{U}_{0}^{\pm}}\right)_{\mathcal{M}_{\text{Lor}},\mathcal{J}_{\text{Lor}},\psi_{[a]0}^{\pm}},\label{eq:betamuWmicrocan-1-1}\\
\beta\mu_{\psi_{[a]}^{\pm}} & =-\left(\frac{\partial S}{\partial\psi_{[a]0}^{\pm}}\right)_{\mathcal{M}_{\text{Lor}},\mathcal{J}_{\text{Lor}},\mathcal{U}_{0}^{\pm},\psi_{[-a]0}^{\pm}},
\end{align}
with $\mathcal{M}_{\text{Lor}}=\frac{2\pi}{\ell}\left(\mathcal{L}^{+}+\mathcal{L}^{-}\right)$
and $\mathcal{J}_{\text{Lor}}=2\pi\left(\mathcal{L}^{+}-\mathcal{L}^{-}\right)$
as above, and where $\psi_{[a]0}^{\pm}=2\pi\psi_{[a]}^{\pm}$ and
$\mathcal{U}_{0}^{\pm}=2\pi\mathcal{U}^{\pm}$ are charges with spin
3/2 and 1 respectively.

One finds 
\begin{align}
\beta & =\frac{\ell}{2}\left(\hat{\xi}_{+}+\hat{\xi}_{-}\right),\\
\beta\mu_{\mathcal{J}_{\text{Lor}}} & =-\frac{1}{2}\left(\hat{\xi}_{+}-\hat{\xi}_{-}\right),\\
\beta\mu_{\mathcal{U}^{\pm}} & =\nu_{\pm},\\
\beta\mu_{\psi_{[a]}^{\pm}} & =\vartheta_{\left[a\right]}^{\pm},
\end{align}
where $\hat{\xi}_{\pm}$, $\nu_{\pm}$ and $\vartheta_{\left[a\right]}^{\pm}$
are given in terms of the charges $\mathcal{\hat{L}}^{\pm},\mathcal{U}^{\pm}$,
and $\psi_{[a]}^{\pm}$ by the same expressions (\ref{eq:xihatintermscharges})-(\ref{Phi-W32})
giving $\hat{\xi}_{\pm}$, $\nu_{\pm}$ and $\vartheta_{\left[a\right]}^{\pm}$
in terms of $\mathcal{\hat{L}}^{\pm},\mathcal{U}^{\pm}$, and $\psi_{[a]}^{\pm}$.
This shows that indeed, the parameters introduced in the temporal
components of the connection have the anticipated physical interpretation
of being the temperature and chemical potentials.

\section{Extension to higher $N$\label{sec:Extension-to-higher}}

We have considered in sections \ref{HS} and \ref{W32-section} above
$sl(3,\mathbb{R})$ black holes. The extension from $N=3$ to a generic
$N$ is straightforward and will only be sketched here. In the principal
embedding of $sl(2,\mathbb{R})$ into $sl(N,\mathbb{R})$, the algebra
$sl(N,\mathbb{R})$ decomposes as $\oplus_{s=1}^{N-1}D_{s}$, where
$D_{s}$ is the irreducible $sl(2,\mathbb{R})$-spin $s$ representation.

The Euclidean-Lorentzian continuation for generic $N$ is discussed
in Appendix (\ref{sub:Euclidean-Lorentzian-generic-N}).

The boundary conditions that ensure that the principal embedding is
enforced generalize (\ref{eq:Aphiapin3}) and take the form \cite{Henneaux-HS,Theisen-HS},
\begin{equation}
A_{\varphi}=L_{1}-\frac{2\pi}{k}\mathcal{L}L_{-1}-\frac{\pi}{2k}\sum_{s=2}^{N-1}\mathcal{W}^{(s)}W_{-s}^{(s)}\;,\label{eq:AphiapinNP}
\end{equation}
or, on the Lorentzian side, 
\begin{equation}
A_{\varphi}^{\pm}=L_{\pm1}-\frac{2\pi}{k}\mathcal{L}^{\pm}L_{\mp1}-\frac{\pi}{2k}\sum_{s=2}^{N-1}\mathcal{W}^{(s)\pm}W_{\mp s}^{(s)}.\label{eq:AphiapinNPpm}
\end{equation}
Here, the $W_{j}^{(s)}$ ($j=-s,-s+1,\cdots,s-1,s$) are the $2s+1$
generators of the $sl(2,\mathbb{R})$-spin $s$ representation $D_{s}$,
so that $W_{-s}^{(s)}$ is the lowest weight state of $D_{s}$. The
functions $\mathcal{L}$, $\mathcal{W}^{(s)}$ (complex) and $\mathcal{L}^{\pm}$,
$\mathcal{W}^{(s)\pm}$ (real) depend on $\varphi$ (and $t$) and
(\ref{eq:AphiapinNP}) and (\ref{eq:AphiapinNPpm}) give only the
leading asymptotic form in the general case. However, for the black
hole in the rest frame, $\mathcal{L}$, $\mathcal{W}^{(s)}$ and $\mathcal{L}^{\pm}$,
$\mathcal{W}^{(s)\pm}$ are constant and the expressions (\ref{eq:AphiapinNP})
and (\ref{eq:AphiapinNPpm}) are exact. If the angular components
of the connection do not fulfill these asymptotic conditions (or equivalent
conditions written in a different gauge, see appendix \ref{AppendixB}),
it will be a different embedding with a different spin content that
will be selected. In particular, these boundary conditions are not
fulfilled by the chemical potential terms in the connection given
in the work \cite{Tan:2011tj} for $sl(4,\mathbb{R})$, which therefore
does not describe a principal embedding black hole but rather, a black
hole endowed with lower spin charges. The corresponding embedding
and asymptotic symmetries are discussed in \cite{FHK2013}.

As shown in \cite{Henneaux-HS,Theisen-HS}, the boundary conditions
(\ref{eq:AphiapinNPpm}) are preserved by asymptotic symmetries that
form a nonlinear $W_{N}$-algebra. The most general (``improper'')
gauge transformation that preserves the boundary conditions is characterized
at infinity by $N-1$ arbitrary functions $\xi$, $\eta^{(s)}$ multiplying
the highest weight generators $L_{1}$ and $W_{s}^{(s)}$ (plus terms
that are determined by them). This is the standard Hamiltonian reduction
\cite{HR1}.

In particular, the temporal component of the connection must define
an asymptotic symmetry. In the black hole case, where $\mathcal{L}$
and $\mathcal{W}^{(s)}$ are constant, the functions $\xi$ and $\eta^{(s)}$
entering $A_{\tau}$ are also constant. They are, as above, the temperature
and chemical potentials conjugate to the charges $\mathcal{L}$ and
$\mathcal{W}^{(s)}$.

The thermodynamical analysis proceeds then as above. The entropy is
determined by (\ref{sataphi-2-1-1}), and the chemical potentials
are determined by the regularity condition (\ref{HoloGenN}) generalized
to an arbitrary $N$. The analysis is direct, although somewhat intricate.
It will not be tackled herein.

\section{Concluding remarks \label{sec:Concluding-remarks}}

In this article we have investigated the generalized black holes appearing
in extensions of three-dimensional anti-de Sitter gravity which include
higher and lower spins. In the absence of available gauge invariant
causality concepts, our approach has been to develop the analysis
entirely from the Euclidean formulation, a black hole solution being
one that has thermal properties. This point of view was first expressed
in the present context in \cite{GK}. We have systematically adhered
to it throughout without using any further input. We have for instance
refrained from giving a metric associated to the black hole solutions.
Such metrics exist but have gauge-dependent geometrical properties
and so may be misleading. We have also based the derivation of the
entropy entirely on the action, and showed that it can be expressed
as a ``boundary term at the horizon'' along the lines developed
in \cite{Teitelboim_action_and_entropy,BTZ_Gauss_Bonnet,Hawking:1994ii}.

Our approach also provides throughout a definite control of the boundary
conditions along the lines of \cite{HPTT}. We have analyzed thoroughly
both, higher spin and also lower spin black holes. The higher spin
black hole solution given here is the first black hole with the required
asymptotics for higher-spin charges. In contrast, the black hole solutions
given earlier in the literature do not have the required asymptotics
and instead, possess only lower-spin charges. The existence of a black
hole with $W_{3}$ asymptotics indicates that, contrary to some opinions
previously expressed in the literature, there is no need to break
the asymptotic behaviour of the connection when discussing the thermodynamics
of solutions carrying higher spin charges.

\noindent \acknowledgments We thank G. Barnich and C. Mart\'{i}nez
for helpful discussions. C.B. and M.H. thank the Alexander von Humboldt
Foundation for Humboldt Research Awards. MH is grateful to the Institute
for Advanced Study (Princeton) where this work was partly completed.
R.T. thanks the International Solvay Institutes and the ULB for warm
hospitality. The work of A.P., D.T. and R.T. is partially funded by
the Fondecyt grants N${^{\circ}}$ 11130262, 11130260, 1130658, 1121031.
The Centro de Estudios Cient\'{i}ficos (CECS) is funded by the Chilean
Government through the Centers of Excellence Base Financing Program
of Conicyt. The work of M.H. is partially supported by the ERC through
the ``SyDuGraM\textquotedblright{}\ Advanced Grant, by IISN - Belgium
(conventions 4.4511.06 and 4.4514.08) and by the ``Communauté Française
de Belgique\textquotedblright{}\ through the ARC program.

\appendix
\global\long\def\thesubsection{\Alph{section}.\arabic{subsection}}

\section{Background}

\subsection{Chern-Simons formulation of gravitation theory in three spacetime
dimensions}

\label{Appendix0} \setcounter{equation}{0}

\subsubsection{Lorentzian formulation}

As shown in \cite{Achucarro:1987vz,Witten:1988hc}, the standard theory
of gravitation with a negative cosmological constant in 2+1 spacetime
dimensions can be reformulated as a Chern-Simons theory by using instead
of the metric variables, one $so(2,2)$-connection. This is because
$so(2,2)$ is the isometry algebra of anti-de Sitter space. The $so(2,2)$-connection
may be written as 
\begin{equation}
A=\omega^{a}J_{a}+e^{a}P_{a},
\end{equation}
where $\omega^{a}$ is the spin connection and $e^{a}$ the dreibein.
Here, the $J_{a}$'s and $P_{a}$'s are the generators of $so(2,2)$,
\begin{equation}
[J_{a},J_{b}]=\eta^{cd}\epsilon_{abc}J_{d},\;\;\;[J_{a},P_{b}]=\eta^{cd}\epsilon_{abc}P_{d},\;\;\;[P_{a},P_{b}]=\eta^{cd}\epsilon_{abc}J_{d},\quad a,b,c,d=0,1,2\;,
\end{equation}
$\eta_{ab}=diag(-1,1,1)$ and $\epsilon_{abc}$ is the Levi-Civita
symbol with $\epsilon_{120}=1$. The $J_{a}$'s generate the local
$so(2,1)$ subalgebra acting in the tangent spaces, which is extended
to $so(2,2)$ by the generators $P_{a}$.

Now, $so(2,2)$ is the direct sum of two copies of $so(2,1)\simeq sl(2,\mathbb{R})$.
A basis where this property is manifest is given by 
\begin{equation}
Y_{a}^{\pm}=\frac{1}{2}\left(J_{a}\pm P_{a}\right),
\end{equation}
which fulfill independently the $sl(2,\mathbb{R})$-algebra. The $4\times4$
matrices $Y_{a}^{\pm}$ commute with each other, 
\begin{equation}
[Y_{a}^{+},Y_{b}^{-}]=0.
\end{equation}
For each $sl\left(2,\mathbb{R}\right)$, we shall use from now on
the fundamental (defining) representation of the algebra, which is
formed by $2\times2$ matrices. Furthermore, we will realize $Y_{a}^{+}$
and $Y_{a}^{-}$ as 
\begin{equation}
Y_{a}^{+}=\begin{pmatrix}X_{a}^{+} & 0\\
0 & 0
\end{pmatrix},\;\;\; Y_{a}^{-}=\begin{pmatrix}0 & 0\\
0 & X_{a}^{-}
\end{pmatrix},
\end{equation}
where $X_{a}^{\pm}$ are $2\times2$ $sl(2,\mathbb{R})$-matrices.
The connection can thus be rewritten as 
\begin{equation}
A=A^{+}\oplus A^{-},
\end{equation}
where 
\begin{equation}
A^{\pm}=\left(\omega^{a}\pm\frac{e^{a}}{\ell}\right)X_{a}^{\pm}\;,
\end{equation}
are two independent connections $A^{\pm}$ for $sl\left(2,\mathbb{R}\right)$.
The $J_{a}$'s and $P_{a}$'s are then realized by the $4\times4$
matrices 
\begin{equation}
J_{a}=\begin{pmatrix}X_{a}^{+} & 0\\
0 & X_{a}^{-}
\end{pmatrix},\;\;\; P_{a}=\begin{pmatrix}X_{a}^{+} & 0\\
0 & -X_{a}^{-}
\end{pmatrix}.
\end{equation}

We shall take the same basis for each $sl\left(2,\mathbb{R}\right)$,
$X_{a}^{+}=X_{a}^{-}$. One choice for both $X_{a}^{+}$ and $X_{a}^{-}$
is given by 
\begin{equation}
T_{0}=-\frac{i\sigma_{2}}{2}=\begin{pmatrix}0 & -\frac{1}{2}\\
\frac{1}{2} & 0
\end{pmatrix}\quad;\quad T_{1}=\frac{\sigma_{3}}{2}=\begin{pmatrix}\frac{1}{2} & 0\\
0 & -\frac{1}{2}
\end{pmatrix}\quad;\quad T_{2}=\frac{\sigma_{1}}{2}=\begin{pmatrix}0 & \frac{1}{2}\\
\frac{1}{2} & 0
\end{pmatrix}\;,\label{Tsl2matrices}
\end{equation}
where $\sigma_{a}$'s are the Pauli matrices, and correspondingly,
\begin{equation}
J_{a}=\begin{pmatrix}T_{a} & 0\\
0 & T_{a}
\end{pmatrix},\;\;\; P_{a}=\begin{pmatrix}T_{a} & 0\\
0 & -T_{a}
\end{pmatrix}.
\end{equation}
These generators $T_{a}$'s obey 
\begin{equation}
\left[T_{a},T_{b}\right]=\eta^{cd}\epsilon_{abc}T_{d}.\label{eq:Tsl2R}
\end{equation}

Another choice of basis in the Lie algebra $sl(2,\mathbb{R})$ is
\begin{equation}
L_{-1}=\begin{pmatrix}0 & 0\\
1 & 0
\end{pmatrix}\quad;\quad L_{0}=\begin{pmatrix}-\frac{1}{2} & 0\\
0 & \frac{1}{2}
\end{pmatrix}\quad;\quad L_{1}=\begin{pmatrix}0 & -1\\
0 & 0
\end{pmatrix}\;,\label{sl2matrices}
\end{equation}
which obeys 
\begin{equation}
\left[L_{i},L_{j}\right]=\left(i-j\right)L_{i+j}\quad,\quad i,j=-1,0,1\;.\label{eq:sl2R}
\end{equation}
The two bases are related by 
\begin{equation}
L_{i}=T_{a}\Lambda_{\;\; i}^{a}\,,
\end{equation}
with 
\begin{equation}
\left(\Lambda_{\;\; i}^{a}\right)=\begin{pmatrix}1 & 0 & 1\\
0 & -1 & 0\\
1 & 0 & -1
\end{pmatrix}.
\end{equation}
We shall carry out the study of the Euclidean-Lorentzian continuation
in the $T_{a}$-basis because the analysis is then expressed in terms
of standard familiar geometrical objects (spin connection and dreibein)
whose behaviour under the continuation is simple and well controlled,
whereas the $L_{i}$-basis is well adapted to the formulation of the
asymptotic conditions.

\subsubsection{Euclidean formulation}

Euclidean three-dimensional gravity also admits a Chern-Simons formulation
\cite{Witten:1988hc},\cite{Witten}. When the cosmological constant
is negative, the role played by anti-de Sitter space in the Lorentzian
case is now played by the hyperbolic space $H_{3}$, or three-dimensional
Lobachevsky space. The isometry algebra of $H_{3}$ is $so(3,1)$.
The identity component of the isometry group is obtained by exponentiation
and is the proper, orthochronous Lorentz group $SO^{+}(3,1)$. This
group is isomorphic to $SL(2,\mathbb{C})/\mathbb{Z}_{2}$, where $\mathbb{Z}_{2}=\{+\mathbbm{1}_{2\times2},-\mathbbm{1}_{2\times2}\}$
is the center of $SL(2,\mathbb{C})$. The full isometry group is the
orthochronous Lorentz group $O^{+}(3,1)$ obtained by adding the spatial
reflection $P=\textrm{diag}(1,-\mathbbm{1}_{3\times3})$. Only the
identity component $SO^{+}(3,1)$ will be relevant here as we will
be considering only gauge transformations connected with the identity.

The Chern-Simons connection of Euclidean three-dimensional gravity
with a negative cosmological constant is thus a $so(3,1)$-connection,
\begin{equation}
A'=\omega^{a}\tilde{J}_{a}+e^{a}\tilde{P}_{a},\label{EuclConn1}
\end{equation}
where $\omega^{a}$ is the Euclidean spin connection and $e^{a}$
the Euclidean dreibein ($a=1,2,3$). The $\tilde{J}_{a}$ and $\tilde{P}_{a}$
generate $so(3,1)$, 
\begin{equation}
[\tilde{J}_{a},\tilde{J}_{b}]=\delta^{cd}\epsilon_{abc}\tilde{J}_{d}\,,\quad[\tilde{J}_{a},\tilde{P}_{b}]=\delta^{cd}\epsilon_{abc}\tilde{P}_{d}\,,\quad[\tilde{P}_{a},\tilde{P}_{b}]=-\delta^{cd}\epsilon_{abc}\tilde{J}_{d}\,,\quad a,b,c,d=1,2,3\;.\label{so31}
\end{equation}
Here, $\epsilon_{abc}$ is the Levi-Civita symbol with $\epsilon_{123}=1$
The $\tilde{J}_{a}$ generate the local $so(3)$ subalgebra acting
on the tangent spaces and is extended to $so(3,1)$ by the generators
$\tilde{P}_{a}$.

Now, the real Lie algebra $so(3,1)$, which is six-dimensional, is
isomorphic to $sl(2,\mathbb{C})$ viewed as a real algebra, which
is also six-dimensional. This is the infinitesimal version of the
group isomorphism recalled above. One way to exhibit this isomorphism
is to use the set $\{\hat{J}^{a},\hat{P}^{a}=i\hat{J}^{a}\}$ as basis
of $sl(2,\mathbb{C})$, where $\{\hat{J}^{a}\}$ is a basis of $su(2)\simeq so(3)$,
\begin{equation}
\left[\hat{J}_{a},\hat{J}_{b}\right]=\delta^{cd}\epsilon_{abc}\hat{J}_{d}\quad,\quad a,b,c,d=1,2,3\;.\label{eq:Jsu2}
\end{equation}
The $\hat{J}^{a},\hat{P}^{a}$ fulfill the commutation relations (\ref{so31})
of the $\tilde{J}_{a}$, $\tilde{P}_{a}$. The matrices $\hat{J}_{a}$
are antihermitian, $\left(\hat{J}_{a}\right)^{\dagger}=-\hat{J}_{a}$.
Another basis of $sl(2,\mathbb{C})$ that exhibits the isomorphism
is $\{\hat{J}^{a},\hat{P}'^{a}=-i\hat{J}^{a}\}$ since these generators
fulfill the same commutation relations.

In order to compare the Lorentzian and Euclidean formulations, it
is convenient to realize the generators $\tilde{J}_{a}$ and $\tilde{P}_{a}$
in terms of block-diagonal, $4\times4$ matrices, since the $J_{a}$
and $P_{a}$ of the Lorentzian theory have been brought to that form
above. This can be achieved by a complex linear transformation. We
take 
\begin{equation}
\tilde{J}_{a}=\begin{pmatrix}\hat{J}_{a} & 0\\
0 & \hat{J}_{a}
\end{pmatrix},\;\;\;\tilde{P}_{a}=\begin{pmatrix}i\hat{J}_{a} & 0\\
0 & -i\hat{J}_{a}
\end{pmatrix}.
\end{equation}
This choice puts the two bases $\{\hat{J}^{a},\hat{P}^{a}=i\hat{J}^{a}\}$,
$\{\hat{J}^{a},\hat{P}'^{a}=-i\hat{J}^{a}\}$ on an equal footing
and is such that the matrices $\tilde{J}_{a}$ and $\tilde{P}_{a}$
remain linearly independent over the complex numbers. The matrices
$\tilde{J}_{a}$ and $\tilde{P}_{a}$, as well as any real linear
combination of them, are of the form 
\begin{equation}
\begin{pmatrix}C & 0\\
0 & -C^{\dagger}
\end{pmatrix},
\end{equation}
with $C\in sl(2,\mathbb{C})$. The map $C\mapsto-C^{\dagger}$ is
an (antilinear) automorphism of the Lie algebra, i.e., it preserves
the commutator.

One can rewrite the connection (\ref{EuclConn1}) as 
\begin{equation}
A'=\begin{pmatrix}A & 0\\
0 & -A^{\dagger}
\end{pmatrix},
\end{equation}
with 
\begin{equation}
A=\left(\omega^{a}+\frac{i}{\ell}e^{a}\right)\hat{J}_{a}.\label{eq:Aspinviel0}
\end{equation}
For the analysis that follows, it is convenient to choose the $\hat{J}_{a}$'s
as 
\begin{equation}
\hat{J}_{1}=-\frac{i\sigma_{3}}{2}=\begin{pmatrix}\frac{-i}{2} & 0\\
0 & \frac{i}{2}
\end{pmatrix}\quad;\quad\hat{J}_{2}=-\frac{i\sigma_{1}}{2}=\begin{pmatrix}0 & \frac{-i}{2}\\
\frac{-i}{2} & 0
\end{pmatrix}\quad;\quad\hat{J}_{3}=-\frac{i\sigma_{2}}{2}=\begin{pmatrix}0 & -\frac{1}{2}\\
\frac{1}{2} & 0
\end{pmatrix}\;.\label{Jsu2matrices}
\end{equation}

\subsubsection{Euclidean-Lorentzian continuation for $N=2$ (pure gravity)}

\medskip{}
 \medskip{}
 \emph{Rules in the metric formulation}

\medskip{}

In order to spell out the Euclidean-Lorentzian continuation rules
in the Chern-Simons formulation, we first write them in the metric
formulation. To do that, we consider first the explicit case of the
2+1 black hole under study in this paper and then write the rules
in the general case.

In ``Schwarzschild coordinates'' the Lorentzian metric for the standard
2+1 black hole \cite{BTZ,BHTZ} reads

\begin{equation}
ds_{\text{Lor}}^{2}=-N_{\text{Lor}}^{2}f_{\text{Lor}}^{2}dt^{2}+f_{\text{Lor}}^{-2}dr^{2}+r^{2}\left(d\varphi+N_{\text{Lor}}^{\varphi}dt\right)^{2},\label{BTZLor}
\end{equation}
with 
\begin{align}
f_{\text{Lor}}^{2} & =\frac{\left(r^{2}-r_{+}^{2}\right)\left(r^{2}-r_{-}^{2}\right)}{\ell^{2}r^{2}},\nonumber \\
N_{\text{Lor}} & =N_{\text{Lor}}\left(\infty\right),\nonumber \\
N_{\text{Lor}}^{\varphi} & =N_{\text{Lor}}^{\varphi}\left(\infty\right)-\frac{r_{+}r_{-}}{\ell r^{2}}N_{\text{Lor}}\left(\infty\right),\label{fNNphi}
\end{align}
where 
\begin{equation}
M_{\text{Lor}}=\frac{r_{+}^{2}+r_{-}^{2}}{\ell^{2}}\quad,\quad J_{\text{Lor}}=\frac{2r_{+}r_{-}}{\ell}.\label{MJLor}
\end{equation}

One usually sets $N_{\text{Lor}}\left(\infty\right)=1$ and $N_{\text{Lor}}^{\varphi}\left(\infty\right)=0$
by a rescaling of $t$ and a transformation of $\varphi$ to $\varphi'=\varphi+N_{\text{Lor}}^{\varphi}\left(\infty\right)t$.
However it will be important for conceptual, and practical purposes
to keep $N_{\text{Lor}}\left(\infty\right)$ and $N_{\text{Lor}}^{\varphi}\left(\infty\right)$
as adjustable parameters. Although this is a matter of choice in the
Lorentzian formulation, it is not so in the Euclidean one, where regularity
conditions at the horizon appear.

The Euclidean continuation for the metric (\ref{BTZLor}) is obtained
by setting 
\begin{equation}
f_{\text{Lor}}^{2}=f_{\text{E}}^{2}\quad,\quad N_{\text{Lor}}=N_{\text{E}}\quad,\quad N_{\text{Lor}}^{\varphi}=iN_{\text{E}}^{\varphi}\:,
\end{equation}

\begin{equation}
M_{\text{Lor}}=M_{\text{E}}\quad,\quad J_{\text{Lor}}=iJ_{\text{E}}\:,
\end{equation}
and demanding that the Euclidean parameters be real. These formulas
may be obtained by setting $t=-i\tau$ in the line element and taking
$\tau$ to be real. More generally, for a generic field configuration
in Hamiltonian form, 
\begin{equation}
\pi_{\text{Lor}}^{ij}=-i\pi_{E}^{ij},\; g_{ij}^{\text{Lor}}=g_{ij}^{\text{E}},\; N_{\text{Lor}}=N_{\text{E}},\; N_{\text{Lor}}^{i}=iN_{\text{E}}^{i},\label{MinkEuc}
\end{equation}
and the Euclidean action is defined by 
\begin{equation}
iI_{\text{Lor}}=I_{\text{E}}.\label{ActionLorEucl22}
\end{equation}
When we deal with the Euclidean continuation below, we will drop the
subscript ``E'' whenever no confusion may arise.\medskip{}

\medskip{}
 \emph{Rules in the Chern-Simons formulation \label{E-C-Rules}}

\medskip{}

From the metric continuation rules, one derives the relationship between
the Euclidean and Lorentzian dreibeins and spin connections. It is
\begin{equation}
e^{1E}=e^{1L},\;\;\; e^{2E}=e^{2L},\;\;\; e^{3E}=-ie^{0L},\;\;\;\omega^{1E}=i\omega^{1L},\;\;\;\omega^{2E}=i\omega^{2L},\;\;\;\omega^{3E}=\omega^{0L}.
\end{equation}
The continuation rules $e^{1E}=e^{1L}$, $e^{2E}=e^{2L}$, $e^{3E}=-ie^{0L}$
for the dreibein are rather direct. The ones from the connection follow
then from $de^{a}+\omega_{\;\; b}^{a}e^{b}=0$ and the definition
of $\omega^{a}$ in terms of $\omega_{\;\; c}^{b}$, i.e. $\omega^{a}=\frac{1}{2}\varepsilon^{abc}\omega_{bc}$.

From the continuation rules for the dreibein and the spin connection,
one derives 
\begin{equation}
A^{+1}=-iA^{1},\;\;\; A^{+2}=-iA^{2},\;\;\; A^{+0}=A^{3},
\end{equation}
and 
\begin{equation}
A^{-1}=-i(A^{1})^{*},\;\;\; A^{-2}=-i(A^{2})^{*},\;\;\; A^{-0}=(A^{3})^{*},
\end{equation}
where the $^{*}$ denotes the complex conjugate. The previous formulas
are summarized in the simple relations 
\begin{eqnarray}
 &  & A^{+}=A,\label{ContinuationA}\\
 &  & A^{-}=-A^{\dagger},\label{ContinuationB}
\end{eqnarray}
where 
\begin{equation}
A=A^{a}\hat{J}_{a},\;\;\; A^{+}=A^{+a}T_{a},\;\;\; A^{-}=A^{-a}T_{a}.\label{AA}
\end{equation}
The relationship between $\hat{J}_{a}$ and $T_{a}$ is the following,
\begin{equation}
T_{1}=i\hat{J}_{1},\;\;\; T_{2}=i\hat{J}_{2},\;\;\; T_{0}=\hat{J}_{3}.
\end{equation}

The Euclidean-Lorentzian continuation rule is remarkable. The two
independent $sl\left(2,\mathbb{R}\right)$ connections are merged
into a single complex connection. The merging could not be simpler,
one simply takes $A^{+}$ and allows it to be complex. The other connection
$A^{-}$ then follows according to (\ref{ContinuationB}). The prescription
takes care automatically of the change in the algebra when going from
(\ref{eq:Tsl2R}) to (\ref{eq:Jsu2}), that is, it replaces $\eta^{ab}$
by $\delta^{ab}$.

\medskip{}
 \medskip{}
 \emph{Reals forms of $sl(2,\mathbb{C})\oplus sl(2,\mathbb{C})$ and
conjugations}

\medskip{}

One may view the analytic continuation as the passage from one real
form of $sl(2,\mathbb{C})\oplus sl(2,\mathbb{C})$ to another. Indeed,
the 6-dimensional real Lie algebras $sl(2,\mathbb{R})\oplus sl(2,\mathbb{R})$,
\begin{equation}
sl(2,\mathbb{R})\oplus sl(2,\mathbb{R})=\left\{ \begin{pmatrix}E & 0\\
0 & F
\end{pmatrix}:E,F\in sl(2,\mathbb{R})\right\} ,\label{Def1}
\end{equation}
and $sl(2,\mathbb{C})$, 
\begin{equation}
sl(2,\mathbb{C})=\left\{ \begin{pmatrix}C & 0\\
0 & -C^{\dagger}
\end{pmatrix}:C\in sl(2,\mathbb{C})\right\} ,\label{Def2}
\end{equation}
are two distinct real forms of the 6-dimensional complex Lie algebra
$sl(2,\mathbb{C})\oplus sl(2,\mathbb{C})$. By this it is meant that
if one complexifies these algebras (consider linear combinations with
complex coefficients of Lie algebra elements), one gets the full $sl(2,\mathbb{C})\oplus sl(2,\mathbb{C})$.

Let 
\begin{equation}
\begin{pmatrix}M & 0\\
0 & N
\end{pmatrix},
\end{equation}
be an arbitrary element of $sl(2,\mathbb{C})\oplus sl(2,\mathbb{C})$.
One defines: 
\begin{equation}
\tau\left(\begin{pmatrix}M & 0\\
0 & N
\end{pmatrix}\right)=\begin{pmatrix}M^{*} & 0\\
0 & N^{*}
\end{pmatrix},\;\;\;\;\sigma\left(\begin{pmatrix}M & 0\\
0 & N
\end{pmatrix}\right)=\begin{pmatrix}-N^{\dagger} & 0\\
0 & -M^{\dagger}
\end{pmatrix}.
\end{equation}
The conjugations $\tau$ and $\sigma$ (antilinear involutions that
preserve the Lie algebra structure) commute and fix the real Lie subalgebras
$sl(2,\mathbb{R})\oplus sl(2,\mathbb{R})$ and $sl(2,\mathbb{C})$,
respectively, i.e., $x\in sl(2,\mathbb{C})\oplus sl(2,\mathbb{C})$
belongs to $sl(2,\mathbb{R})\oplus sl(2,\mathbb{R})$ if and only
if $\tau(x)=x$, while $x$ belongs to $sl(2,\mathbb{C})$ if and
only if $\sigma(x)=x$.

One goes from $sl(2,\mathbb{R})\oplus sl(2,\mathbb{R})$ to $sl(2,\mathbb{C})$
by decomposing any element $A\in sl(2,\mathbb{R})\oplus sl(2,\mathbb{R})$
as $A=A_{0}+A_{1}$ where $\sigma(A_{0})=A_{0}$ and $\sigma(A_{1})=-A_{1}$.
The corresponding $sl(2,\mathbb{C})$-element is $A_{0}+iA_{1}$.
Conversely, one can decompose any element $B\in sl(2,\mathbb{C})$
as $B=B_{0}+B_{1}$ with $\tau(B_{0})=B_{0}$ and $\tau(B_{1})=-B_{1}$.
The corresponding $sl(2,\mathbb{R})\oplus sl(2,\mathbb{R})$-element
is $B_{0}-iB_{1}$.

\medskip{}
 \medskip{}
 \emph{Euclidean-Lorentzian continuation of the asymptotic boundary
conditions\label{sub:Eucl-Lor-cont}}

\medskip{}

In order to apply the Euclidean-Lorentzian continuation rules to the
connection (\ref{B-H-CONN1b}), 
\begin{equation}
A_{\varphi}^{\pm}\left(r,\varphi\right)=L_{\pm1}-\frac{2\pi}{k}\mathcal{L}^{\pm}\left(r,\varphi\right)L_{\mp1},
\end{equation}
it is convenient to decompose the Virasoro generators $\mathcal{L}^{\pm}$
into its even and odd parts under the exchange of the two $sl(2,\mathbb{R})$-factors,
\begin{equation}
\mathcal{L}^{\pm}=\mathcal{A}\pm\mathcal{B}.
\end{equation}
The corresponding complex connection is then 
\begin{equation}
A_{\varphi}\left(r,\varphi\right)=L_{1}-\frac{2\pi}{k}\mathcal{L}\left(r,\varphi\right)L_{-1},
\end{equation}
with 
\begin{equation}
\mathcal{L}=\mathcal{A}+i\mathcal{B}.
\end{equation}
One thus says that the complex Virasoro generator $\mathcal{L}$ is
related to its Lorentzian counterparts $\mathcal{L}^{\pm}$ through
the continuation rules $\mathcal{L}^{+}=\mathcal{L}$, $\mathcal{L}^{-}=\mathcal{L}^{*}$,
with the understanding that the imaginary part $i\mathcal{B}$ of
$\mathcal{L}$ is continued to $\mathcal{B}$, something that one
sometimes writes as $\mathcal{B}_{\textrm{E}}=-i\mathcal{B}_{\textrm{Lor}}$.
With these continuation rules, $A_{\varphi}$ becomes $A_{\varphi}^{+}$
while $-A_{\varphi}^{\dagger}$ becomes $A_{\varphi}^{-}$.

For the temporal component (\ref{BHAt-2b}), 
\begin{equation}
A_{t}^{\pm}=\pm\xi_{\pm}\left(r,\varphi\right)\left(L_{\pm1}-\frac{2\pi}{k}\mathcal{L}^{\pm}\left(r,\varphi\right)L_{\mp1}\right),
\end{equation}
one first continues $t$ into $t=-i\tau$ to get 
\begin{equation}
A_{\tau}^{\pm}=\mp i\xi_{\pm}\left(r,\varphi\right)\left(L_{\pm1}-\frac{2\pi}{k}\mathcal{L}^{\pm}\left(r,\varphi\right)L_{\mp1}\right).
\end{equation}
Setting $\xi_{\pm}=a\pm b$, one then complexifies as above and gets
\begin{equation}
A_{\tau}=-i\xi\left(r,\varphi\right)\left(L_{1}-\frac{2\pi}{k}\mathcal{L}\left(r,\varphi\right)L_{-1}\right),
\end{equation}
with $\xi=a+ib$. The rule $b_{\textrm{E}}=-ib_{\textrm{Lor}}$ yields
$A_{\tau}^{+}$ from $A_{\tau}$ and $A_{\tau}^{-}$ from $-A_{\tau}^{\dagger}$.

\subsubsection{Euclidean-Lorentzian continuation for generic $N$\label{sub:Euclidean-Lorentzian-generic-N}}

We derived the continuation rule by translating into the Chern-Simons
language the known rules for the metric formulation and using a special
basis for the gauge algebra. However, the answer makes no reference
to: (i) the metric, (ii) the need to identify the dreibein and the
spin connection from among the connection components, (iii) the basis
in the gauge algebra and (iv) the gauge algebra itself!. The rules
(\ref{ContinuationA}) and (\ref{ContinuationB}) (``continue $A^{+}$
to complex values and take $A^{-}=-A^{\dagger}$'') will be taken
as the {\em definition} of the Euclidean-Lorentzian continuation
for the generalized case where $sl(2,{\mathbb{R}})$ is replaced by
$sl(N,{\mathbb{R}})$ and $sl(2,{\mathbb{C}})$ is replaced by $sl(N,{\mathbb{C}})$.

The Lorentzian action is of the form 
\begin{equation}
I_{\text{Lor}}=I_{\text{CS}}\left[A^{+}\right]-I_{\text{CS}}\left[A^{-}\right].\label{IlorCS}
\end{equation}
It is immediate to verify that if one inserts in (\ref{IlorCS}) the
definitions (\ref{ContinuationA}) and (\ref{ContinuationB}), one
finds 
\begin{align}
iI_{\text{Lor}} & =I_{E}=-2\text{Im}\left[I_{\text{CS}}\left[A\right]\right].
\end{align}
For a generic $N$ the regularity condition is 
\begin{equation}
H_{\tau}=\left(-1\right)^{N+1}\mathbbm{1},\label{HoloGenN}
\end{equation}
where one employs the representation in terms of smallest matrices
($2\times2$ for $N=2$, $3\times3$ for $N=3$).

The entropy is given by

\begin{align}
S & =-2k_{N}\text{Im}\left(\text{tr}\left[A_{\tau}^{\text{on-shell}}\left(r_{+}\right)A_{\varphi}^{\text{on-shell}}\left(r_{+}\right)\right]\right)\;,\label{sataphi-2-1-1}
\end{align}
with $k_{N}=6k/N\left(N^{2}-1\right)=3\ell/2GN\left(N^{2}-1\right)$.

\subsection{Thermodynamics of the pure gravity 2+1 black hole in the metric formulation}

\label{AppendixA}

\subsubsection{Geometry of the 2+1 Euclidean black hole}

The geometry of the Euclidean 2+1 black hole was investigated in \cite{Carlip:1994gc}.
It was shown there that the topology induced by the metric on the
three dimensional Euclidean space is that of a solid torus, or equivalently
$\mathbb{R}^{2}\times S^{1}$, as illustrated in figure 1.

\hspace{-0.5cm}\includegraphics[width=14cm,height=20cm,keepaspectratio]{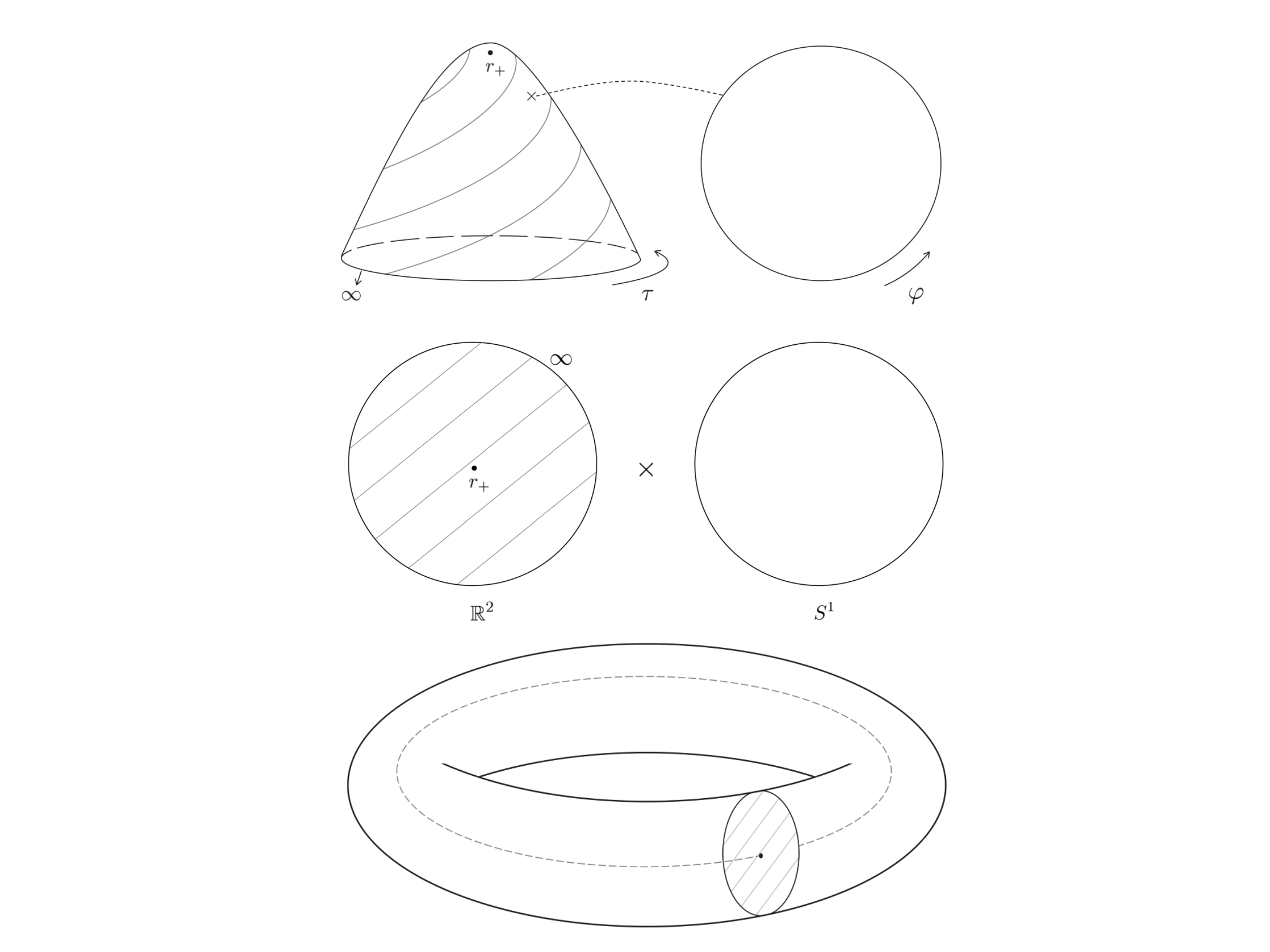}

\begin{scriptsize}Figure 1: Topology of the Euclidean black hole
in three-dimensional spacetime. The sequence of images illustrates
how $\mathbb{R}^{2}\times S^{1}$ is identical to a solid torus. The
``Euclidean horizon'' $r_{+}$ is the origin of a system of polar
coordinates $r,\tau$ in $\mathbb{R}^{2}$. The Euclidean time $\tau$
is the polar angle. On the other hand, the $S^{1}$ is parametrized
by the angle $\varphi$. The points $\left(r,\tau,\varphi\right)$
and $\left(r,\tau+1,\varphi+2\pi\right)$ are identified. In the metric
formulation, which is available only in the pure gravity case, the
opening angle of an ``off-shell'' conical singularity at $r_{+}$
is conjugate to the area of $S^{1}$ at $r_{+}$ because the variation
of the action with respect to the area gives $\Theta$ . On-shell
one has $\Theta=2\pi$, and there is no conical singularity. In the
Chern-Simons formulation, which is available for the pure gravity
black hole and also for its generalizations, the holonomy of the contractible
$\tau$ cycle is conjugate to that of the non-contractible $\varphi$
cycle, in the sense that the variation of the action with respect
to $A_{\varphi}\left(r_{+}\right)$ gives $A_{\tau}\left(r_{+}\right)$.
On-shell the holonomy of the $\tau$ cycle is trivial and the solution
is regular.

\end{scriptsize}

\vspace{3em}

The coordinate $\tau$ is an angle in the $\mathbb{R}^{2}$ factor,
and $\varphi$ is an angle in the $S^{1}$. The periods of $\tau$
and $\varphi$ may be fixed once and for all to any nominal value,
the physical restriction of interest below will be formulated in terms
of $N$ and $N^{\varphi}$. This is why we have allowed from the onset
for the appearance of $N\left(\infty\right)$ and $N^{\varphi}\left(\infty\right)$
in (\ref{BTZLor}). In this way, when dealing with the action integral
further below one can vary the fields without having to worry about
the variation of the range of integration. In order to make easy contact
with the standard conventions, we will take 
\begin{align}
0 & <\tau\leq1,\\
0 & <\varphi\leq2\pi.
\end{align}
Note that with this conventions $\tau$ is dimensionless and the ``Killing
lapse'' $N$ has dimensions of length. If in our formulas we replace
$\tau$ by $N(\infty)^{-1}\tau$ we obtain those of ref. \cite{Carlip:1994gc}.

\subsubsection{Euclidean action and entropy\label{sub:Euclidean-action-and}}

The black hole entropy, as well as other thermodynamic functions such
as the Helmholtz and Gibbs free energies, are obtained by evaluating
the appropriate Euclidean action on the black hole solution. Here
the word ``appropriate'' means that the chosen action must be such
that if one demands that it be stationary with some boundary conditions
at infinity, then the equations of motion should hold everywhere.
If one fixes at infinity the mass and the angular momentum, which
corresponds to the microcanonical ensemble, then the value of the
corresponding action is the entropy. If instead one fixes $N\left(\infty\right)$
and the angular momentum $\mathcal{J}$, then the value of the corresponding
action is $-\beta F$, where $F$ is the Helmholtz free energy $F=\mathcal{M}-TS$,
with the inverse temperature $\beta=N\left(\infty\right)$. If one
fixes $N\left(\infty\right)$ and $N^{\varphi}\left(\infty\right)$,
then the value of the corresponding action is $-\beta G$, where $G$
is the Gibbs free energy $G=\mathcal{M}-TS-\mu_{\mathcal{J}}\mathcal{J}$,
and $\mu_{\mathcal{J}}=-\beta^{-1}N^{\varphi}\left(\infty\right)$.

To construct the desired action we start with the canonical form of
the Lorentzian action 
\begin{equation}
I_{\text{can}}=\int d^{3}x\left(\pi^{ij}\dot{g}_{ij}-N\mathcal{H}-N^{i}\mathcal{H}_{i}\right).\label{Hamiltactionmetric}
\end{equation}
After performing the Euclidean continuation with the prescriptions
(\ref{MinkEuc}) given above, the Euclidean action (\ref{ActionLorEucl22}),
expressed in term of the Euclidean variables, takes exactly the same
form, with the only change that in the Hamiltonian generator $\mathcal{H}$
the term that is quadratic in the momenta $\pi^{ij}$ reverses his
sign with respect to the Lorentzian case. Next, we use a polar system
of coordinates in the $\mathbb{R}^{2}$ plane with $r$ being the
radial coordinate and $\tau$ being the polar angle, and we call $\varphi$
the coordinate that runs along the $S^{1}$. We will call $r_{+}$
the value of $r$ at the origin of the polar coordinate system in
$\mathbb{R}^{2}$. If one performs the variation of the action (\ref{Hamiltactionmetric}),
one obtains three terms: (i) a volume integral over $r,t$ and $\varphi$,
which vanishes when the equations of motion hold for $r_{+}<r<\infty$,
(ii) a boundary term at $r_{+}$ which is an integral over the $S^{1}$
at that point, and (iii) a boundary term which is an integral over
$S^{1}$ at infinity. The boundary term at infinity will be dealt
with afterwards because its form need to be adjusted according to
which variables are fixed at infinity, that is, as explained above,
it depends on the thermodynamic ensemble that is chosen. On the other
hand, in dealing with the boundary term at $r_{+}$ one only has to
demand that the equations of motion should hold at $r_{+}$ since
no variable is fixed there as a boundary condition. Since the equations
of motion already hold for $r$ greater than $r_{+}$, the requirement
on the boundary term at $r_{+}$ is that it should vanish when the
fields are regular at that point, otherwise there would be a source
at the origin. If the boundary term does not vanish, the action must
be amended by adding to it a term whose variation cancels the boundary
term coming from the variation of the canonical action. As discussed
in \cite{BTZ_Gauss_Bonnet,Teitelboim_action_and_entropy}, the boundary
term at the origin takes the form 
\begin{equation}
\delta I_{\text{can}}\left(r_{+}\right)=-\int_{r_{+}}d\varphi\left[\frac{1}{8\pi G}\Theta\left(\varphi\right)\delta\sqrt{g_{\varphi\varphi}}+2N^{i}\left(\varphi\right)\delta\pi_{i}^{\: r}\right].
\end{equation}

If one demands that the variation with respect to $\pi_{i}^{\: r}\left(r_{+},\phi\right)$
should vanish, one obtains the condition 
\begin{equation}
N^{i}\left(r_{+},\varphi\right)=0.\label{Nphizero}
\end{equation}
As it will be discussed below, this condition can always be imposed,
and it fixes the chemical potential to its correct value. However,
the situation with respect to $\Theta\left(\varphi\right)$ is different.
The precise form of $\Theta\left(\varphi\right)$ in terms of $N$
and $g_{ij}$ will be written down below (\ref{eq:Theta}), but it
is of not needed here. What is key is that unless, in addition to
(\ref{Nphizero}) one has 
\begin{equation}
\Theta\left(\varphi\right)^{\text{on-shell}}=2\pi,\label{thetaigual2pi}
\end{equation}
there is a singularity at the origin and therefore the equations of
motion are not satisfied at that point. However, extremization of
$I_{\text{can}}$ with respect to $g_{\varphi\varphi}$ at $r_{+}$
yields $\left.\Theta\left(\varphi\right)\right|_{\text{on-shell}}=0$.
This means that the canonical action needs to be modified so that,
extremization of the corrected action with respect to $g_{\varphi\varphi}$
should yield $\Theta\left(\varphi\right)^{\text{on-shell}}=2\pi$.
One must add therefore to the action the term 
\begin{equation}
\frac{1}{4G}\int_{r_{+}}\sqrt{g_{\varphi\varphi}}d\varphi.\label{bdryterm1}
\end{equation}
The correct action then reads 
\begin{align}
I & =\frac{1}{8\pi G}\Theta^{\text{on-shell}}\int_{r_{+}}\sqrt{g_{\varphi\varphi}}d\varphi+I_{\text{can}}+B_{\infty},\nonumber \\
 & =\frac{1}{4G}\int_{r_{+}}\sqrt{g_{\varphi\varphi}}d\varphi+I_{\text{can}}+B_{\infty}.\label{eq:Itheta}
\end{align}
Extremization of this action under variation of $g_{\varphi\varphi}\left(r_{+}\right)$
gives the equations of motion for $r_{+}\leq r<\infty$ . We will
also see below that, just as $N^{i}\left(r_{+}\right)=0$ fixes the
chemical potential, $\Theta\left(\varphi\right)^{\text{on-shell}}=2\pi$,
fixes the inverse temperature $\beta$. An important comment in this
context, one may interpret the case $\Theta=0$ as corresponding to
the ``closed cone'', that is as an infinitely long throat that becomes
narrower and narrower as far as one approaches the origin. Topologically,
this means that the origin $r_{+}$ is excised from the manifold and
therefore the topology is no longer $\mathbb{R}^{2}\times S_{1}$
but rather $\left[\mathbb{R}^{2}-\left\{ r_{+}\right\} \right]\times S_{1}$.
In this case, no improvement of the canonical action at the origin
is needed. This happens for the extreme black hole \cite{Teitelboim_action_and_entropy}.

It should be emphasized that the only amendment of the action at the
origin is the addition of (\ref{bdryterm1}) which ensures that the
equations of motion hold there, independently of what one chooses
to fix at infinity. This is a reflection of a profound fact: in thermodynamics,
the entropy - and its integrating factor in the first law, the temperature-
are quite distinct from the charges and the chemical potentials, and
thus enter in a very different footing in its construction through
the action.

Now, for any solution that is time independent, and so are the black
holes, the value $I_{\text{can}}$ evaluated on the solution (``on-shell'')
is zero, because $\dot{g}_{ij}=0$, and the constraint equations $\mathcal{H}=0,\mathcal{H}_{i}=0$.
This is the reason why it is so convenient to use the canonical action
in this context. Therefore, one has 
\begin{equation}
I=\frac{1}{4G}A+B_{\infty},\;\text{ (on-shell)}.\label{AmasBinfty}
\end{equation}

The value of the boundary term at infinity depends on the choice of
the ensemble. For the microcanonical ensemble $B_{\infty}=0$. One
therefore finds that the entropy is given by 
\begin{equation}
S=\frac{1}{4G}A.
\end{equation}
If one works in the microcanonical ensemble, one needs to express,
through the solutions of the constraint equations, the horizon ``area'',
$A=2\pi r_{+}$ in terms of the mass and the angular momentum which
are defined at infinity. Similarly for the other ensembles. One cannot
focus only on the horizon, or only on infinity, one needs both to
construct the thermodynamics in whatever ensemble one chooses to work
in.

For the 2+1 black hole, the asymptotic symmetries have been studied
in \cite{Brown:1986nw,BHTZ}. It is found that when ``asymptotically
Anti-de Sitter'' boundary conditions are given for $g_{ij}$ and
$\pi^{ij}$, the improving boundary term takes the form 
\begin{equation}
\delta B_{\text{Lor}}^{\infty}=-\frac{1}{2\pi}\left(t_{2}-t_{1}\right)\sum_{n}\left[\frac{1}{\ell}N_{n\text{Lor}}\delta\left(\mathcal{L}_{n}^{+}+\mathcal{L}_{n}^{-}\right)-N_{n\text{Lor}}^{\varphi}\delta\left(\mathcal{L}_{n}^{+}-\mathcal{L}_{n}^{-}\right)\right],\label{bdrytermvirasoro}
\end{equation}
where the $\mathcal{L}_{n}^{\pm}$ are build out of the $g_{ij}$
and $\pi^{ij}$, and where at infinity, the ``Killing lapse'' $N$
tends to a function of $\varphi$ whose Fourier components are $N_{n}$,
and similarly for $N^{\varphi}$. Furthermore, it is shown that, in
terms of the Poisson bracket, the $\mathcal{L}_{n}^{\pm}$ are two
independent copies of the Virasoro algebra. The expression (\ref{bdrytermvirasoro})
for the boundary term shows that the most general permissible motion,
is obtained when for large $r$ both $N$ and $N^{\varphi}$ tend
to arbitrary functions of $\varphi$. This motion is not a gauge transformation,
but it is a global symmetry transformation at infinity.

In the ``rest frame'' of the black hole, the only surviving mode
of ${\mathcal{L}}^{\pm}$ is the zero mode. Now, $N\left(\infty\right)=\frac{1}{2\pi}N_{0}$
corresponds to making a displacement in ``proper Killing time''
of magnitude $N\left(\infty\right)\left(t_{2}-t_{1}\right)$. The
corresponding generator then deserves to be called the negative of
the mass, and similarly $N^{\varphi}\left(\infty\right)=\frac{1}{2\pi}N_{0}^{\varphi}$
is a spatial rotation of magnitude $N^{\varphi}\left(\infty\right)\left(t_{2}-t_{1}\right)$,
and the corresponding generator deserves to be called the angular
momentum. Indeed one finds for the black hole metric (\ref{BTZLor})
\begin{align}
\textrm{(Rest mass)} & :=\mathcal{M_{\text{Lor}}}=\frac{1}{\ell}\left(\mathcal{L}_{0}^{+}+\mathcal{L}_{0}^{-}\right)=\frac{M_{\text{Lor}}}{8G},\nonumber \\
\text{(Angular momentum)} & :=\mathcal{J}_{\text{Lor}}=\mathcal{L}_{0}^{+}-\mathcal{L}_{0}^{-}=\frac{J_{\text{Lor}}}{8G}.\label{mass-angularmomBTZ}
\end{align}

\subsubsection{Thermodynamics}

When studying black hole thermodynamics we will assume that we are
in that ``rest frame''. As indicated in the main text, there is
no more loss of generality in doing this than the one incurred in
if one studies the thermodynamics of a gas in a box assuming that
the box is at rest. Thus, the only extensive parameters present will
be the mass and the angular momentum. To construct the thermodynamics
one may work in any ensemble. If one chooses to work in the microcanonical
ensemble, then one fixes at infinity those extensive parameters. If
they are fixed, the variation (\ref{bdrytermvirasoro}) vanishes,
and so thus its Euclidean continuation. The surface term $B_{\infty}$
in (\ref{AmasBinfty}) then vanishes, and as already stated, the Euclidean
action on-shell is equal to the entropy given by $S=\frac{1}{4}A$.
All the thermodynamics is captured once one expresses the area $A$
in terms of the mass and the angular momentum. From (\ref{BTZLor})
it follows that 
\begin{equation}
r_{+}=\ell\sqrt{\frac{M_{\text{Lor}}}{2}}\left[1+\left(1-\frac{J_{\text{Lor}}^{2}}{M_{\text{Lor}}^{2}\ell^{2}}\right)^{\frac{1}{2}}\right]^{\frac{1}{2}}=2\ell\sqrt{G\mathcal{M}_{\text{Lor}}}\left[1+\left(1-\frac{\mathcal{J}_{\text{Lor}}^{2}}{\mathcal{M}_{\text{Lor}}^{2}\ell^{2}}\right)^{\frac{1}{2}}\right]^{\frac{1}{2}},
\end{equation}
and therefore 
\begin{align}
S & =\pi\ell\sqrt{\frac{\mathcal{M}_{\text{Lor}}}{G}}\left[1+\left(1-\frac{\mathcal{J}_{\text{Lor}}^{2}}{\mathcal{M}_{\text{Lor}}^{2}\ell^{2}}\right)^{\frac{1}{2}}\right]^{\frac{1}{2}}.\label{entropymj}
\end{align}
From the entropy (\ref{entropymj}), one may evaluate the inverse
temperature $\beta$ and the chemical potential for the angular momentum
$\mu_{\mathcal{J}_{\text{Lor}}}$ 
\begin{align}
\beta & =\left(\frac{\partial S}{\partial\mathcal{M}_{\text{Lor}}}\right)_{\mathcal{J}_{\text{Lor}}},\label{eq:betamicrocan}\\
\beta\mu_{\mathcal{J}_{\text{Lor}}} & =-\left(\frac{\partial S}{\partial\mathcal{J}_{\text{Lor}}}\right)_{\mathcal{M}_{\text{Lor}}}.\label{eq:betamumicrocan}
\end{align}
On the other hand, we have from the Euclidean version of (\ref{bdrytermvirasoro})
\begin{align}
\beta & =N\left(\infty\right),\label{eq:betanphi}\\
\beta\mu_{\mathcal{J}} & =-N^{\varphi}\left(\infty\right).
\end{align}

Now, one can determine directly the value of $N\left(\infty\right)$
and $N^{\varphi}\left(\infty\right)$ from the line element (\ref{BTZLor},
\ref{fNNphi}, \ref{MJLor}) through equations (\ref{Nphizero}) and
(\ref{thetaigual2pi}). One needs to bring in the expression for $\Theta$,
which has not been given yet. It reads \cite{Teitelboim_action_and_entropy},
\begin{equation}
\Theta=\frac{1}{2}N\left(\infty\right)\left(f^{2}\right)^{'}\left(r_{+}\right).\label{eq:Theta}
\end{equation}
Both calculations must agree, since they come from the same action
principle. They do, and either way one obtains 
\begin{align}
\beta & =\frac{2\pi r_{+}\ell^{2}}{r_{+}^{2}-r_{-}^{2}},\label{betabtz}\\
\mu_{\mathcal{J}_{\text{Lor}}} & =\frac{r_{-}}{\ell r_{+}}.\label{betamubtz}
\end{align}

We have done this analysis in detail because it shows blatantly that
one must allow for the most general permissible Lagrange multipliers
even if one works in the microcanonical ensemble. The boundary conditions
for both, the canonical variables and the Lagrange multipliers are
part of the definition of the theory, and are given once and for all.
The ensemble one works in is a matter of choice, and the form of the
action must be chosen in tune with the ensemble, but the asymptotic
symmetries remain the same for all ensembles. For example, if one
wants to work in the grand canonical ensemble, one must choose the
boundary term $B_{\infty}$ in (\ref{AmasBinfty}) so that the action
has an extremum when $\beta$ and $\mu_{\mathcal{J}}$ are fixed at
infinity, instead of $\mathcal{M}$ and $\mathcal{J}$. So, the variation
of the canonical action must be cancelled by the variation of $B_{\infty}$.
In that case one must take 
\begin{equation}
B_{\infty}^{\text{Lor}}=\left(t_{2}-t_{1}\right)\left[-N_{\text{Lor}}\left(\infty\right)\mathcal{M}_{\text{Lor}}+N_{\text{Lor}}^{\varphi}\left(\infty\right)\mathcal{J}_{\text{Lor}}\right].\label{Binfty}
\end{equation}
When one performs the Euclidean continuation (\ref{Binfty}) becomes
\begin{align}
B_{\infty} & =-\beta\mathcal{M}+\beta\mu_{\mathcal{J}}\mathcal{J},
\end{align}
which is precisely what is needed to replace the entropy $S$ by the
Gibbs free energy $G$, as it should be the case for the grand canonical
ensemble. One may then go through the same steps as before to derive
the thermodynamics.

\subsection{Conformal weight and $sl\left(2,\mathbb{R}\right)$ spin}

\label{WeightVersusSpin}

For completeness, we recall here a few concepts related to our use
of the terminology ``higher spin''.

\subsubsection{Conformal weight}

A field $\phi(z)$ is defined to have conformal weight or conformal
dimension $J$ if under coordinate transformation $z\rightarrow z'(z)$,
the field $\phi$ transforms as: 
\begin{equation}
\phi'(z')=\left(\frac{dz}{dz'}\right)^{J}\phi(z).
\end{equation}
The metric $g_{zz}$ has conformal weight 2 since $g'_{zz}dz'dz'=g_{zz}dzdz$.
A tensor of rank 3 has conformal weight $3$.

For infinitesimal transformations $z\rightarrow z'=z+\epsilon(z)$,
this implies 
\begin{equation}
\delta\phi=-\epsilon\frac{d\phi}{dz}+J\phi\frac{d\epsilon}{dz},
\end{equation}
or, in terms of Fourier modes and Poisson brackets (and given that
the $L_{n}$'s generate the transformations), 
\begin{equation}
i\{L_{m},\phi_{n}\}=\left(m(J-1)-n\right)\phi_{m+n}.
\end{equation}

For the $L_{m}$'s themselves, one has this law with $J=2$ but the
bracket is modified in this case by the central charge. For the $W_{m}$'s,
one has $J=3$, 
\begin{equation}
i\{L_{m},W_{n}\}=\left(2m-n\right)W_{m+n},
\end{equation}
and so the $W_{m}$'s have conformal weight $3$.

\subsubsection{Relation with $sl(2,\mathbb{R})$-spin}

The generators $\{L_{-1},L_{0},L_{1}\}$ span $sl(2,\mathbb{R})$.
By the above relations, the generators \\
 $\{W_{-2},W_{-1},W_{0},W_{1},W_{2}\}$ form a representation of $sl(2,\mathbb{R})$.
The finite dimensional irreducible representations of $sl(2,\mathbb{R})$
are characterized by their ``spin'' (like for the compact version
$su(2)$). The dimension $D$ of a representation is related to its
spin $k$ by $D=2k+1$. The $\{W_{-2},W_{-1},W_{0},W_{1},W_{2}\}$
transform in the spin-$2$ representation. There is a shift by one
between the $sl(2,\mathbb{R})$-spin and the conformal weight, 
\begin{equation}
J=k+1.
\end{equation}
This can be understood from the fact that the gauge field $A_{\mu}^{a}$
carries a vector index $\mu$ in addition to the internal index $a$.
This index carries its own spin $1$.

\section{Relationship with previous results}

\setcounter{equation}{0} \label{AppendixB}

\subsection{Permissible gauge transformations}

The subject of generalized black holes in higher-spin three-dimensional
gravity was started in \cite{GK,AGKP}. However, the black hole connections
discussed by these authors -- which we shall term ``GK black holes''
or ``GK connections'' after the initials of the author's last names
-- have angular components $A_{\varphi}^{\pm}$ which fulfill neither
the boundary conditions (\ref{asymptw3phi-1}) of $W_{3}$ gravity
nor the boundary conditions (\ref{aThetaW3(2)}) of $W_{3}^{(2)}$
gravity. Since a theory is defined by equations of motion and boundary
conditions, and since the GK connections satisfy the equations of
motion but not the boundary conditions, one might just take the point
of view that these are simply not solutions of the theory. Rejecting
the connections of \cite{GK,AGKP} on the grounds that they do not
fulfill the boundary conditions (\ref{asymptw3phi-1}) or (\ref{aThetaW3(2)})
might be a bit excessive, however, as one might with the same strict
attitude reject the Schwarzschild solution written in polar coordinates
on the grounds that it does not fulfill the asymptotically flat boundary
conditions written in cartesian coordinates. So one might ask the
question: can the GK connections be made to fulfill the boundary conditions
(\ref{asymptw3phi-1}) or (\ref{aThetaW3(2)}) by a permissible change
of gauge?

The issue is a bit subtle because any singularity-free flat connection
on the solid torus with given holonomies is related to any other one
with the same holonomies (up to conjugation) by a gauge transformation.
What makes a gauge transformation permissible in this context?. The
criterion for admissibility is that \emph{the gauge transformation
should not interfere with the asymptotic algebra}. In order for this
to happen, the gauge transformation should be independent of the asymptotic
charges. This is because then the gauge transformations commute with
the variation of the charges and, under it, the image of the variation
is the variation of the image. This admissibility criterion certainly
holds for the analogy with the Schwarzschild metric mentioned above,
because the passage from cartesian to spherical coordinates is independent
of the mass.

\subsection{The GK black hole is a $W_{3}^{\left(2\right)}$ black hole}

We now pass to show that the GK connection can be brought asymptotically
to the $W_{3}^{(2)}$ form (\ref{aThetaW3(2)})-(\ref{eq:thetar})
by a permissible gauge transformation, and so it should be thought
of as a black hole belonging to the diagonal embedding family investigated
in section \ref{W32-section}.

The gauge transformation is constructed directly by demanding 
\begin{equation}
A^{GK}=g^{-1}A^{\text{diag-emb}}g+g^{-1}dg,\label{eq:FromW32toGK}
\end{equation}
for each of the two copies $A^{\pm}$. Here, $A^{\text{diag-emb}}$
is given by our eqs. (\ref{aThetaW3(2)})-(\ref{eq:thetar}), and
$A^{GK}$ is given in \cite{GK}.

One finds 
\begin{equation}
g=e^{\lambda},\label{eq:gfromw32togk}
\end{equation}
with 
\begin{equation}
\lambda^{\pm}=\pm\frac{1}{2}\log\left(4\mu_{\pm}\right)\left(L_{0}\mp\frac{2\mu_{\pm}+\sqrt{\mu_{\pm}}}{2\mu_{\pm}\left(1-4\mu_{\pm}\right)}\left[W_{\mp1}\mp\left(\partial_{\pm}\mu_{\pm}\right)L_{\mp1}\right]+\frac{\mu_{\pm}^{\prime}}{4\mu_{\pm}\left(1-4\mu_{\pm}\right)}W_{\mp2}\right)\ .\label{gtw32-1}
\end{equation}
 The gauge transformation (\ref{eq:FromW32toGK})-(\ref{gtw32-1})
depends only on the parameters $\mu_{\pm}$ appearing in the GK connection,
and it is independent of the charges, and it is thus permissible%
\footnote{When (\ref{eq:gfromw32togk}) and (\ref{gtw32-1}) are inserted in
the right-hand side of (\ref{eq:FromW32toGK}) one finds that the
$A^{\text{GK}}$ appearing in (\ref{eq:FromW32toGK}) differs from
the one given in eqs. (3.2) and (3.3) of ref \cite{GK} by $t^{\text{here}}=2\left(\hat{\xi}_{+}+\hat{\xi}_{-}\right)^{-1}\frac{t^{\text{there}}}{\ell}$,
$\varphi^{\text{here}}=-\left(\varphi^{\text{there}}+\frac{\hat{\xi}_{+}-\hat{\xi}_{-}}{\hat{\xi}_{+}+\hat{\xi}_{-}}\frac{t^{\text{there}}}{\ell}\right)$,
and $\mu_{-}^{\text{here}}=-\mu_{-}^{\text{there}}$. The multipliers
$\hat{\xi}_{\pm}$ are taken to be independent of $t$ and $\varphi$.%
}.

The relationship between the $W_{3}^{\left(2\right)}$ charges and
the ``charges'' appearing in the GK solution is given by 
\begin{align}
\mathcal{U}^{\pm} & =\pm\frac{2}{3}\mu_{\pm}\mathcal{\tilde{L}}^{\pm}\mp\frac{k}{16\pi}\left[\mu_{\pm}^{-1}-\mu_{\pm}^{-1}\left(\partial_{\pm}\mu_{\pm}\right)^{2}+\frac{4}{3}\partial_{\pm}^{2}\mu_{\pm}\right]\ ,\label{U-GK-1}\\
\mathcal{\hat{L}}^{\pm} & =\mathcal{\tilde{L}}^{\pm}\pm3\mu_{\pm}\mathcal{\tilde{W}}^{\pm}-\frac{2}{3}\mu_{\pm}^{2}\left(\partial_{\pm}^{2}\mathcal{\tilde{L}}^{\pm}-\frac{16\pi}{k}\left(\mathcal{\tilde{L}}^{\pm}\right)^{2}\right)+\frac{5}{3}\mathcal{\tilde{L}}^{\pm}\left(\partial_{\pm}\mu_{\pm}\right)^{2}-\frac{5}{6}\left(\partial_{\pm}\mathcal{\tilde{L}}^{\pm}\right)\left(\partial_{\pm}\mu_{\pm}^{2}\right)\nonumber \\
 & -\frac{10}{3}\mu_{\pm}\mathcal{\tilde{L}}^{\pm}\partial_{\pm}^{2}\mu_{\pm}+\frac{k}{24\pi}\left[\left(\partial_{\pm}^{2}\mu_{\pm}\right)^{2}-2\left(\partial_{\pm}\mu_{\pm}\right)\left(\partial_{\pm}^{3}\mu_{\pm}\right)+2\mu_{\pm}\left(\partial_{\pm}^{4}\mu_{\pm}\right)\right.\nonumber \\
 & \left.+\frac{9}{4}\mu_{\pm}^{-2}\mu_{\pm}^{\prime2}-\frac{3}{2}\mu_{\pm}^{-1}\left(\partial_{\mp}^{2}\mu_{\pm}+2\partial_{\pm}^{2}\mu_{\pm}-3\partial_{-}\partial_{+}\mu_{\pm}\right)\right]\ ,\label{L-GK-1}
\end{align}
and 
\begin{align}
\psi_{\left[a\right]}^{+} & =\frac{a}{3}\sqrt{2\mu_{+}}\left[\hat{\mathcal{L}}^{+}-\frac{24\pi}{k}\left(\mathcal{U}^{+}\right)^{2}+\frac{3k}{32\pi}\mu_{+}^{-2}\mu_{+}^{\prime}\left(\ell\dot{\mu}_{+}-2a\right)-\frac{k\ell}{16\pi}\mu_{+}^{-1}\dot{\mu}_{+}^{\prime}\right]\ ,\label{spinor GK+-1}\\
\psi_{\left[a\right]}^{-} & =-\frac{1}{3}\sqrt{2\mu_{-}}\left[\hat{\mathcal{L}}^{-}-\frac{24\pi}{k}\left(\mathcal{U}^{-}\right)^{2}-\frac{3k}{32\pi}\mu_{-}^{-2}\mu_{-}^{\prime}\left(\ell\dot{\mu}_{-}-2a\right)+\frac{k\ell}{16\pi}\mu_{-}^{-1}\dot{\mu}_{-}^{\prime}\right]\ .\label{spinor GK--1}
\end{align}

Note that one can produce the eight independent charges of $W_{3}^{\left(2\right)}$
out of the four $\mathcal{\tilde{L}}^{\pm},\mathcal{\tilde{W}}^{\pm}$
because the right hand side of (\ref{L-GK-1}) contains one first
time derivative and second time derivative of $\mathcal{\tilde{L}}^{\pm}$
for each copy the algebra thus providing the right number of independent
initial data.

The GK Lagrange multipliers correspond to the particular case, 
\begin{equation}
\nu_{\pm}:=\mp\frac{3}{2}\left(\hat{\xi}_{+}+\hat{\xi}_{-}\right)\mu_{\pm}^{-1}\ \ ,\ \ \vartheta_{\left[a\right]}^{+}:=\frac{\hat{\xi}_{+}+\hat{\xi}_{-}}{\sqrt{2\mu_{+}}}\ \ ,\vartheta_{\left[a\right]}^{-}:=-a\frac{\hat{\xi}_{+}+\hat{\xi}_{-}}{\sqrt{2\mu_{-}}}\,\,.\label{CP-GK}
\end{equation}

It follows from this analysis that the GK black hole solution, for
which $\mu_{\pm}$, $\mathcal{\tilde{L}}_{\pm}$ and $\mathcal{\tilde{W}}_{\pm}$
are assumed to be constants, corresponds to a particular case of the
one described in section \ref{W32-section}, whose Euclidean version
is given by eqs. (\ref{eq:aphiw32}), (\ref{eq:atw32}) and (\ref{OnShellConds-W32})
. This means that the GK black holes do not carry fundamental higher
spin charges, but instead, according to (\ref{eq:Binftyw32}), that
they are endowed with spin-$2$, spin-$\frac{3}{2}$, and $U\left(1\right)$
spin-$1$ charges given by 
\begin{align}
\mathcal{\hat{L}}^{\pm} & =\mathcal{\tilde{L}}^{\pm}\pm3\mu_{\pm}\mathcal{\tilde{W}}^{\pm}+\frac{32\pi}{3k}\mu_{\pm}^{2}\left(\mathcal{\tilde{L}}^{\pm}\right)^{2}\ ,\nonumber \\
\psi_{\left[a\right]}^{+} & =a\sqrt{2\mu_{+}}\left(\mathcal{\tilde{L}}^{+}+\mu_{+}\mathcal{\tilde{W}}^{+}-\frac{k}{32\pi\mu_{+}^{2}}\right)\ ,\nonumber \\
\psi_{\left[a\right]}^{-} & =-\sqrt{2\mu_{-}}\left(\mathcal{\tilde{L}}^{-}-\mu_{-}\mathcal{\tilde{W}}^{-}-\frac{k}{32\pi\mu_{-}^{2}}\right)\ ,\label{GK-BH-Charges}\\
\mathcal{U}^{\pm} & =\pm\frac{2}{3}\mu_{\pm}\mathcal{\tilde{L}}^{\pm}\mp\frac{k}{16\pi\mu_{\pm}}\ ,\nonumber 
\end{align}
respectively.

\subsection{Entropy paradox resolved}

The following paradox has appeared in the literature in connection
with the entropy of the GK black hole: depending on the method of
evaluation, two different results for it have been proposed. The paradox
already appears in the simplest case of a static GK black hole for
which the Euclidean charges $\mathcal{\tilde{L}}$ and $\mathcal{\tilde{W}}$
are real. In that case, the two conflicting proposals, each of which
have been endorsed by a number of authors \cite{GK,AGKP,KP,Gaberdiel:2012yb,PTT1,CFPT2,PTT2,David:2012iu,deBoer:2013gz,KU,dBJ2,ACI,CS,CJS,PTTreview}
read: 
\begin{equation}
S_{1}^{GK}=4\pi\sqrt{2\pi k\mathcal{\tilde{L}}}\sqrt{1-\frac{3}{4C}}\left(1-\frac{3}{2C}\right)^{-1}\ \quad\text{(correct)},\label{eq:scorrect}
\end{equation}
and 
\begin{equation}
S_{2}^{GK}=4\pi\sqrt{2\pi k\mathcal{\tilde{L}}}\sqrt{1-\frac{3}{4C}}\ \quad\text{(incorrect)},\label{eq:sincorrect}
\end{equation}
where $C$ is defined through 
\begin{equation}
\mathcal{\tilde{W}}=\sqrt{\frac{32\pi}{k}\mathcal{\tilde{L}}^{3}}\frac{C-1}{C^{3/2}}.\label{eq:C}
\end{equation}

We shall now elucidate how the paradox arose, and establish that,
as anticipated above, the first proposal (\ref{eq:scorrect}) is correct,
while the second one (\ref{eq:sincorrect}) is incorrect.

The key fact, that has been established in the present article is
that the GK black hole is a $W_{3}^{\left(2\right)}$ black hole.
Its entropy in terms of the $W_{3}^{\left(2\right)}$ charges of spin
2, 3/2 and 1 has been exhibited in eq. (\ref{eq:SW32}). That entropy
yields (\ref{eq:scorrect}) after the following steps: (i) Take for
$\mu_{\pm}$ in (\ref{GK-BH-Charges}) the value 
\begin{equation}
\mu_{+}=\mu_{-}=\frac{3}{4}\sqrt{\frac{kC}{2\pi\mathcal{\tilde{L}}}}\frac{1}{2C-3},
\end{equation}
which follows from the regularity conditions for the $W_{3}^{\left(2\right)}$-case
and the relationship (\ref{CP-GK}) between the chemical potentials,
and actually agrees with ref. \cite{GK} itself. (ii) Express the
$W_{3}^{\left(2\right)}$ charges in terms of $\mathcal{\tilde{L}}$
and $C$ through (\ref{GK-BH-Charges}).

How did $S_{2}^{GK}$ arise then? It was obtained through integration
of the first law of thermodynamics starting from the expression of
the inverse temperature and the chemical potentials obtained from
the regularity condition, \emph{assuming that $\mathcal{\tilde{L}}$
and $\mathcal{\tilde{W}}$ were fundamental charges. }But, -when expressed
in terms of \emph{$\mathcal{\tilde{L}}$} and \emph{$\mathcal{\tilde{W}}$}
and upon use of (\ref{CP-GK}) - the expression for $A_{\tau}$, which
is what is involved in the regularity condition, is the same for the
$W_{3}$ and $W_{3}^{\left(2\right)}$ black holes. Therefore, what
was being calculated was in effect the entropy of the $W_{3}$ black
hole discussed in section \ref{HS} %
\footnote{In the present discussion we have expressed $S_{1}^{GK}$ and $S_{2}^{GK}$
in terms of the variable $C$ that was employed in the original literature
on the subject. In order to compare with equation (\ref{eq:entropyspin3}),
one must use (\ref{eq:C}), and ``remove the tildes''.%
}, which is indeed given by (\ref{eq:sincorrect}), rather than that
of the GK black hole which is a particular case of a $W_{3}^{\left(2\right)}$
black hole. The trap was that a calculation solely based in $A_{\tau}$
could not put in evidence the fact, that can only be revealed by $A_{\varphi}$,
that $\mathcal{\tilde{L}}$ and $\mathcal{\tilde{W}}$ were not fundamental
spin 2 and spin 3 charges but were ``composite charges'' made out
from charges of spin 2, 3/2 and 1.

\subsection{Further comments}

The following comments are in order:

(i) If one performs the direct analysis of the asymptotic symmetries
of the GK connections, without implementing the gauge transformation
(\ref{eq:gfromw32togk}), (\ref{gtw32-1}) that brings them into the
diagonal embedding boundary conditions (\ref{aThetaW3(2)})-(\ref{eq:ArW32}),
one finds, as one should, that these asymptotic symmetries form a
$W_{3}^{(2)}$-algebra in each $\pm$ sector. The calculation is somewhat
cumbersome and, for the sake of brevity it will not be reported here.

(ii) If one tries to match the asymptotic conditions in \cite{GK}
with the ones appropriate to the principal embedding, direct calculation
shows that the group elements $g_{\pm}$ necessary to achieve the
transformation depend on the charges. As explained above, this is
not allowed. Note that, a fortiori, the gauge transformation whose
existence is argued in \cite{Banados:2012ue} would necessarily have
this same impediment.

(iii) The solutions considered in \cite{CM} can also be mapped on
another particular case of our general form (\ref{eq:aphiw32}) by
means of a permissible gauge transformation.

(iv) Some of the preceding issues were dwelled upon in the work \cite{CS}.
It was asserted there that the GK asymptotic conditions could be viewed
as possessing both $W_{3}^{(2)}$ and $W_{3}$ symmetries. According
to the analysis herein they possess only $W_{3}^{(2)}$. One could
perhaps imagine that composite $W_{3}$ charges might be constructed
out of the $W_{3}^{(2)}$ generators by, for example, combining two
spins 3/2 to form a spin 3. Whether a construction of such sort could
be realized through a gauge transformation depending on the charges,
or by a some other mechanism, remains at present pure speculation.

\end{document}